\documentclass[11pt]{article}

    \usepackage[dvipsnames]{xcolor}
    \usepackage[T1]{fontenc}
	\usepackage[utf8]{inputenc}
	\usepackage[english]{babel}
	\usepackage{amsfonts}
	\usepackage{amssymb}
        \usepackage{dsfont}
	    \usepackage{amsmath}
	
	\usepackage{enumerate}
	\usepackage{bbm}
	\usepackage{mathtools}
	\usepackage{booktabs}	
	\usepackage{setspace}
	\usepackage{graphicx}
	\usepackage{fullpage}
	
	\usepackage{soul}
	\usepackage{xspace}
	\usepackage{csquotes}
	\usepackage{placeins}

	\usepackage{lscape}
	\usepackage{longtable}
	\usepackage{mathabx}
    \usepackage{accents}
    \usepackage{float}
    \usepackage{afterpage}
    \usepackage{listings}

    \usepackage{tikz}
    \usetikzlibrary{calc,intersections}
    \usepackage{appendix}
    \usepackage{algorithm}
    \usepackage{algpseudocode}
     \usepackage{adjustbox}

	\usepackage{natbib}

\usepackage{geometry}
\usepackage{caption} 
\usepackage{subcaption}

\usepackage{pgfplotstable}

\usepackage[explicit]{titlesec}

\titleformat{\section}
  {\normalfont  \fontsize{16}{14} \bf}
  {\thesection \,}
  {0.25em}
  {#1}
\titleformat{\subsection}
  {\normalfont \fontsize{14}{12} \bf}
  {\thesubsection}
  {1em}
  {#1}

\titleformat{\subsubsection}
  {\normalfont \fontsize{12}{10} \bf}
  {\thesubsubsection}
  {1em}
  {#1}

	\usepackage{hyperref}
	\hypersetup{
    colorlinks=true,
    linkcolor=Red,
    filecolor=magenta,      
    urlcolor=blue,
    citecolor = Blue
}
	\usepackage{cleveref}

    \usepackage{rotating}
	\usepackage[framemethod=tikz]{mdframed}
	\usepackage{todonotes}
	\presetkeys{todonotes}{color=black!5,inline}{} 

    \usepackage{tcolorbox}
    \newtcbox{\feedback}{nobeforeafter,colframe=black,colback=white,boxrule=0.5pt,arc=2pt,
      boxsep=0pt,left=2pt,right=2pt,top=2pt,bottom=2pt,tcbox raise base}
      
    \usepackage{amsthm}

    \theoremstyle{definition}

\usepackage{array}
\newcolumntype{L}[1]{>{\raggedright\let\newline\\\arraybackslash}m{#1}}
\newcolumntype{C}[1]{>{\centering\let\newline\\\arraybackslash\hspace{0pt}}m{#1}}
\newcolumntype{R}[1]{>{\raggedleft\let\newline\\\arraybackslash\hspace{0pt}}m{#1}}

\usepackage{ragged2e}
\newlength\ubwidth

\def\eea{\end{eqnarray*}}
\def\bea{\begin{eqnarray*}}
\renewcommand{\[}{\begin{equation}}
\renewcommand{\]}{\end{equation}}
\newcommand{\bi}{\begin{itemize}}
\newcommand{\ei}{\end{itemize}}

\pgfplotsset{compat=1.15}

\usepackage{comment}

\newcommand{\sym}[1]{$^{#1}$}

\newcommand{\tablesize}{\footnotesize}
\newenvironment{tablebox}
  {\tablesize\begin{adjustbox}{max width=\textwidth, max totalheight=0.75\textheight}}
  {\end{adjustbox}}

\newcommand{\figurenotes}[1]{%
  {\footnotesize\linespread{0.95}\selectfont\raggedright\textit{Notes}:\ #1\par}%
  \addvspace{12pt plus 2pt}%
}
\newcommand{\tablenotes}[1]{%
  {\footnotesize\linespread{0.95}\selectfont\textit{Notes}:\ #1\par}%
  \addvspace{12pt plus 2pt}%
}

\begin{document}
\thispagestyle{empty}

\vspace{20pt}
\begin{center}
\textcolor{white}{fill}\\
\vspace{20pt}

\LARGE{AI-Assisted Programming and Labor Demand$^{*}$}\\\mbox{ }
\end{center}

\begin{singlespace}
\begin{center}
\begin{tabular}{ccccccc}
\large{Lindsey Raymond} & \mbox{ } & \mbox{ } & \large{Eleanor Dillon} \\
\large{MIT } &  \mbox{ } & \mbox{ } & \large{Microsoft Research} \\
\end{tabular}
\end{center}

\end{singlespace}

\begin{center}
\vspace{.20in}
\normalsize{November 6, 2026}

\vspace{10pt}
\end{center}
\vspace{10pt}

\begin{abstract}
\vspace{-11pt}
\singlespacing
\noindent

Generative AI has demonstrated human-level performance on technical work tasks, yet evidence on how these task-level capabilities are changing labor demand remains scarce. We study the firm-level adoption of GitHub Copilot, an AI coding assistant, across 5,414 multinational firms accounting for 38\% of the U.S. programming workforce. To identify causal effects, we exploit a natural experiment generated by firms' predetermined software procurement cycles. Early AI-adopting firms hired more workers before the launch of GitHub Copilot and had larger, more technical, and older workforces. After one year, GitHub Copilot adoption reduces quarterly technical hiring by 29\% among review-induced adopters, with no effect on separations, overall technical headcount, or vacancy duration. The hiring slowdown is concentrated in the subset of junior technical roles that involve the highest share of coding work. Overall, AI coding assistance causes firms to shift demand towards more experienced workers and roles that pair some coding with broader design responsibilities.
\end{abstract}


\bigskip

\hspace{-25pt}\rule[0.1ex]{0.33\textwidth}{0.2mm}\vspace{0.05in}\\\indent
\begin{singlespace}
\vspace{-25pt}
\noindent \scriptsize{\mbox{  }\mbox{  }\mbox{  }\mbox{  }\mbox{}$^{*}$Correspondence to lraymond@mit.edu and eleanor.dillon@microsoft.com.
We are enormously grateful to Lisa Simon, Brian Xu, Kevin Xu, and Ben Zweig for invaluable data assistance, to Bee Xiong for helpful conversations about the GitHub sales process. We thank Daron Acemoglu, Isaiah Andrews, David Autor, Erik Brynjolfsson, Luis Garicano, Jon Gruber, Namrata Kala, Danielle Li, Sendhil Mullainathan, Ashesh Rambachan, David Reshef and Anna Russo for helpful comments. Will Wang provided excellent research assistance. Raymond acknowledges support from the Schmidt Sciences AI2050 Fellowship and the Microsoft Research Fellowship. The content is solely the responsibility of the authors and does not necessarily represent the official views of Microsoft Research or the Massachusetts Institute of Technology.}
\end{singlespace}

\thispagestyle{empty}

 \setcounter{page}{0}

\clearpage

\newpage

\onehalfspacing

 \newpage
 \pagebreak

\section{Introduction}
\label{section:intro}

LLMs now rival or exceed human performance on some tasks in mathematics, legal and financial analysis, and software development \citep{rein2023gpqagraduatelevelgoogleproofqa, luong2025robustmathematicalreasoning, guha2023legalbenchcollaborativelybuiltbenchmark}. Progress has been especially rapid in programming, where AI coding assistants can generate, explain, test, and debug code. As organizations incorporate these task-level improvements and redesign workflows, firms may reduce demand for human work that can now be done more effectively by machines while expanding opportunities for human work that complements AI. Understanding where these changes will be felt helps workers, firms, and policymakers adjust efficiently. However, the mapping from task-level technical capabilities to role- and firm-level demand changes is difficult to predict and difficult to measure. Field experiments can randomize access to AI across workers, but equivalent experiments across firms are generally infeasible. Absent randomization, firms that adopt AI early may already follow different growth trajectories or employ distinct management practices. Observational studies cannot reliably separate this endogenous choice by firms to adopt AI, other concurrent economic shocks, and the direct effects of AI on labor demand.

This paper exploits a natural experiment in firm-level AI adoption to provide causal evidence on how AI coding assistants affect hiring and employment for technical workers. While AI coding assistants are our broader object of study, we focus on GitHub Copilot because it is the setting for which we have data access and a clean empirical design. We study its rollout across 5,414 multinational firms employing 45 million workers worldwide, including 38\% of the U.S. programming workforce. Our identification strategy rests on the centralized and complex nature of enterprise software procurement. In large firms, individual developers cannot independently deploy new software tools; instead, adoption must proceed through a formal procurement process involving dedicated legal, security, compliance, and purchasing personnel. All firms in our sample were GitHub customers using GitHub's enterprise product for code management before GitHub Copilot's January 2023 launch. These enterprise customers follow a regular schedule for contract reviews with GitHub, during which the relevant teams for coordinating contract evaluations and budgetary decisions are assembled. Firms faced a lower organizational cost of adopting GitHub Copilot during these existing GitHub reviews relative to initiating a separate procurement process. Because these schedules were established before GitHub Copilot existed, their timing provides plausibly exogenous variation in the timing of adoption, unrelated to firms' underlying employment trajectories.

One year post-review, treated firms are 20 percentage points more likely to have adopted GitHub Copilot than not-yet-reviewed firms. Using a staggered difference-in-differences design that compares earlier- and later-reviewed firms, we estimate both the first-stage effect of review timing on GitHub Copilot adoption and its reduced-form effect on employment. Under additional assumptions, the Wald ratio of these two estimates identifies the causal effect of GitHub Copilot among firms induced to adopt earlier by the instrument of their pre-scheduled contract reviews.

We link GitHub's administrative records on coding assistant adoption and contract reviews to employment data from Revelio Labs, which tracks firm-level headcount, hiring, and separations by role and seniority. This matched dataset allows us to observe actual enterprise AI adoption rather than relying on indirect proxies or occupational exposure scores. We focus on firms headquartered in the Americas and Europe, where GitHub's enterprise customer base is concentrated and Revelio provides comprehensive coverage. The resulting monthly panel runs from January 2021 through June 2025, which spans two years before and two and a half years after GitHub Copilot's January 2023 enterprise launch. Over this window, GitHub Copilot was the dominant enterprise coding assistant, predating the rollout of the enterprise versions of competing tools such as Cursor and Claude Code.

Within a firm, AI coding assistance affects labor demand through three channels: changes in the scale of software production, the allocation of tasks across roles and seniority levels, and productivity gains on existing tasks. While expanding software production raises overall labor demand, productivity gains on existing tasks reduce the labor required for a given level of output. Changes in how tasks are bundled and allocated across jobs further shape which roles and seniority levels firms require. Our empirical estimates capture the net effect of these potentially offsetting forces on employment, hiring, and separations.

Because we observe GitHub Copilot take up directly, we can cleanly identify the characteristics of early AI adopters and non-adopters. Consistent with \citet{acemoglu2023advanced}, early AI-adopting firms in our sample hired more workers each month before the launch of GitHub Copilot. They were also larger overall with a higher share of technical workers and a lower share of junior workers. These compositional differences illustrate the econometric challenge of separating the drivers of AI adoption from the effects of adoption. Our data also highlight that generative AI tools arrived during a tumultuous period of adjustment for technical workers that affected both adopting and non-adopting firms. From January, 2022 to December, 2024, monthly mean hiring rates among firms that never adopted GitHub Copilot during our sample decreased from 2.7\% to 1.8\%.

Using our procurement instrument, we find that GitHub Copilot adoption reduces technical hiring without increasing separations. One year after scheduled contract review, monthly hiring for technical workers, those who build, test, deploy, secure, or support software systems, is 7.6\% lower than at firms whose reviews have not yet occurred. The corresponding Wald estimates imply a 29\% reduction in quarterly technical hiring among complying firms that were induced to adopt GitHub Copilot earlier by their scheduled contract reviews. From a pre-period baseline of roughly 55 quarterly technical hires per firm, the reduced-form and IV estimates imply hiring slows to about 51 and 39 workers per quarter, respectively. Against an average firm workforce of 8,000 employees, these reductions represent less than 0.2\% of headcount per quarter. By contrast, we find no meaningful change in separations, suggesting that layoffs have not been a meaningful tool of adjustment for these firms. We also find no change in overall technical headcount over the first year, consistent with the small scale of the hiring changes relative to the stock of workers at these established firms.

Technical job postings also fall while posting durations remain unchanged, suggesting that the drop in hiring reflects lower labor demand rather than reduced matching efficiency. Several placebo and robustness checks weigh against the possibility that this decline is driven by corporate budgeting cycles, broader hiring freezes or AI tools adoption. Scheduled contract reviews have no effect on technical hiring prior to GitHub Copilot's launch or post-launch hiring for non-technical professionals such as accountants, and yield similar estimates when controlling for firm-size and industry trends.

Decomposing this decline in overall technical hiring, we find GitHub Copilot reduces hiring primarily for junior workers in code-intensive roles. We measure each role's coding intensity using pre-GitHub Copilot job descriptions and distinguish between junior and experienced workers. Neither coding intensity nor seniority alone predicts the hiring response: more code-intensive roles do not experience larger overall hiring declines, and junior hiring does not fall uniformly across all technical roles. Rather, the hiring decline is concentrated at the intersection. Four quarters after contract review, a one-standard-deviation increase in coding intensity is associated with 6.8\% lower quarterly hiring among junior workers, with no comparable gradient among experienced workers. Comparing the two ends of the coding-intensity distribution illustrates this pattern: among compliers, quarterly hiring of junior software engineers falls by roughly 40\%, whereas hiring of experienced software engineers and of systems and infrastructure engineers is unaffected. 

This contraction in entry-level coding positions reshapes both where junior workers enter firms and the seniority mix of technical teams. Before GitHub Copilot's launch, 72\% of junior technical hires entered code-intensive roles; our estimates imply that adoption reduces this share to 52\%, leaving entry-level hiring nearly evenly split between more and less code-intensive roles. 
This shift away from entry-level workers is striking given experimental evidence that novices realize the largest task-level productivity gains from generative AI, demonstrating that larger individual productivity effects need not translate into greater labor demand.

Although the volume of experienced software engineering hires remains stable, firms raise the bar on their qualifications and firm-specific tenure. Three to four quarters after contract review, new hires into experienced software engineering positions have roughly 1.3 additional years of potential experience and are 4.7 percentage points more likely to hold a master’s or doctoral degree. Experienced hiring also shifts toward workers with firm-specific human capital: external hires into experienced software engineering roles fall by approximately 13\%, whereas internal fills remain unchanged. Among junior software engineering hires, by contrast, observable education, experience, and internal sourcing do not detectably change. 

Finally, we examine whether GitHub Copilot alters the geographic distribution of hiring between firms' home countries and India, their primary offshore engineering hub. In principle, AI coding tools could make lower-cost offshore talent more attractive by bridging time-zone and communication gaps and helping remote developers get up to speed on proprietary systems. Conversely, if offshore teams are disproportionately assigned well-defined implementation work, their roles may be more directly exposed to AI substitution. Empirically, we find no evidence that adoption shifts hiring toward India; rather, the point estimates point to a somewhat steeper slowdown in offshore hiring than at headquarters.

Together, our results show that AI coding assistance is already reshaping technical labor demand, with adjustments emerging within a year. Even within technical roles, the early impacts of AI fall unevenly, with marked decreases in hiring in the most code-intensive junior roles but no clear changes in hiring for other technical junior roles. For experienced workers, the impacts show up in the demand for specific skills rather than a change in heads hired.  Two considerations are important when interpreting these findings. First, our estimates capture firm-level employment responses rather than economy-wide equilibrium effects. Reduced entry-level software engineering hiring at large adopting firms need not translate into lower aggregate technical employment, as displaced candidates may find work in other technical occupations or at smaller employers \citep{raymondjmp, acemogluRobotsJobsEvidence2020}. Conversely, if adopting firms expand output at the expense of competitors, our estimates may capture business-stealing effects that do not translate into aggregate employment gains, potentially making market-wide effects more negative than the firm-level estimates \citep{acemoglu_competing_2020}. Second, our estimates capture an initial adjustment period rather than a long-run steady state. As firms continue to understand the potential of AI tools, both to automate tasks and to support new work, hiring in different areas of technical work could continue to adjust in either direction.

This paper makes three contributions. First, we introduce a portable empirical strategy for identifying the firm-level causal effects of rapidly evolving technologies like enterprise AI. Because AI capabilities advance continuously across new tools and model upgrades, any single empirical study captures a specific product generation, such as our analysis of GitHub Copilot between 2023 and mid-2025. However, our identification strategy is not specific to a single tool or release. It exploits a pervasive feature of corporate software purchasing: enterprise vendors and AI providers routinely sell new products, capability upgrades, and expanded token budgets through existing customer contracts that follow predetermined review schedules. These recurring procurement cycles therefore provide a generalizable research design that researchers can use to provide causal estimates in successive waves of AI across firms, occupations, and industries.

Second, we provide the first credible causal evidence on how firm-level adoption of generative AI affects hiring, separations, and workforce composition. Existing studies report conflicting findings, documenting reduced demand for AI-exposed workers, null employment effects, or faster growth.\footnote{See \citet{brynjolfsson2025canaries, humlum2025large, babina2024artificial, acemoglu2022artificial, westby2026impact, chen_stratton_2026, yotzov_firm_2026, fairlie_wu_2026, hampole_artificial_2025, chandar_how_2026, baird2024early, kharazian2026new, hosseinimaasoum2025generative, tucker_youre_2026, kleinteeselink_generative_2025, daniotti_who_2026, dey_ai_exposure_2026, lodefalk2026same}.} Our analysis shows how two distinct identification challenges may explain this divergence. First, studies comparing occupations by AI exposure may risk picking up occupation-specific macroeconomic contractions, which our estimates show account for at least half of the post-2022 drop in technical hiring. Second, studies comparing adopting and non-adopting firms may face positive selection into treatment: in our sample, eventual adopters were already growing faster before our AI coding assistant launched. Our research design overcomes both challenges, separating the causal effect of adoption from firm-level selection and aggregate shocks. Combining administrative adoption records with predetermined procurement cycles overcomes both challenges to isolate the causal effect of adoption.

Third, we contribute to the literature on how technology shapes skill demand and workforce adjustment by showing how adjustment to generative AI coding assistants unfolds across hiring, separations, and skill requirements.\footnote{On technology and skill demand, see \citet{rosen1981, katz1992changes, autorComputingInequalityHave1998, autor2003skill, bartel_how_2007, goldin_race_2008, michaelsHasIctPolarized2014, akerman_skill_2015, acemogluRobotsJobsEvidence2020, deming_noray_2020, taniguchiICTCapitalSkillComplementarity2022, autor_expertise_2025}; on technology and the organization of work inside firms, see \citet{BrynHittJEP2000, atheystern2002, BBH2002, garicano_organization_2006, acemoglu2007, bloom2014, garicano2015, hoffman2018, dillon2025shifting, ide_talamas_2025}.} Firms can reduce labor input by slowing hiring or increasing separations, with the margin of adjustment varying across technologies and institutions \citep{abowd_corbel_kramarz_1999, davis_faberman_haltiwanger_2012, carlsson_messina_skans_2021}. Whereas industrial automation in the Netherlands displaced incumbent workers, robot adoption in Germany shifted the burden of adjustment onto young labor market entrants \citep{bessen2025what, dauth2021adjustment, feigenbaum2024answering}. We find that firms' initial response to generative AI more closely resembles the latter pattern: firms reduce external hiring of early-career workers without increasing separations or reducing internal transfers. On the skill margin, our findings show that even when generative AI boosts novice productivity on individual tasks, firm-level labor demand shifts away from entry-level implementation work and toward experienced workers with advanced degrees, greater industry experience, and firm-specific human capital.

\section{Setting: GitHub Copilot for Enterprise}\label{sec:setting}
\subsection{GitHub Enterprise}\label{sec:github_enterprise}

Commercial software development requires large teams of engineers to collaborate on shared, interdependent codebases. When an engineer writes code to update a product feature or internal system, that change cannot simply be released to users immediately. To prevent outages and security breaches, proposed code must first be reviewed by peers, run through automated tests to ensure it works with existing systems, scanned for security vulnerabilities, and approved for release. Large organizations must also satisfy strict security and regulatory standards by controlling which employees can modify live systems, maintaining audit logs of every change, and governing the geographic region where software code is stored and accessed.

GitHub Enterprise, owned by Microsoft and used by 90\% of Fortune 100 companies, is the leading platform for managing this workflow at scale \citep{github_customer_2026, wessel2023githubdevelopmentworkflowautomation, githubenterprise2026}. Importantly, GitHub Enterprise, which we refer to as GitHub, is not an AI tool. Rather, it is the platform firms use to store their code, collaborate on software development, review changes, and manage who has access to it. Focusing on pre-existing GitHub customers therefore isolates established employers with substantial software engineering operations.

\subsection{GitHub Copilot}\label{sec:copilot}
GitHub Copilot for Enterprise is an AI coding assistant powered by large language models (LLMs) that operates directly within a developer's code editor.\footnote{For the remainder of the paper, we refer to GitHub Copilot for Enterprise as ``GitHub Copilot'' to differentiate the coding assistant launched for enterprises in 2023 from other Microsoft Copilot AI tools and other versions of the coding assistant designed for individuals. Appendix Section~\ref{asec:codingtools} provides a brief history of ML-based coding assistants.} Initially powered by OpenAI's Codex, GitHub Copilot later integrated other frontier or customer-specific LLMs \citep{zhao_github_2023, zhao_github_2024, dohmke_github_2024}. GitHub Copilot offered two main features during our study period. In-line completion draws on surrounding code and comments to suggest single lines, entire functions, or unit tests, reducing the need to write boilerplate code or look up syntax (Panel A of Appendix Figure~\ref{afig:copilot_autocomplete}). The chat window provides a natural-language interface to the codebase for understanding the existing codebase, looking up business logic and debugging errors (Panel B). The fact that both types of assistance appear in the same application also reduces the need to switch applications to search documentation, or use email or Slack messages to ask questions, limiting performance losses from task switching \citep{goncalves_collaboration_2011, meyer_work_2017, leroy_attention_2009, kumar2025timewarpgapdevelopers}.

From its January 2023 launch through mid-2025, GitHub Copilot faced little competition in the enterprise market \citep{nadella_microsoft_2025}. In Stack Overflow's 2023 Developer Survey, 56.0\% of respondents regularly used GitHub Copilot, compared with 4.9\% for the main enterprise alternative, Amazon CodeWhisperer \citep{stackoverflow2023developersurvey, wiggers_amazon_2024}. Other prominent competitors launched enterprise tiers only after our sample period ended, including Anthropic's Claude Code in August 2025 and Cursor in October 2025 \citep{anthropic_claude_2025,cursor_introducing_2025}. Observing GitHub Copilot licensing captures the primary margin of enterprise AI coding adoption during our study window, limiting the concern that non-adopting firms were adopting other enterprise AI coding assistants. Even if present, unmeasured adoption in the control group would attenuate our estimates toward zero. Panel~A of Figure~\ref{fig:adoption_and_reviews} shows that sample GitHub Copilot adoption rose gradually after January 2023, reaching 62\% by mid-2025. Once firms adopt, they generally provision licenses across most of their technical workforce rather than restricting access to a small number of developers.

\subsection{Enterprise Procurement Cycles and Contract Review}\label{sec:instrument}

In large firms, individual developers and managers cannot simply download and use new software tools on their own. Because enterprise software often interacts with proprietary systems and sensitive corporate data, companies block employees from installing unapproved software applications on work computers \citep{nist2018framework,weill2004it}. Instead, adopting a new tool must go through a centralized procurement process: legal, security, and procurement teams must vet the vendor, verify that proprietary code and customer data are protected, and negotiate contract terms and pricing. Initiating this process outside of regular purchasing cycles requires incurring organizational frictions and coordination costs. A scheduled contract review already coordinates the legal, security, and finance teams for a planned evaluation. By contrast, an off-cycle purchase requires managers to assemble a separate business justification, wait in legal and security review queues, secure budgetary approval, and schedule meetings with all the relevant decision-makers.

Our identification strategy exploits recurring, pre-scheduled contract reviews built into firms' existing contracts for GitHub's source code software product. During these check-ins, a firm's procurement team and GitHub's sales representatives meet to adjust license counts, evaluate new features, and update contract terms. Even before GitHub Copilot's release, account adjustments were heavily concentrated around these dates. In the pre-period, the share of firms increasing their core GitHub licenses quadrupled from roughly 10\% to 40\% in the two months surrounding a review, while the share reducing licenses rose from 4\% to 14\% (Appendix Figure~\ref{afig:seat_change_magnitudes_mtnr}).\footnote{While firms adjust license counts in both directions, complete platform exits are extremely rare.} This concentration shows that enterprise software procurement is sticky. Even for routine license adjustments on an existing product, whether adding seats or cutting unused ones, firm modifications cluster around scheduled review windows. A natural concern is that these contract reviews coincide with firms' broader budgeting and hiring cycles. However, as we show in Section~\ref{sec:robustness}, scheduled reviews prior to GitHub Copilot's launch had no effect on technical hiring or separations (Appendix Table~\ref{atab:placebo_2022_cycle_q0_q2}).

\begin{figure}[H]
\centering
\caption{GitHub Copilot Adoption and Scheduled Contract Reviews among Sample Firms}
\label{fig:adoption_and_reviews}

\begin{subfigure}[t]{0.42\textwidth}
    \centering
    \includegraphics[width=\textwidth]{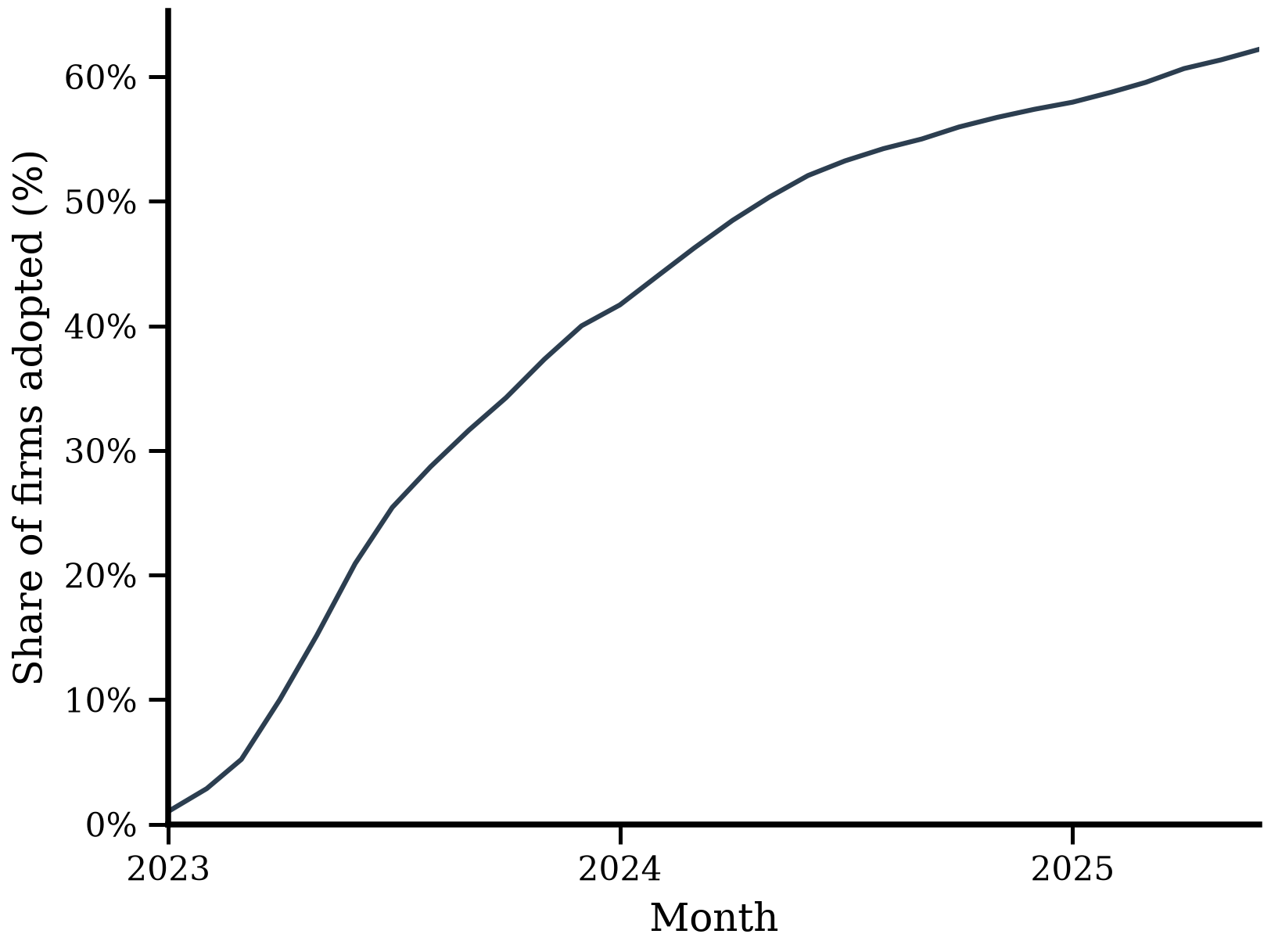}
    \subcaption{Cumulative GitHub Copilot adoption}
\end{subfigure}
\hfill
\begin{subfigure}[t]{0.56\textwidth}
    \centering
    \includegraphics[width=\textwidth]{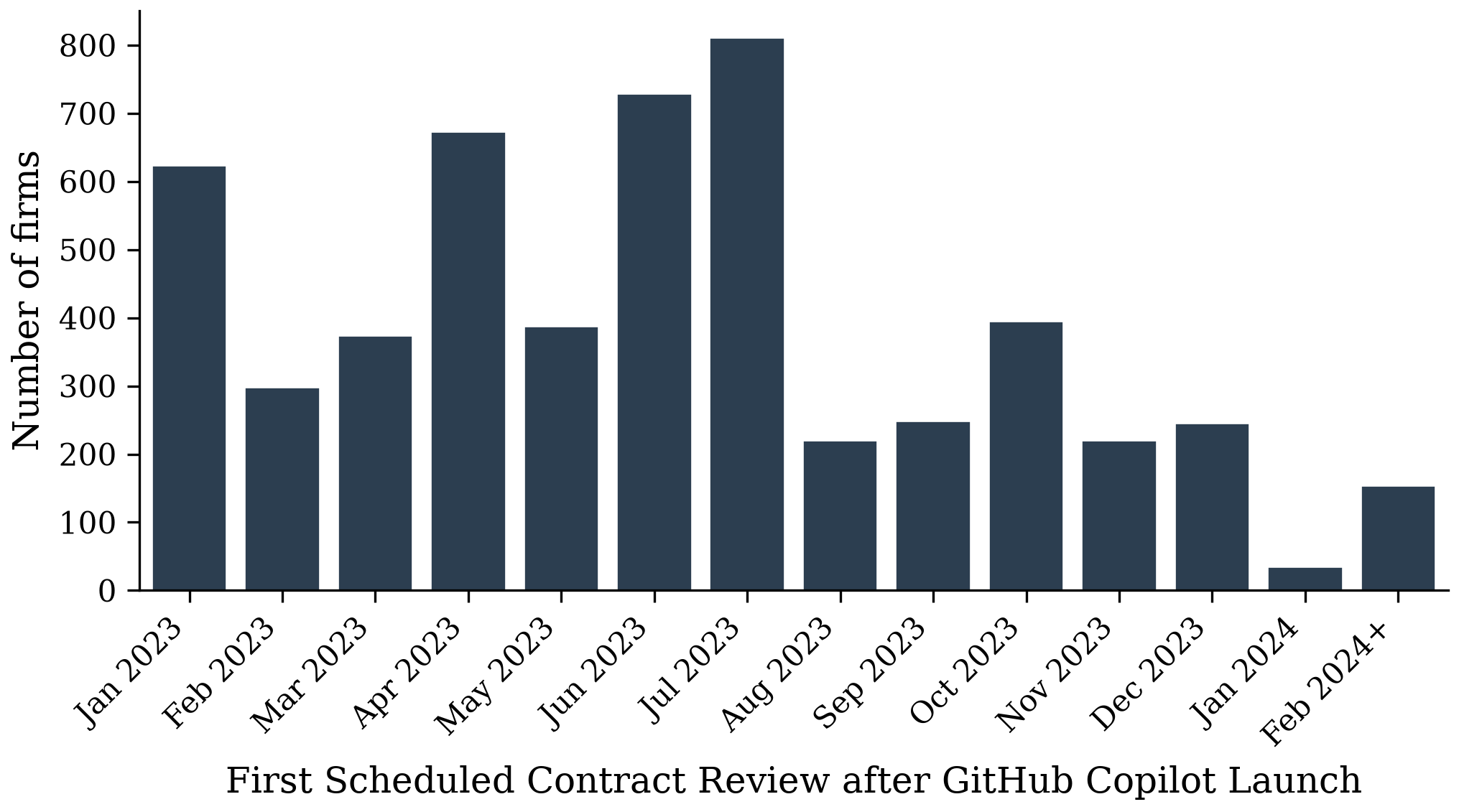}
    \subcaption{Distribution of scheduled contract reviews}
\end{subfigure}

\par\vspace{1ex}
\figurenotes{Panel A plots the share of sample firms that have adopted GitHub Copilot by each month from January 2023 through June 2025. Adoption begins in the first month a firm purchases at least one GitHub Copilot license and is treated as an absorbing state. Panel B shows the distribution of sample firms’ next scheduled GitHub contract review month, using the schedule in place as of December 2022.}
\end{figure}

Because review dates were set before GitHub Copilot's launch, they could generate variation in when firms adopt that is plausibly unrelated to their underlying employment trajectories. We therefore use each firm's first scheduled post-launch contract review, as recorded in December 2022, as an instrument for GitHub Copilot adoption.\footnote{We use only the first post-launch review because subsequent review dates could be affected by earlier adoption decisions.} Because these scheduled reviews are staggered across the calendar year rather than bunched in a single quarter, firms reach their first post-launch evaluation in different months, allowing us to compare firms arriving at a review to not-yet-reviewed firms at the same point in calendar time (Panel B of Figure~\ref{fig:adoption_and_reviews}).

While procurement stickiness clearly concentrates adjustments to core GitHub licenses around review dates, how contract reviews shape the timing and speed of GitHub Copilot adoption is an empirical question. On one hand, reaching a pre-scheduled review substantially lowered the administrative friction of adopting the tool compared to initiating an off-cycle procurement process. Because introducing a new product like GitHub Copilot is more complex than adjusting seat counts for an already integrated product, these administrative frictions could be even more binding. On the other hand, strong demand for generative AI might lead firms to adopt GitHub Copilot immediately, although the slow path of adoption in Panel A of Figure \ref{fig:adoption_and_reviews} weighs against this. We examine this in our first-stage analysis in Section~\ref{sec:first_stage}.

\section{Data}\label{sec:data}
Our analysis combines three data sources: GitHub licensing data, GitHub customer relationship management (CRM) data, and Revelio Labs employment data.

\subsection{GitHub Data}\label{sec:github_data}

The GitHub data consist of monthly firm-level snapshots from January 2021 through November 2025 linked via internal identifiers. The licensing records report the number of GitHub and GitHub Copilot licenses held by each firm. We define GitHub Copilot adoption as the first month a firm holds a positive number of GitHub Copilot licenses and treat adoption as an absorbing state. To avoid endogenous sample entry, we restrict the sample to active GitHub customers as of September 2022. We obtain each firm's first post-launch contract review date from the CRM records as of December 2022, one month prior to GitHub Copilot's January 2023 launch.

\subsection{Revelio Labs Employment Data}\label{sec:revelio_data}

We measure employment stocks and flows using Revelio Labs, which aggregates public professional profiles and online job postings into standardized worker-employer histories. From individual employment spells and educational records, we construct monthly firm-level counts of headcount, new hires, and separations, both overall and by role-by-seniority cell, alongside monthly counts and average durations of job postings. Because Revelio draws primarily on LinkedIn profiles, coverage is strongest among large firms and technical workers in North America and Europe \citep{baird2024early}. We therefore restrict our sample to private-sector firms headquartered in the Americas and Europe (see Appendix Section~\ref{asec:data_construction} for details on variable construction and coverage).

\subsection{Sample Characteristics}\label{sec:sample}

We link GitHub customers to Revelio using automated name matching supplemented by AI-assisted LinkedIn searches and human review, applying pre-2023 quality filters to exclude false matches and firms with thin profile coverage (Appendix Sections~\ref{asec:matching_procedure}--\ref{asec:sample_construction} and Appendix Table~\ref{atab:sample_construction}). The resulting balanced panel tracks 5,414 firms monthly from January 2021 through June 2025. Benchmarking Revelio against external records confirms strong alignment: total headcount correlates closely with Compustat 10-K employment in log levels ($r = 0.756$) and annual log changes ($r = 0.576$), while pre-launch software engineering headcount correlates strongly with purchased GitHub licenses in log levels ($r = 0.684$; Appendix Section~\ref{asec:data_construction}).

Sample firms are large, established employers: at the September 2022 baseline (Appendix Table~\ref{atab:summary_stats}), the average firm is 47 years old and employs 8,284 workers globally, with 61.6\% headquartered in the United States and the largest concentrations in information technology (24.8\%) and financial services (12.9\%). We classify eight roles as technical: software, data, quality-assurance, infrastructure, systems, cybersecurity, and design engineers, alongside application developers. Technical roles represent 12.6\% of sample employment (5.7\% in software engineering alone) and 1.6 million U.S.\ workers, covering 38\% of the U.S.\ programming workforce (Appendix Section~\ref{asec:sample_coverage} and Appendix Table~\ref{atab:bls_market}).\footnote{As a benchmark, computer and mathematical occupations represent 3.4\% of U.S.\ employment \citep{bls_oews_2025}.}

\section{Empirical Strategy}\label{sec:empirical_strategy}

\subsection{Technical Labor-Market Trends and Pre-Adoption Differences}\label{sec:selection}

\begin{figure}[H]
\centering
\caption{Technical Labor-Market Outcomes Around GitHub Copilot Adoption}
\label{fig:selection_descriptives}

\begin{subfigure}[t]{0.48\textwidth}
    \centering
    \includegraphics[width=\textwidth]{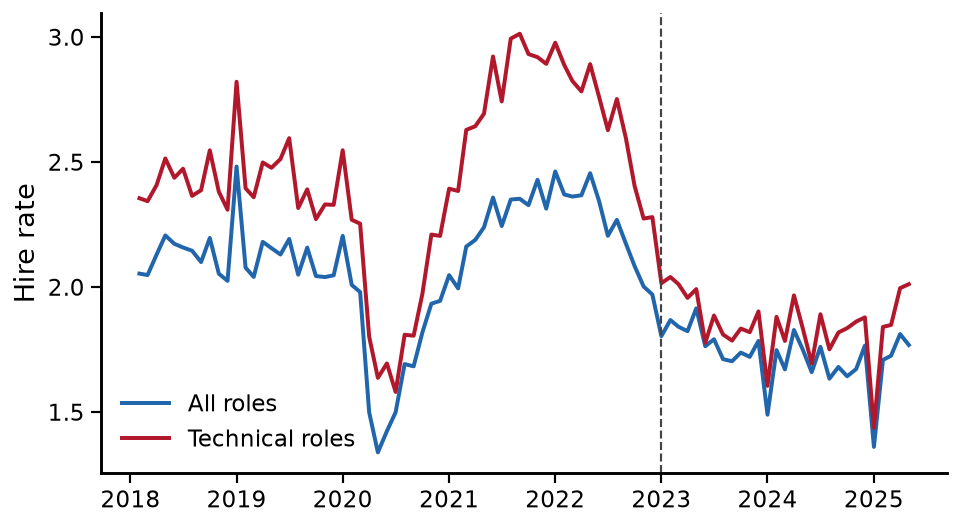}
    \subcaption{Aggregate technical hiring}
\end{subfigure}
\hfill
\begin{subfigure}[t]{0.48\textwidth}
    \centering
    \includegraphics[width=\textwidth]{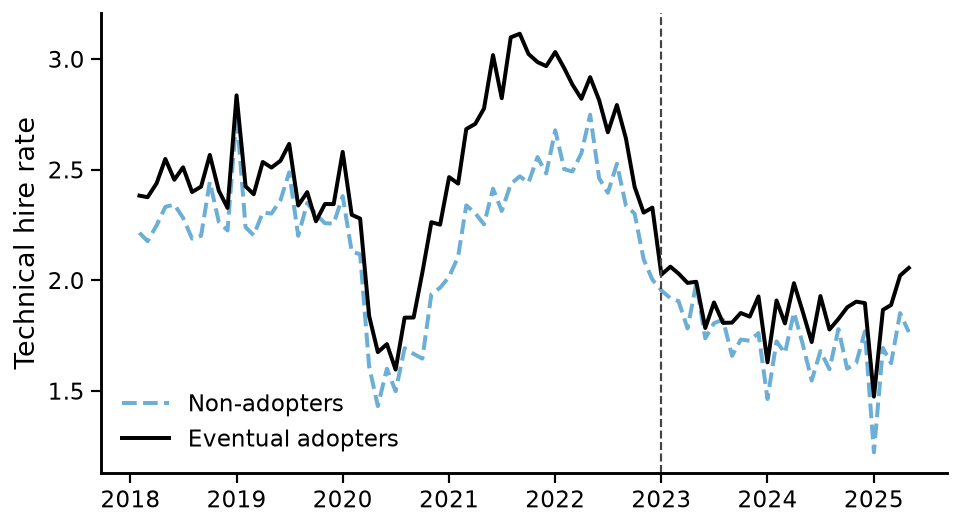}
    \subcaption{Technical hiring by eventual adoption}
\end{subfigure}

\vspace{0.5em}

\begin{subfigure}[t]{0.48\textwidth}
    \centering
    \includegraphics[width=\textwidth]{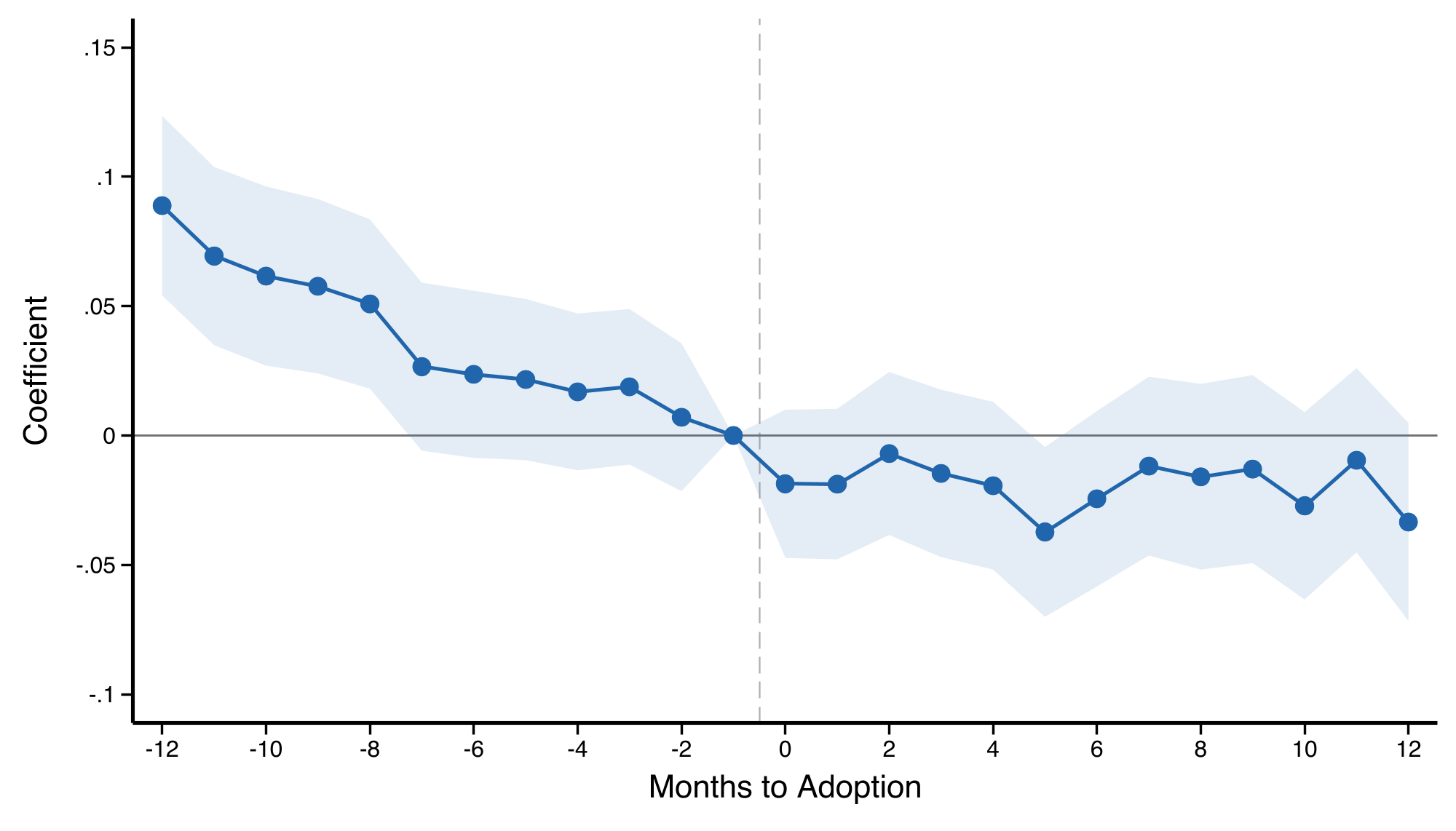}
    \subcaption{Log technical hiring (adoption-centered)}
\end{subfigure}
\hfill
\begin{subfigure}[t]{0.48\textwidth}
    \centering
    \includegraphics[width=\textwidth]{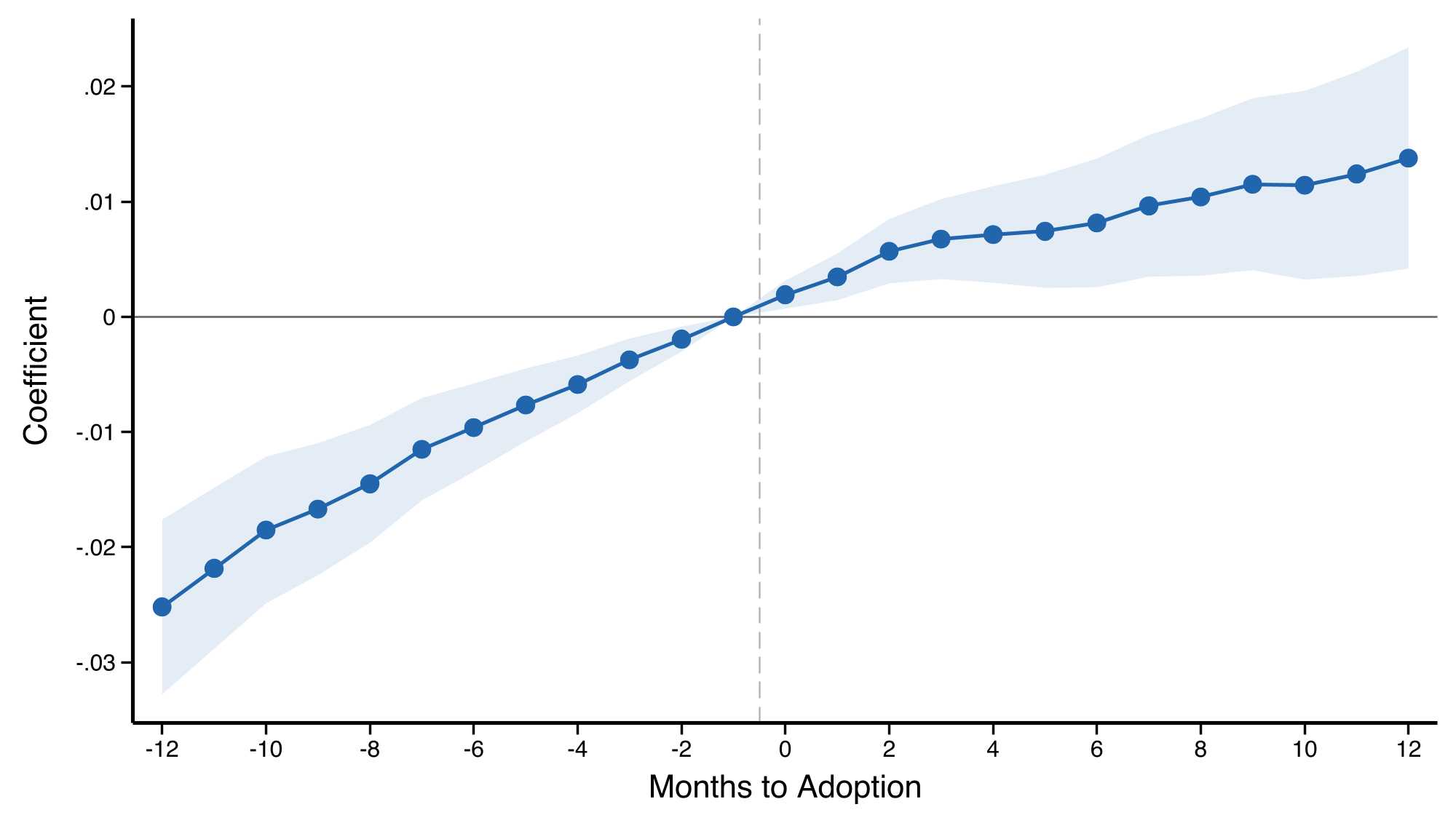}
    \subcaption{Log technical employment (adoption-centered)}
\end{subfigure}

\vspace{0.5em}

\begin{subfigure}[t]{0.48\textwidth}
    \centering
    \includegraphics[width=\textwidth]{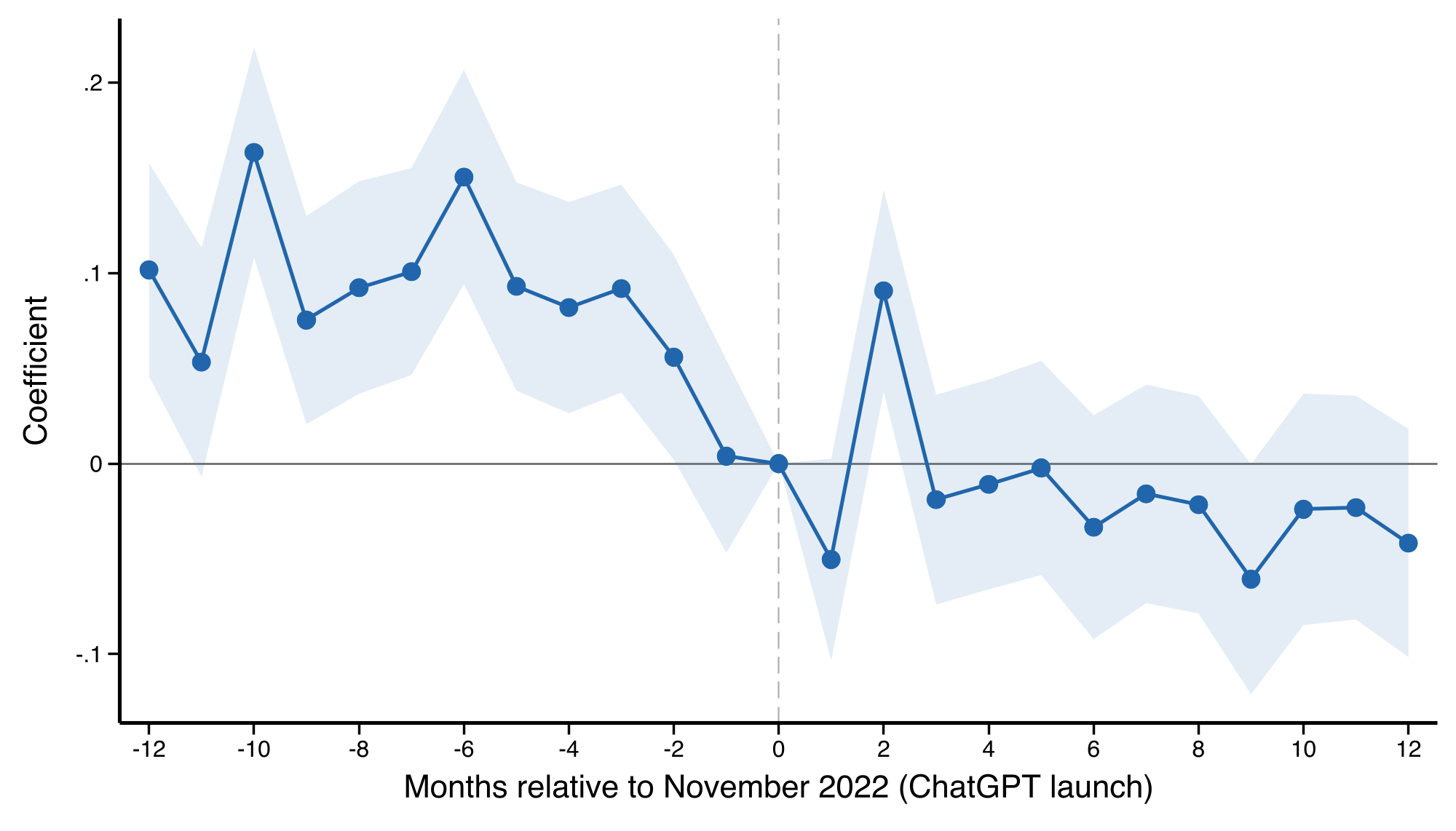}
    \subcaption{Log technical hiring (ChatGPT-launch)}
\end{subfigure}
\hfill
\begin{subfigure}[t]{0.48\textwidth}
    \centering
    \includegraphics[width=\textwidth]{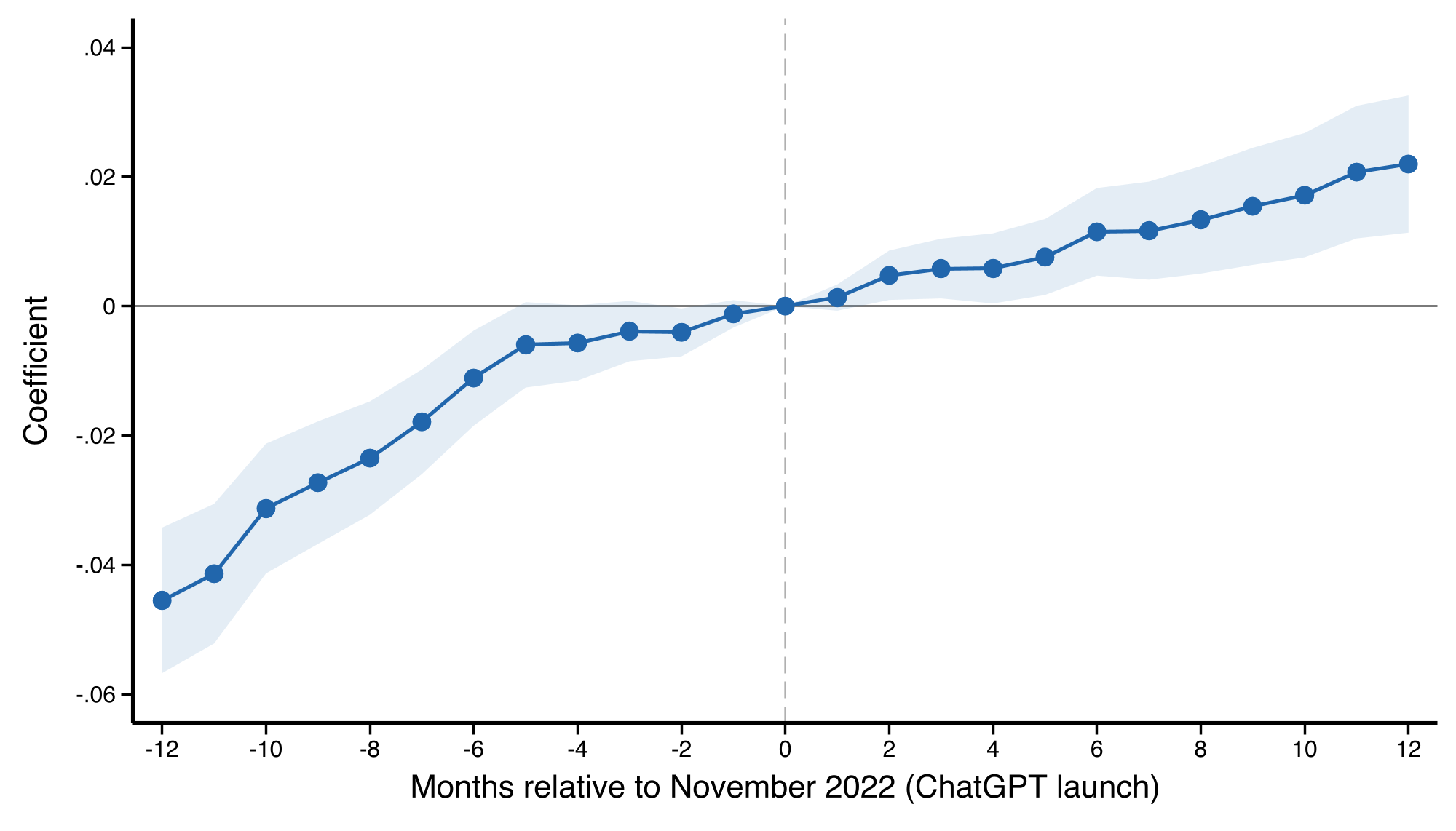}
    \subcaption{Log technical employment (ChatGPT-launch)}
\end{subfigure}

\figurenotes{Panels A and B plot seasonally adjusted monthly hiring rates, measured as hires divided by lagged employment, from January 2018 through June 2025. Panel A compares all and technical roles; Panel B compares eventual adopters and non-adopters. The dashed line marks GitHub Copilot's January 2023 enterprise launch. Panels C and D plot adoption-centered event-study coefficients and 95\% confidence intervals from Equation~\eqref{eq:ols_event_study}. Panels E and F plot event-study estimates around the November 2022 release of ChatGPT from Equation~\eqref{eq:chatgpt_event_study}. Standard errors clustered at the firm level.}
\end{figure}

Simple comparisons between GitHub Copilot adopters and non-adopters face two empirical challenges. GitHub Copilot launched during a sharp macroeconomic contraction in technical labor demand. Panel~A of Figure~\ref{fig:selection_descriptives} plots the mean monthly technical hiring rate (new technical hires divided by lagged technical headcount). After rising from 2.5\% to 3.0\% in 2021, technical hiring began falling in March 2022, ten months before GitHub Copilot's enterprise release, coinciding with the Federal Reserve's rapid monetary tightening cycle \citep{iscenko_looking_2026}. Eventual adopters and non-adopters followed different trajectories in the pre-period. First, eventual adopters experienced a much steeper hiring surge in 2021 and contraction throughout 2022 (Panel B). Second, eventual adopters employed a higher share of technical workers, may have been differentially exposed to the broader tech slowdown, and may not have followed parallel trends with non-adopters. As a result, standard observational benchmarks could fail in two distinct ways: centering on firm adoption dates may confound treatment with pre-existing firm growth trends (Panels C and D), whereas comparing adopters with non-adopters around late 2022 may reflect differential exposure to the broader tech cycle (Panels E and F).

We first examine the adoption-centered benchmark by estimating a two-way fixed effects event study around each firm's observed adoption month $A_i$:
\begin{equation}
Y_{it} = \alpha_i + \lambda_t
+ \sum_{\substack{k=-12 \\ k\neq -1}}^{12}
\beta_k \mathbf{1}\{t-A_i=k\}
+ \varepsilon_{it}
\label{eq:ols_event_study}
\end{equation}
where $A_i$ is the first month in which firm $i$ holds a positive number of GitHub Copilot licenses, $\alpha_i$ are firm fixed effects, and $\lambda_t$ are calendar-month fixed effects, with standard errors clustered at the firm level.

Panels C and D of Figure~\ref{fig:selection_descriptives} plot the resulting estimates for log technical hiring and headcount, measured relative to the month prior to adoption ($k = -1$). Both panels exhibit pronounced pre-trends reflecting the pre-adoption expansion and deceleration. On the hiring margin (Panel C), pre-adoption coefficients slope steadily downward as recruiting cools from its earlier surge, yielding a negative pooled OLS association of -5.1\% (Appendix Table~\ref{atab:ols_adoption}). On the headcount margin (Panel D), the accumulation of those pre-adoption hires produces an upward pre-trend in log employment that continues rising at a slower pace after adoption, yielding positive pooled OLS associations of 1.8\% for technical headcount and 1.9\% for software engineering headcount (Appendix Table~\ref{atab:ols_adoption}).

We next examine the second benchmark, which avoids endogenous adoption dates by comparing eventual adopters with never-adopters before and after the November 2022 release of ChatGPT ($t_0$):\footnote{This is the firm-level analogue to regressions that compare employment dynamics around the launch of ChatGPT using task or occupational exposure as a proxy for adoption.}
\begin{equation}
Y_{it} = \alpha_i + \lambda_t
+ \sum_{\substack{k=-12 \\ k\neq 0}}^{12}
\delta_k \mathbf{1}\{t-t_0=k\} \times \mathbf{1}\{\text{Adopter}_i\}
+ \varepsilon_{it}
\label{eq:chatgpt_event_study}
\end{equation}
where $\mathbf{1}\{\text{Adopter}_i\}$ equals one for firms that adopt GitHub Copilot by June 2025 and $t_0$ is November 2022 (see Appendix Table~\ref{atab:post_launch_adopter_did} for the pooled post-November 2022 difference-in-differences estimates).

Fixing the calendar date does not resolve adopters' differential exposure to the 2021–2022 tech cycle. Panels E and F reproduce the same opposing pre-trends and opposite-signed pooled estimates (Appendix Table~\ref{atab:post_launch_adopter_did}). Because adopters' hiring surged higher during the 2021–2022 boom before converging toward non-adopter levels (Panel E), comparing post-2022 hiring to the pre-period yields a 10.1\% relative decline (Column 1). Conversely, because that earlier hiring surge expanded adopters' workforce without subsequent layoffs (Column 2), their post-2022 technical headcount remains 6.7\% higher than their pre-2022 baseline (Panel F and Column 3). Standard observational regressions can therefore imply either that AI reduces employment (when measured using hiring flows) or increases employment (when measured using headcount stocks).

\subsection{Instrumented Differences-in-Differences with Staggered Timing}
To isolate exogenous variation in adoption timing, we instrument for GitHub Copilot adoption using the timing of each firm's first scheduled GitHub contract review after GitHub Copilot's launch ($R_i$). Let $\ell = t - R_i$ index event time relative to this first scheduled review. We estimate the first-stage effect of reaching a contract review on GitHub Copilot adoption and the reduced-form effect on employment outcomes using the following event-study specifications:

\begin{equation}
    D_{it} = \alpha_i + \lambda_t + \sum_{\ell \in \mathcal{L}} \pi_{\ell}\mathbf{1}\{t-R_i=\ell\} + \varepsilon_{it}
    \label{eq:first_stage}
\end{equation}

\begin{equation}
    Y_{it} =\alpha_i + \lambda_t + \sum_{\ell \in \mathcal{L}}  \rho_{\ell}\mathbf{1}\{t-R_i=\ell\} + u_{it}
    \label{eq:reduced_form}
\end{equation}

where $D_{it}$ is an absorbing indicator equal to one once firm $i$ holds a positive number of GitHub Copilot licenses and $Y_{it}$ is an employment outcome such as headcount, hiring, or separations. For hiring and separations, which may be zero in a firm-month, we separately examine the extensive margin, an indicator for a positive value, and the intensive margin, the log outcome conditional on a positive value \citep{chen_logs_2024}. We estimate firm-wide and broad occupational specifications on a balanced monthly panel; for more disaggregated role-by-seniority and within-firm analyses, where monthly hiring in individual cells is sparser, we aggregate the data to the quarterly level.

In both equations, firm fixed effects $\alpha_i$ absorb time-invariant firm differences such as baseline size and industry, while calendar-period fixed effects $\lambda_t$ absorb common macroeconomic and seasonal shocks. The coefficients $\pi_{\ell}$ and $\rho_{\ell}$ measure the effect of contract review on GitHub Copilot adoption and employment at horizon $\ell$ relative to firms that have not yet reached their scheduled review. Because contract reviews are staggered over time, conventional two-way fixed-effects estimators can be biased when treatment effects vary across cohorts or over time \citep{miyaji_two-way_2024}. We therefore estimate Equations~\eqref{eq:first_stage} and~\eqref{eq:reduced_form} using the imputation estimator of \citet{borusyak_revisiting_2024}, which is robust to treatment effect heterogeneity. Cohorts with contract reviews after 2023 serve as our last-treated comparison group, entering the sample only prior to their review date ($R_i$). All regressions are unweighted and thus represent effects across firms rather than employment-weighted effects. Standard errors are clustered at the firm level.

Under parallel trends and no anticipation in employment outcomes, the reduced form coefficient $\rho_{\ell}$ identifies the intention-to-treat (ITT) effect of reaching a scheduled GitHub contract review on employment. Parallel trends requires that, absent a contract review, employment trajectories do not vary systematically with a firm's scheduled review date. No anticipation requires that firms do not adjust employment ahead of an approaching contract review. The reduced form is informative even without interpreting contract review as an instrument: it captures the intention-to-treat effect of reaching a scheduled opportunity to evaluate and procure GitHub Copilot on employment outcomes.

To move from the intent-to-treat effect to the effect of adopting GitHub Copilot itself, we scale the reduced-form estimates by the first-stage adoption response using the horizon-specific Wald ratio:
\begin{equation}
\widehat{\theta}_{\ell}
= \frac{\widehat{\rho}_{\ell}}{\widehat{\pi}_{\ell}}
\label{eq:wald_did}
\end{equation}

We construct 95\% percentile confidence intervals for the Wald estimates using a firm-block bootstrap with 1{,}000 draws that re-estimates both legs and their ratio on the same subsample.

Interpreting $\widehat{\theta}_{\ell}$ as a causal treatment effect requires standard instrumental variables assumptions adapted to a staggered panel setting \citep{miyaji_instrumented_2024, miyaji_two-way_2024}. Parallel trends and no anticipation must also hold for adoption.\footnote{No anticipation does not rule out adoption before review; it requires that firms not time adoption in response to an approaching review.} Contract review must increase adoption on average (relevance), must not delay adoption for any firm (monotonicity), and must affect employment only through GitHub Copilot adoption (exclusion). Finally, employment outcomes at horizon $\ell$ depend on current adoption status rather than the specific duration of prior exposure (no carryover) \citep{miyaji_instrumented_2024}. 
Under these conditions, $\widehat{\theta}_{\ell}$ identifies the horizon-specific local average treatment effect on the treated (LATET) among compliers at horizon $\ell$ \citep{miyaji_instrumented_2024, miyaji_two-way_2024}. In our setting, compliers are firms that adopted GitHub Copilot once their scheduled review lowered the administrative hurdle, but would not yet have adopted off-cycle absent the review. For these firms, $\widehat{\theta}_{\ell}$ captures the effect of having adopted GitHub Copilot by horizon $\ell$ relative to not yet having adopted. We evaluate the plausibility of these assumptions in Sections~\ref{sec:first_stage}, \ref{sec:balance}, and~\ref{sec:robustness}.

\subsection{First Stage: Contract Review and GitHub Copilot Adoption}
\label{sec:first_stage}
\begin{figure}[H]
\centering
\caption{Contract Review and GitHub Copilot Adoption}
\label{fig:fs_treat_baseline_BJS}
    \centering
    \includegraphics[width=.8\textwidth]{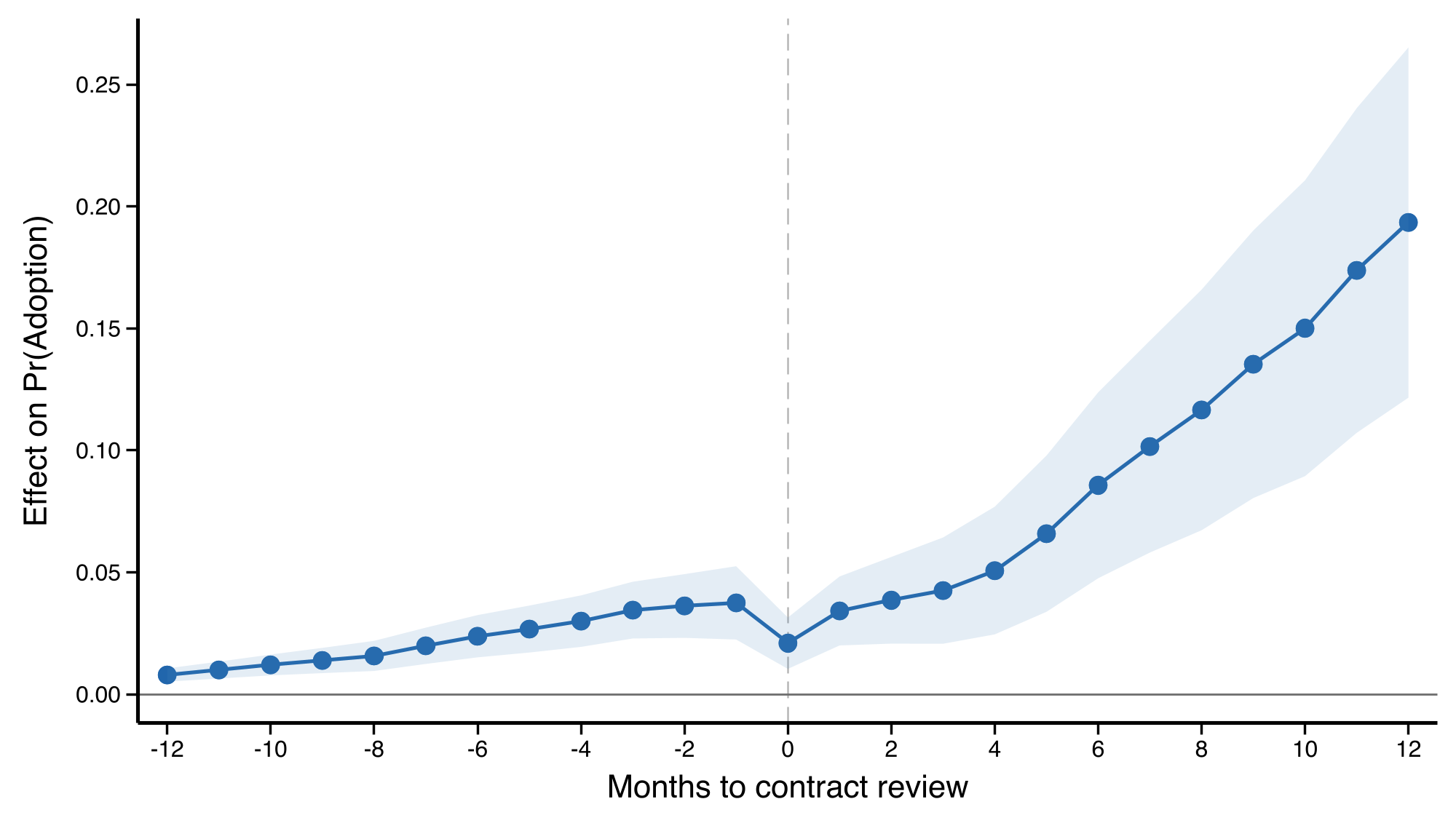}

\par\vspace{1ex}
\figurenotes{The figure reports event time coefficients from Equation \ref{eq:first_stage}. Estimates use the imputation estimator of \citet{borusyak_revisiting_2024} on the balanced firm-month panel with firm and calendar-month fixed effects. Shaded bands are 95\% confidence intervals; standard errors are clustered by firm.}
\end{figure}

Figure~\ref{fig:fs_treat_baseline_BJS} plots the first stage event study from Equation~\eqref{eq:first_stage} (for quarterly estimates, see Column 2 Appendix Table~\ref{atab:tech_flows_quarterly_latet}). Reaching a scheduled contract review steadily increases GitHub Copilot adoption, raising the probability of adoption by 8.6 percentage points after six months and 19.3 percentage points after twelve months, both significant increases relative to the 62\% end-of-sample adoption rate. At the four-quarter horizon, the first-stage $F$-statistic is 30.7, well above standard weak-instrument thresholds.\footnote{In our just-identified setting with firm-clustered standard errors, the squared first stage $t$-statistic equals the \citet{montiel_olea_robust_2013} effective $F$-statistic, for which 23.1 is the critical value for 10\% worst-case bias.} Whereas adjustments to existing GitHub Enterprise license counts are clustered in the two months after the contract review month, the first-stage effect on GitHub Copilot adoption builds more gradually. Updating seat counts on an existing agreement is a routine administrative change, whereas a scheduled review initiates the evaluation of GitHub Copilot as a new AI product, which requires setting up the software and rolling it out to employees.

In a standard event study, pre-period coefficients ($\ell < 0$) provide a test of anticipation. In our setting, however, two distinct forces might generate the upward slope prior to a scheduled contract review in Figure~\ref{fig:fs_treat_baseline_BJS}. First, there is a mechanical composition effect from GitHub Copilot's January 2023 launch. Before January 2023, GitHub Copilot was not available, so off-cycle adoption was not possible and adoption is identically zero. After January 2023, however, always-takers could begin adopting off-cycle prior to their scheduled review. Thus, as event time moves from $\ell = -12$ to $\ell = -1$, a growing share of firm-month observations fall after January 2023: at $\ell = -12$, nearly all observations occur in 2022 when adoption is mechanically zero, whereas by $\ell = -2$, most observations occur in 2023 when off-cycle adoption can accumulate.\footnote{This means that while the share of ``always takers'' in the sample may remain constant, the share able to adopt off cycle changes.} This shifting sample composition may by itself create an upward slope in the pre-review leads. Second, firms might genuinely be adopting GitHub Copilot at higher rates in anticipation of their approaching contract review. Because the baseline pre-review leads mix pre-launch zeros (December 2022 and prior) with post-launch observations (January 2023 and after), the pre-period coefficients cannot distinguish anticipation directly.

To isolate and test for anticipation directly, we restrict the sample to post-January 2023 months, when off-cycle adoption was feasible for all firms. Because adoption is an absorbing state,
cumulative adoption in Figure~\ref{fig:fs_treat_baseline_BJS} accumulates all prior adoptions over time. As a result, even a constant rate of off-cycle adoption will cause the cumulative share of adopters to drift upward as the review date approaches. By contrast, the monthly adoption hazard measures the flow into adoption in a given month: among firms that have not yet adopted entering month $t$, it is the fraction that adopt GitHub Copilot for the first time during month $t$ among those who have not yet adopted in the prior month. If firms were timing their adoption in anticipation of an approaching contract review, this monthly hazard rate would rise in the months leading up to the review. However, among firms who have not yet adopted, the pre-review monthly adoption hazard is flat at roughly 4\% across the months leading up to scheduled contract review and jumps sharply in the review month itself ($\ell = 0$). This pattern shows no evidence of anticipation. Estimating hazard ratios in a complementary log-log model that controls for months since GitHub Copilot's launch shows an equally flat pre-review profile (Appendix Figure~\ref{afig:hazard_mtnr_idur}).

\begin{figure}[H]
\centering
\caption{GitHub Copilot Adoption Hazard by Months to Scheduled Contract Review}
\label{fig:hazard_mtnr}
    \centering
    \includegraphics[width=0.5\textwidth]{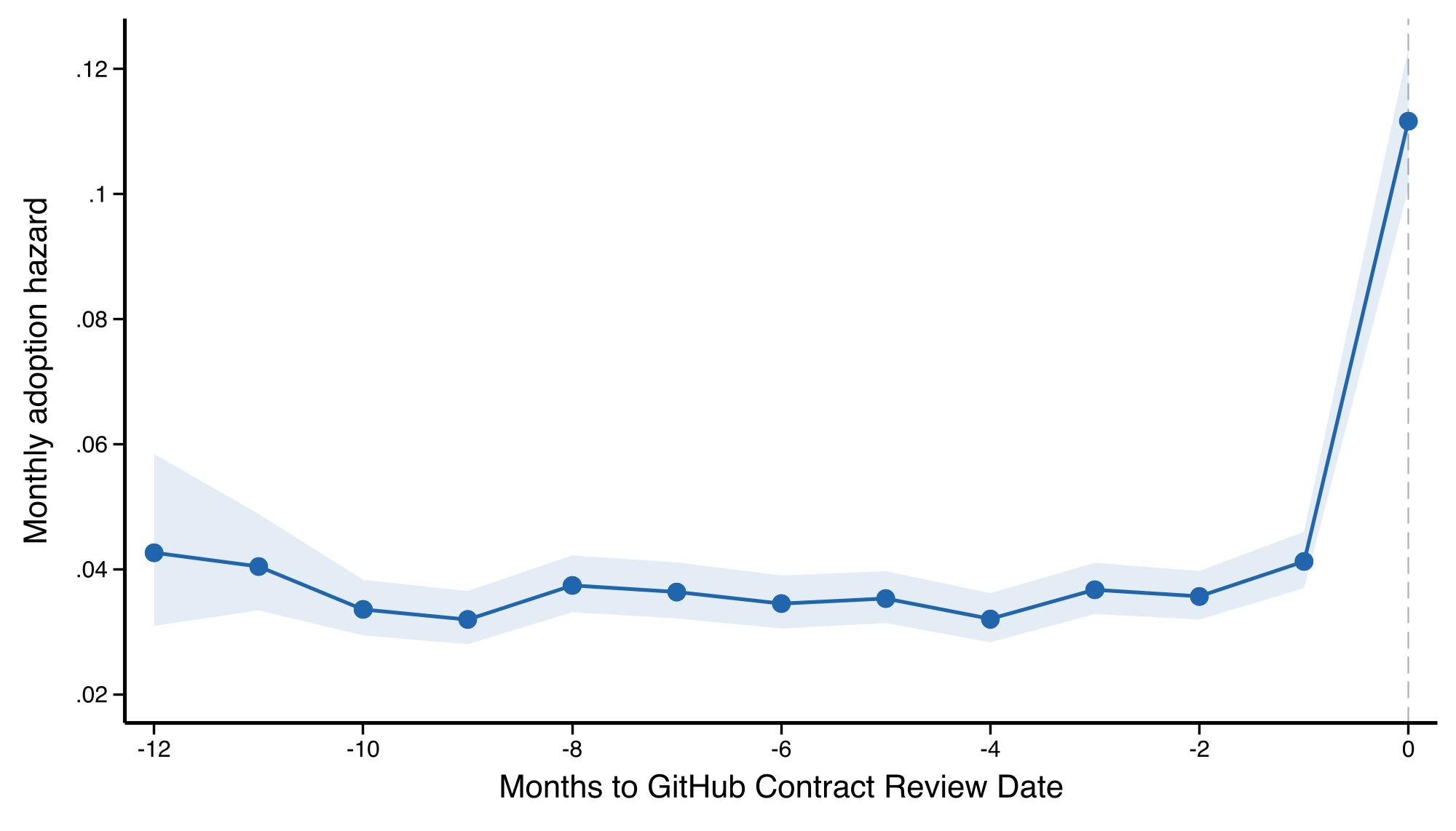}

\par\vspace{1ex}
\figurenotes{The figure plots the monthly hazard of first GitHub Copilot adoption among the risk set of firms that have not yet adopted, restricted to the post-launch period. Month $0$ is the month of a firm's scheduled contract review and negative values are months before it. The shaded band is the 95\% Wilson confidence interval.}
\end{figure}

\subsection{Covariate Balance Across Contract Review Timing}\label{sec:balance}

Our identification strategy requires that, absent a contract review, firms scheduled for review at different dates would have followed parallel employment trajectories. While baseline covariate balance cannot verify parallel trends directly, it can help evaluate whether review timing is systematically related to pre-launch firm characteristics or employment growth. 

\begin{figure}[H]
\centering
\caption{Balance of Baseline Firm Characteristics}
\label{fig:balance_coefplots}

\begin{subfigure}[t]{0.49\textwidth}
    \centering
    \includegraphics[width=\textwidth]{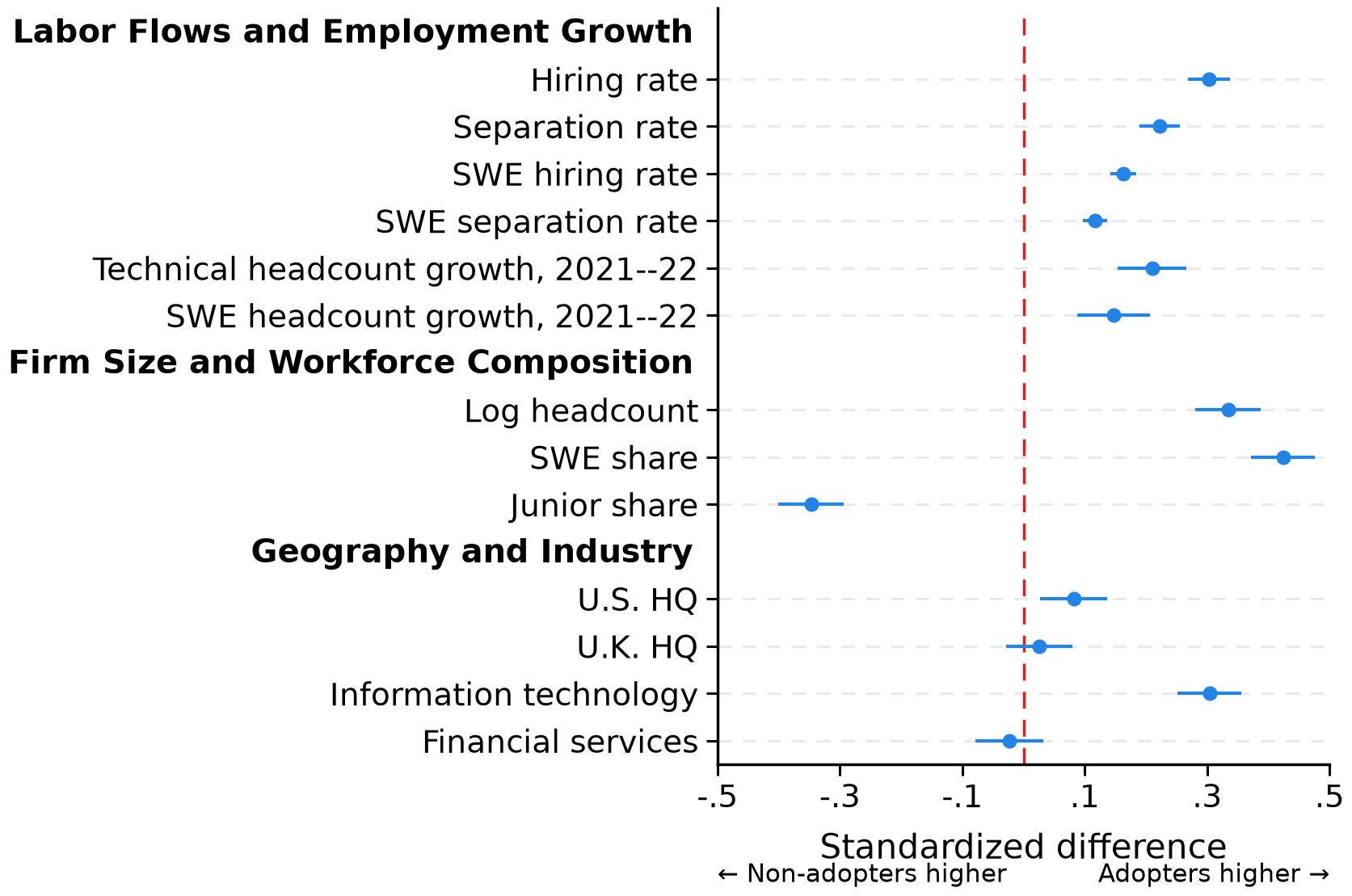}
    \subcaption{Eventual Adopters vs.\ Non-Adopters}
\end{subfigure}
\hfill
\begin{subfigure}[t]{0.49\textwidth}
    \centering
    \includegraphics[width=\textwidth]{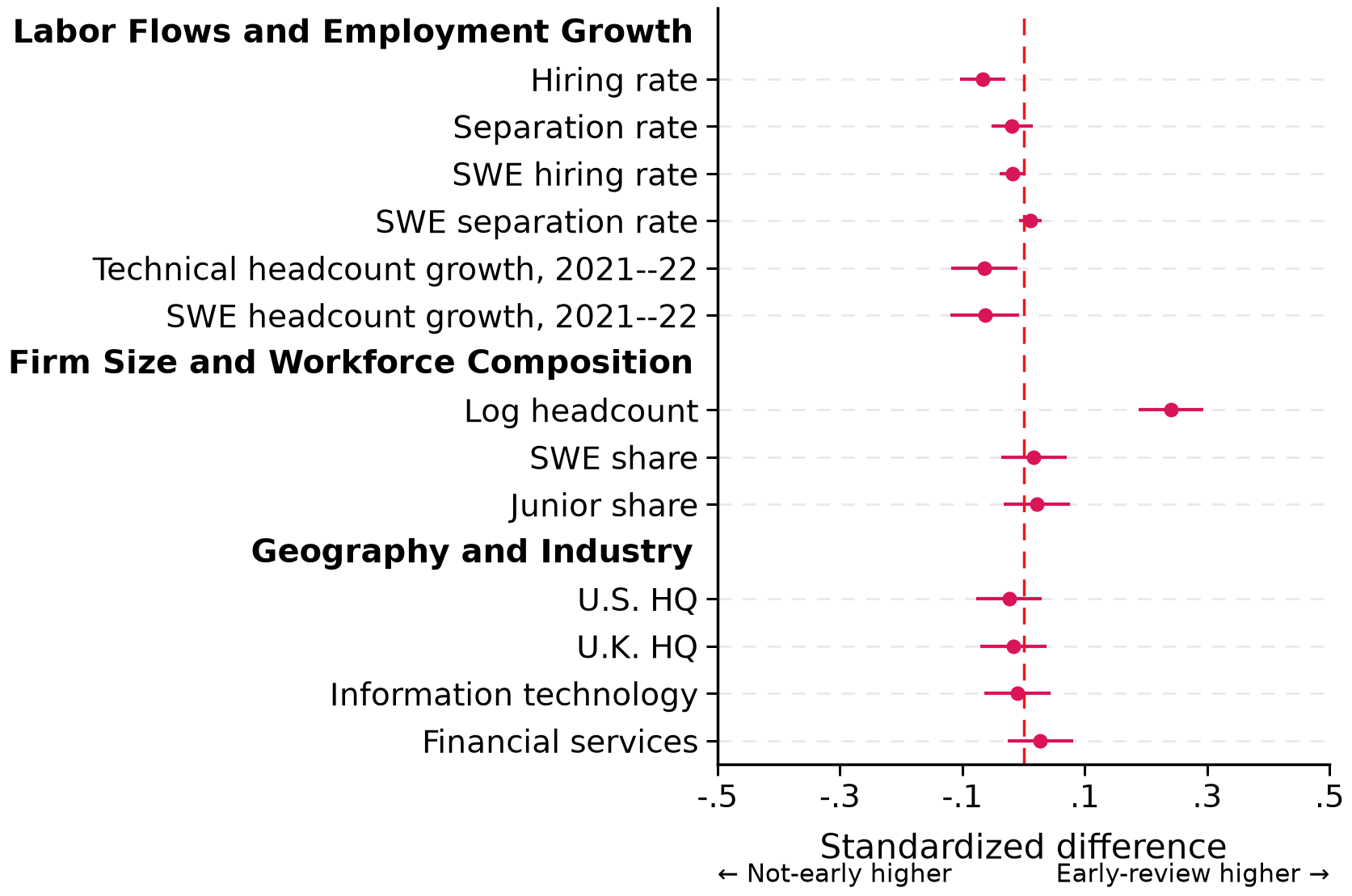}
    \subcaption{Early vs.\ Later Scheduled Contract Review}
\end{subfigure}

\par\vspace{1ex}
\figurenotes{Panel A compares eventual adopters with non-adopters; Panel B compares firms reviewed in January--June 2023 with the remaining sample. Bars are 95 \%confidence intervals; standard errors are clustered by firm. Appendix Table~\ref{atab:balance_regs} reports the underlying regression results.}
\end{figure}

Panel~A of Figure~\ref{fig:balance_coefplots} illustrates the differences by adoption status. Compared to non-adopters, eventual adopters were substantially larger, had higher baseline hiring and separation rates, employed a higher share of software engineers and a lower share of junior workers, and were more concentrated in the information technology sector. In contrast, Panel~B compares firms with an early contract review to firms with a later review, split at the median contract review month of June 2023. Across all workforce composition, hiring, separation, and industry measures, the standardized differences are small and within the conventional 0.10 threshold for covariate balance \citep{austin_balance_2009}.

The exception is firm size: firms with early contract reviews have log headcount that is 0.24 standard deviations higher than firms reviewed later. This pattern reflects how initial enterprise GitHub contracts were signed years prior to GitHub Copilot's launch. Because sales commissions scale with contract value, enterprise sales teams prioritized the largest accounts early in the calendar year, and subsequent annual reviews remained anchored to those initial signing months. In our baseline specification, firm fixed effects absorb time-invariant size differences across firms, though they do not rule out differential employment trends by firm size. In Section~\ref{sec:robustness}, we add baseline firm-size-by-quarter fixed effects, as well as industry-by-quarter fixed effects, and show that our results are unchanged.

\section{Conceptual Framework}\label{sec:conceptual_framework}

To guide our empirical analysis, we introduce a simple accounting framework that decomposes GitHub Copilot's effect on firm employment into three channels: the scale of software activity, the allocation of tasks across roles and seniority levels, and worker productivity.\footnote{Because our empirical design captures short run firm adjustments rather than economy-wide equilibrium responses in wages and employment, we use an accounting identity rather than an equilibrium model.}

Let $T_i$ denote the total volume of software-related tasks at firm $i$. Workers are grouped into cells $(r,s)$ defined by role $r$ and seniority $s$. Let $m_{i,rs}$ be the share of software-related work assigned to cell $(r,s)$, with $\sum_{r,s}m_{i,rs}=1$. Finally, $a_{i,rs}$ is task-level productivity, defined as the number of tasks completed per unit of labor in that cell. Given this structure, the labor the firm requires in a particular role-by-seniority cell is:
\begin{equation}
L_{i,rs} = \frac{m_{i,rs}T_i}{a_{i,rs}}.
\label{eq:conceptual_cell_labor_demand}
\end{equation}
Taking logs and differencing with respect to a firm-level GitHub Copilot adoption ($\Delta$) gives:
\begin{equation}
\Delta\ln L_{i,rs}
=
\underbrace{\Delta\ln T_i}_{\text{scale}}
+
\underbrace{\Delta\ln m_{i,rs}}_{\text{task reallocation}}
-
\underbrace{\Delta\ln a_{i,rs}}_{\text{productivity}}.
\label{eq:conceptual_labor_demand_decomposition}
\end{equation}

Equation~\eqref{eq:conceptual_labor_demand_decomposition} highlights three potentially offsetting forces. First, by lowering software development costs, GitHub Copilot may induce the firm to undertake more projects, expanding total task volume ($\Delta\ln T_i > 0$) and raising labor demand proportionally across all cells. Second, GitHub Copilot may alter task allocation ($\Delta\ln m_{i,rs}$). For example, senior engineers might use GitHub Copilot to handle implementation work previously delegated to junior engineers, or developers might absorb testing tasks previously assigned to quality assurance teams. Third, GitHub Copilot changes task productivity ($\Delta\ln a_{i,rs}$).\footnote{In principle, the effect on task-level productivity could be positive or negative. While GitHub Copilot may speed up code generation, effective productivity could fall ($\Delta\ln a_{i,rs} < 0$).} Completing assigned tasks faster reduces the labor required for a given workload, whereas spending extra time reviewing and debugging AI-generated code lowers effective productivity and increases labor requirements.

This framework clarifies what our empirical estimates can and cannot identify. In the data, we observe employment and hiring in each cell ($L_{i,rs}$), but we do not observe the firm's total volume of software work ($T_i$), how tasks are divided across cells ($m_{i,rs}$), or task-level productivity ($a_{i,rs}$). Consequently, our estimates for any single group capture the combined net effect of all three channels. 
However, comparing employment responses across cells within the same firm, such as junior versus experienced engineers in the same role, differences out the firm-wide scale effect ($\Delta\ln T_i$), which is common to all cells in firm $i$. Any divergence across groups within a firm must therefore reflect differential changes in task allocation ($\Delta\ln m_{i,rs}$) or task-level productivity ($\Delta\ln a_{i,rs}$). Finally, because hiring decisions are forward-looking, our short-run estimates cannot separate responses to realized productivity gains from adjustments driven by managers' beliefs and expectations about AI capabilities. As firms learn about the tool's actual performance and as project scale and task allocation adjust at different speeds, these net employment effects may evolve over longer horizons.

\section{Results}
\subsection{Overall Effects on the Technical Workforce}
\label{sec:technical_hiring}
 
\subsubsection{Hiring, Separations, and Headcount}
\begin{figure}[H]
\centering
\caption{Reduced Form Effects on Technical Hiring and Separations}
\label{fig:technical_flows}

\begin{subfigure}[t]{0.49\textwidth}
    \centering
    \includegraphics[width=\textwidth]{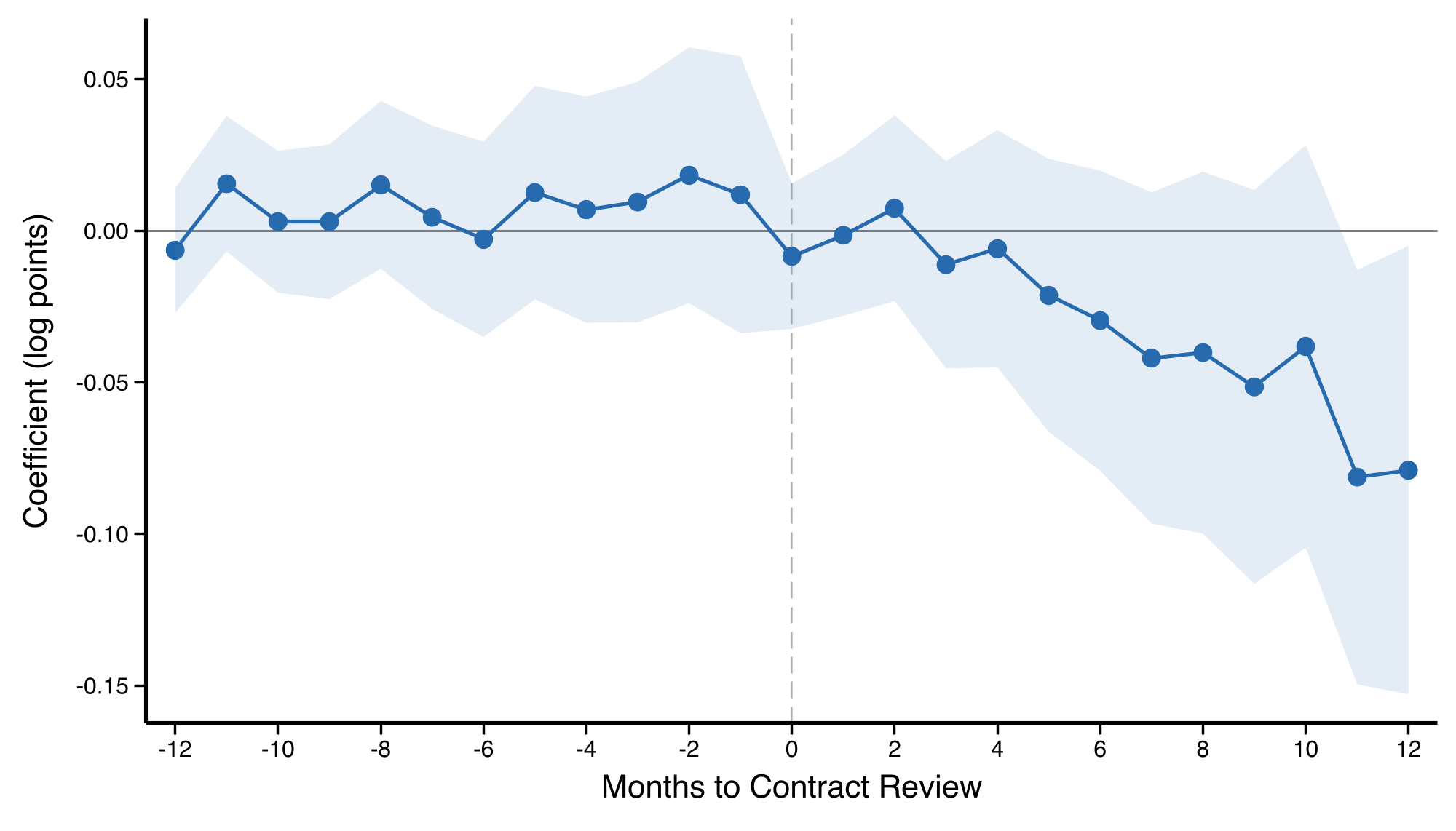}
    \subcaption{Hiring}
\end{subfigure}
\hfill
\begin{subfigure}[t]{0.49\textwidth}
    \centering
    \includegraphics[width=\textwidth]{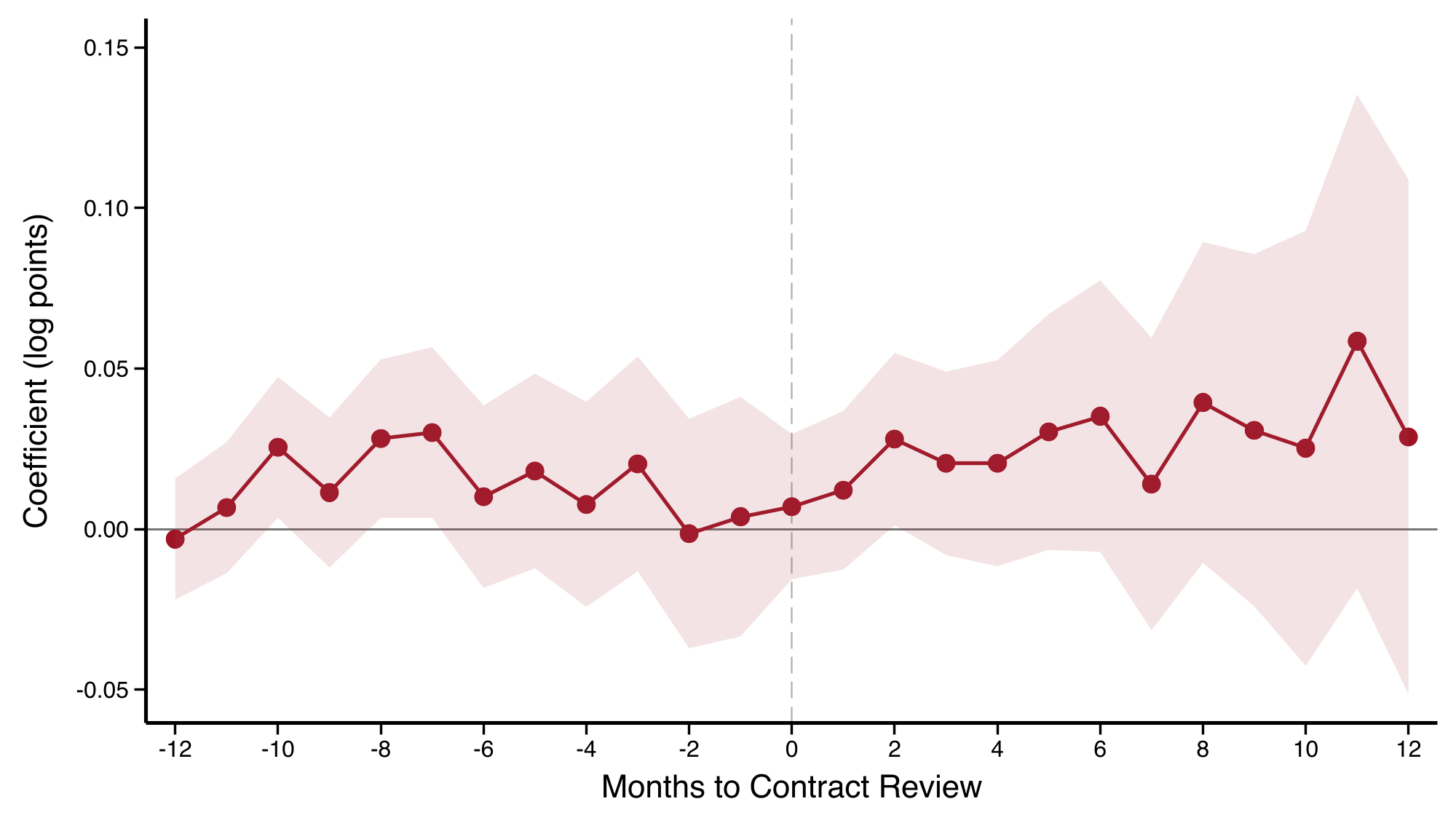}
    \subcaption{Separations}
\end{subfigure}

\par\vspace{1ex}
\figurenotes{The figure reports reduced form event-study coefficients and 95\% confidence intervals on log technical hires and log technical separations. Estimates use the imputation estimator of \citet{borusyak_revisiting_2024} on the balanced panel with firm and calendar-month fixed effects. Standard errors are clustered by firm.}
\end{figure}

We first examine how a scheduled contract review affects overall technical employment flows and stocks. Figure~\ref{fig:technical_flows} plots the monthly reduced form coefficient of technical hiring and separations. Panel~A shows that the reduced form hiring coefficients turn negative approximately five months after a scheduled contract review. By month twelve, technical hiring is 7.6\% ($-0.079$ log points) lower relative to firms that have not yet reached their scheduled review (Table~\ref{tab:technical_flows_m12}, Column~1). Panel~B shows no corresponding increase in separations, which include both voluntary and involuntary worker exits: at month twelve, the point estimate is a statistically insignificant 2.9\% ($0.029$ log points; Column~3). The 95\% confidence interval rules out separation increases larger than roughly 12\%. See Appendix Table~\ref{atab:tech_flows_monthly} for the full monthly coefficients. Together, these results show that firms adjust their technical workforce after a contract review by slowing new hiring rather than increasing separations.

Columns~(2) and~(4) of Table~\ref{tab:technical_flows_m12} show no evidence of a corresponding change on the extensive margin of either hiring or separations (Appendix Figure~\ref{afig:technical_flows_extensive} plots the full extensive margin event studies). On average, firms continue to hire technical workers each month, but in smaller numbers.

\afterpage{
\begin{table}[H]
\centering
\caption{Reduced Form Effects on the Technical Workforce Twelve Months after Contract Review}
\label{tab:technical_flows_m12}
\begin{tablebox}
{
\def\sym#1{\ifmmode^{#1}\else\(^{#1}\)\fi}
\begin{tabular}{lccccc}
\toprule
Dependent variable & \multicolumn{2}{c}{Technical hires} & \multicolumn{2}{c}{Technical separations} & Log technical headcount \\
\cmidrule(lr){2-3} \cmidrule(lr){4-5}
 & Log & Any & Log & Any & \\
 & (1) & (2) & (3) & (4) & (5) \\
\midrule
Month 12 after contract review&      -0.079\sym{**} &      -0.010         &       0.029         &       0.011         &      -0.016         \\
            &     (0.038)         &     (0.017)         &     (0.041)         &     (0.015)         &     (0.018)         \\
\midrule
Firm FE     &\(\checkmark\)         &\(\checkmark\)         &\(\checkmark\)         &\(\checkmark\)         &\(\checkmark\)         \\
Month FE    &\(\checkmark\)         &\(\checkmark\)         &\(\checkmark\)         &\(\checkmark\)         &\(\checkmark\)         \\
Pre-trend joint \(p\)&       0.849         &       0.244         &       0.120         &       0.014         &       0.289         \\
DV mean     &       1.722         &       0.707         &       1.533         &       0.661         &       4.812         \\
Firms       &       5,339         &       5,414         &       5,357         &       5,414         &       5,414         \\
Observations&     149,565         &     221,346         &     143,324         &     221,346         &     221,298         \\
\bottomrule
\end{tabular}
}

\end{tablebox}
\begin{minipage}{\textwidth}
\vspace{3pt}
\tablenotes{The table reports the month-12 reduced form coefficient from the monthly event studies in Appendix Table~\ref{atab:tech_flows_monthly}, estimated with the imputation estimator of \citet{borusyak_revisiting_2024} on the balanced firm-month panel with firm and calendar-month fixed effects. Columns (1) and (3) are log technical hires and separations on firm-months with positive flows; columns (2) and (4) are indicators for any such flow; column (5) is log technical headcount. The pre-trend joint $p$-value tests that the lead coefficients are jointly zero. Regressions are unweighted. Standard errors, clustered by firm, are in parentheses. * $p<0.10$, ** $p<0.05$, *** $p<0.01$.}
\end{minipage}
\end{table}
}

We find no statistically detectable change in total technical headcount twelve months after a contract review, with a point estimate of $-0.016$ log points, or a 1.6\% decline (Column~5 of Table~\ref{tab:technical_flows_m12}). The point estimate on headcount is exactly what would be implied by the slowdown in hiring. The average firm in our sample employs 823 technical workers and hires 17.7 technical workers per month. As an illustrative calculation, applying the estimated 7.6\% hiring decline to this mean, while holding the probability of hiring and separations fixed, implies approximately 1.35 fewer hires per month, or 16 fewer hires over a year. This cumulative shortfall would reduce technical headcount by approximately 2\% after one year, close to the estimated 1.6\% decline. If sustained, the same hiring shortfall would produce a headcount gap of roughly 3.5\% after about two years, approximately the magnitude needed for statistical significance at the 5\% level given the current standard error, although both the effect and its precision may change at longer horizons. These results confirm that because monthly hiring flows represent a minor fraction of overall headcount, even sizable declines in hiring translate into modest cumulative effects on total workforce size over a one-year horizon.

Scaling these reduced form estimates by the first stage yields the implied effect of GitHub Copilot adoption on hiring flows among compliers. The month-twelve Wald estimate indicates that having adopted GitHub Copilot by month twelve reduces monthly technical hiring by approximately 34\% among compliers. In the quarterly specification, which we use for our disaggregated analyses below, the fourth-quarter Wald estimate is $-0.340$ log points, or a 29\% decline in quarterly technical hiring, with a 95\% firm-block bootstrap confidence interval of $[-0.948,\,0.088]$ log points (Appendix Table~\ref{atab:tech_flows_quarterly_latet} reports the full quarterly path). While the horizon-specific Wald ratios are similar in magnitude across quarters, their bootstrap intervals are wide, so the estimates cannot distinguish an immediate, constant effect from one that builds with exposure. Nor can our design separate the two, since identifying the duration of exposure would need additional instruments beyond the timing of contract review. Overall, although the Wald ratios are less precisely estimated than the reduced form because they compound sampling variation from both stages, both sets of estimates point to gradual workforce adjustment through lower hiring rather than worker quits or layoffs.

\subsubsection{Recruiting Activity, Vacancy Duration, and Incidence}

Realized hiring reflects both how many positions firms seek to fill and how quickly they fill open vacancies. A decline in hiring could therefore reflect either lower labor demand or a slower matching process, for instance if AI adoption induced firms to search for scarcer candidates with AI-specific skills. To distinguish between these explanations, Appendix Table~\ref{atab:postings} reports the reduced form effects of a scheduled contract review on three quarterly measures of technical recruiting activity: job postings, duration postings remain open, and any posting activity. Four quarters after a scheduled contract review, the intensive margin of open technical postings falls by 14\% ($-0.148$ log points). In contrast, contract review has no statistically detectable effect on the probability of having any open technical posting or the time a posting is open, although there are some pre-trends in posting duration. The slowdown in technical hiring seems to be driven by firms opening fewer positions rather than taking longer to fill existing vacancies.

Taken together, the employment flow and job posting results show that GitHub Copilot adoption initially slows technical workforce growth by reducing new hiring rather than displacing incumbent workers. Although the direct incidence of this adjustment falls on prospective hires, incumbent workers may still be affected indirectly. As hiring slows across adopting firms, fewer outside job openings can narrow incumbents' outside options, reducing job-to-job mobility and weakening wage bargaining within the firm \citep{caldwell_danieli_2024,jager_roth_roussille_schoefer_2024}.  Concentrating adjustment on the hiring margin could be consistent with models of labor demand with quasi-fixed labor costs or AI coding assistance devaluing the skills of novices \citep{oi_1962,abowd_corbel_kramarz_1999}.

\subsubsection{Placebo and Robustness Checks}\label{sec:robustness}

We evaluate three alternative explanations to verify that the decline in technical hiring reflects GitHub Copilot adoption rather than recurring corporate budgeting cycles, contemporaneous firm-wide hiring freezes or general AI procurement, or differential trends by firm size and sector. First, scheduled contract reviews could in principle coincide with annual corporate budgeting or fiscal planning dates when firms routinely reset hiring targets. To test this directly, we conduct a pre-launch placebo check by assigning each firm its scheduled contract review date prior to GitHub Copilot's 2023 launch. Although this restricts the post-review window to two quarters before GitHub Copilot's launch, a recurring budget-driven contraction decided at the contract review should begin to register in realized hiring within that timeframe. Instead, we find no effects on overall technical hiring, separations, or junior and experienced software engineering hiring (Appendix Table~\ref{atab:placebo_2022_cycle_q0_q2}). Moreover, even after GitHub Copilot's release, the technical hiring decline does not appear as an immediate adjustment at the review date, as an annual budget reset would imply, but instead emerges gradually in the third and fourth quarters after review, tracking the rollout of GitHub Copilot licenses.

Second, even if pre-launch reviews did not affect employment, a firm's 2023 review might coincide with a broader post-2022 white-collar hiring freeze, an IT-wide budget cut, or the simultaneous procurement of general office AI tools (although GitHub Copilot was the only AI product sold by GitHub's dedicated sales force). To test for firm-wide shocks, we examine hiring for accountants, the most common non-technical professional occupation in our data (employed by 98\% of sample firms, with a pre-launch median of 34 per firm) and one exposed to both general AI tools and broader white-collar hiring slowdowns. Appendix Figure~\ref{afig:accountant_placebo} shows no decline in either log accountant hires or the probability of hiring an accountant following contract review. Four quarters after review, the reduced-form estimate for log accountant hiring is a statistically insignificant $0.035$ log points. Furthermore, as we show in Section~\ref{sec:composition}, the slowdown is not uniform even within the technical workforce: hiring is unchanged in less code-intensive technical roles and among experienced engineers, weighing against a broad technical or IT-department hiring freeze.

Third, because firms with earlier contract-review dates have larger pre-launch workforces, our estimates could be confounded if larger and smaller firms, or different sectors, followed distinct post-launch hiring trajectories. In Appendix Table~\ref{atab:additional_controls}, we re-estimate our specifications adding interactions between calendar-quarter fixed effects and pre-launch firm-size terciles, indicators for the information and professional-services sectors, or both. Across all specifications, the four-quarter reduced-form estimates for total technical hiring and separations change by at most $0.008$ log points (Columns~1 and~2), confirming that differential trends by firm size or sector do not drive our findings.

\subsection{Hiring Composition by Code Intensity and Seniority}
\label{sec:composition}
The preceding results show that GitHub Copilot adoption reduces overall technical hiring. We next examine whether this decline is spread evenly across the technical workforce or concentrated among particular roles and seniority levels, changing the composition of new hires.

\subsubsection{Measuring the Coding Content of Technical Work}
\label{sec:code_intensity}

We define a position's code intensity as the share of its described responsibilities devoted to activities within GitHub Copilot's core domain during our study period: writing, completing, testing, documenting, explaining, and debugging code. To measure code intensity across roles and seniority levels, we use the text of LinkedIn position descriptions for technical jobs at sample firms with start dates between 2018 and 2022, prior to GitHub Copilot's launch.\footnote{Because LinkedIn users can edit position descriptions retrospectively, some text may have been written or revised after GitHub Copilot's launch; our sample restriction applies to the position's start date.} Unlike occupation-level task measures that assign a single average to a broad occupational code, position descriptions capture how responsibilities vary both across granular technical roles and across seniority levels within a role. Appendix Figure~\ref{afig:description_examples} illustrates this variation using examples from outside our sample: a software engineer describes building a job-scheduling platform and implementing APIs (Panel~A), whereas a test automation engineer combines programming in Python and LabVIEW with physical design and testing of liquid-nitrogen heat exchangers (Panel~B). Similarly, within infrastructure engineering, a technical lead describes setting team strategy and recruiting, while a technical lead manager combines engineering management with building machine learning infrastructure (Panels~C and~D).

We score each description on this dimension using Claude Sonnet 4.6 (see Appendix Section~\ref{asec:code_intensity_prompt} for the prompt and implementation details). The model observes only the free-text description, without the job title, occupation, seniority, location, firm, or start date. For our baseline cross-role analysis, we average the description-level scores within each of the eight technical roles. Code intensity is thus an \emph{ex ante} measure of task overlap with GitHub Copilot's capabilities rather than a measure of actual time spent coding or realized GitHub Copilot usage.

\begin{figure}[H]
\centering
\caption{Code Intensity of Technical Roles}
\label{fig:coding_intensity}
\includegraphics[width=.8\textwidth]{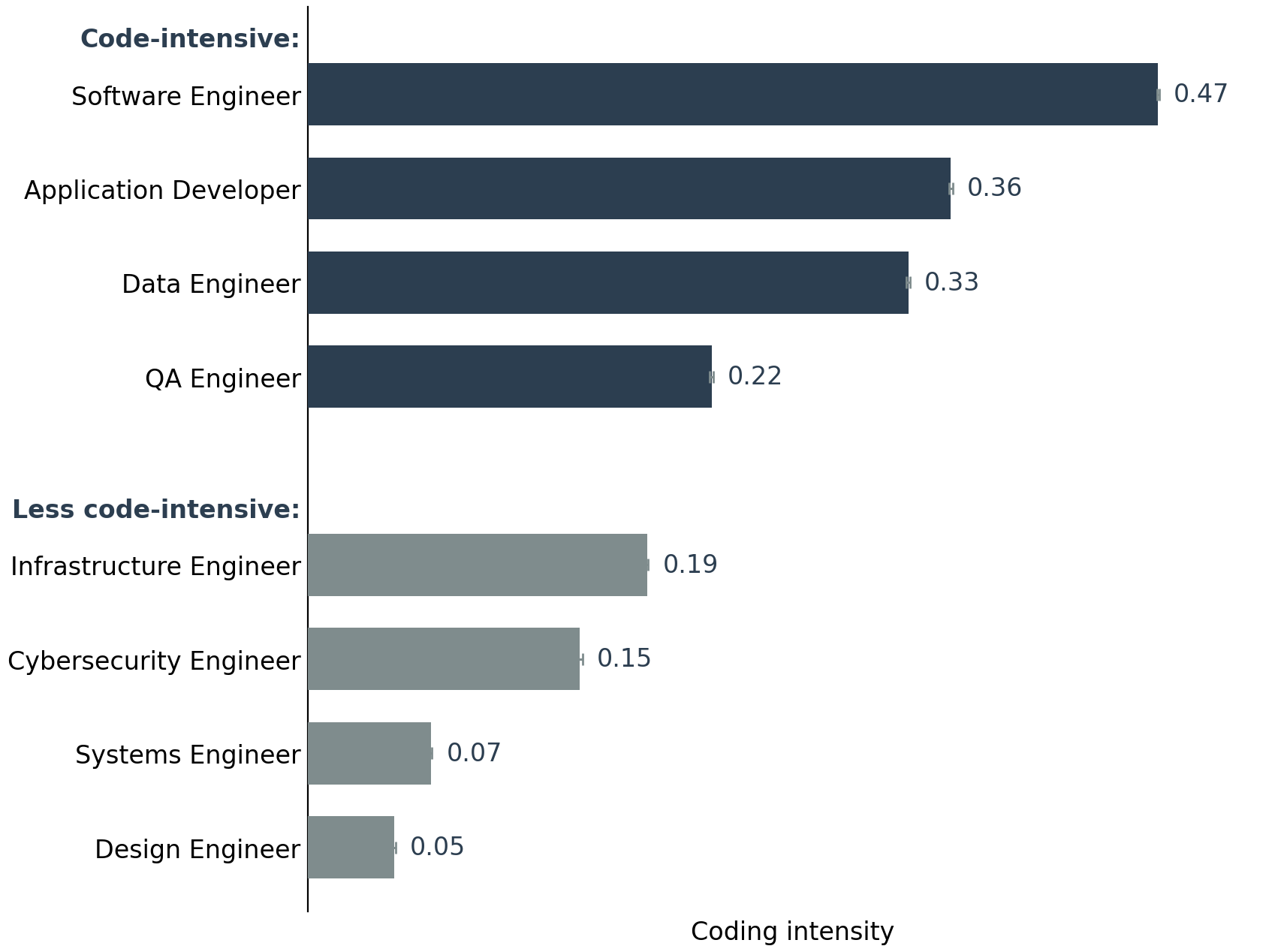}

\figurenotes{The figure reports mean code intensity scores across position descriptions with start dates between 2018 and 2022. Each description receives equal weight. Intervals are normal 95\% confidence intervals based on description-level variation. The four highest-scoring roles form the code-intensive group.}
\end{figure}

Figure~\ref{fig:coding_intensity} reports average code intensity by technical role. Roles whose primary output is software code receive the highest scores, led by software engineering (47\%), application development (36\%), data engineering (33\%), and quality assurance engineering (22\%). We refer to these four roles above the median as code-intensive roles. Coding is less central in infrastructure engineering (19\%), cybersecurity engineering (15\%), systems engineering (7\%), and design engineering (5\%), which, in our sample, primarily captures hardware, silicon, and product design roles where programming supports broader operational, security, or physical engineering responsibilities.

\subsubsection{Hiring Responses by Role and Seniority}
A large empirical literature measures AI exposure at the occupation level by scoring the share of an occupation's tasks that overlap with AI capabilities \citep{eloundou2024gpts, feltenOccupationalHeterogeneityExposure2023}. If occupation-level task overlap directly governs how AI shifts labor demand, hiring responses should vary systematically with a role's pre-launch code intensity. However, pooling all hires within a role may obscure heterogeneity across seniority levels, either because firms reduce junior hiring uniformly across technical roles as in cyclical downturns \citep{forsythe_why_2022}, or because GitHub Copilot substitutes more readily for the coding tasks performed by junior workers than for those bundled into experienced roles. To evaluate how role code intensity and worker seniority shape hiring, we construct a firm-by-role-by-seniority-by-quarter panel and estimate:

\begin{equation} \label{eq:junior_code_interaction}
Y_{irst} = \alpha_{irs} + \lambda_{rst} + \sum_{\ell}\left[ \rho_{\ell} +
\gamma_{\ell}\mathrm{Junior}_{s} +
\theta_{\ell}x_r +
\beta_{\ell}
\left(\mathrm{Junior}_{s}\times x_r\right)
\right] \mathbf{1}\{t-R_i=\ell\} + u_{irst}
\end{equation}

where $Y_{irst}$ is log quarterly hires for firm $i$, role $r$, seniority group $s$, and quarter $t$ in cells with positive hiring. There are two seniority groups: $\mathrm{Junior}_s$ equals one for junior positions (Revelio seniority levels 1 and 2, covering interns, trainees, and early-career workers) and zero for experienced positions (levels 3 and above). The specification includes firm-by-role-by-seniority fixed effects $\alpha_{irs}$ and role-by-seniority-by-quarter fixed effects $\lambda_{rst}$.

Table~\ref{tab:junior_code_interaction} reports the quarter-four estimates across these specifications. Column~(1) begins with a role-level benchmark that pools across seniority on a firm-by-role-by-quarter panel with firm-by-role and role-by-quarter fixed effects. Four quarters after contract review, a one-standard-deviation increase in role code intensity is associated with a statistically insignificant $-0.023$ log-point difference in hiring ($\text{SE} = 0.022$). If \emph{ex ante} task overlap with GitHub Copilot's capabilities were a sufficient summary of labor substitutability, hiring declines should be steepest in the most coding-intensive roles. Instead, role-level code intensity alone does not reliably predict where hiring falls. Column (2) tests for a uniform shift away from junior hiring across all technical roles and similarly finds a small, statistically insignificant junior-experienced differential ($-0.013$ log points, $\text{SE} = 0.040$). Neither occupation-level code intensity nor worker seniority alone predicts where hiring falls.

A sharper pattern emerges in Column~(3), which interacts role code intensity with seniority. Among experienced positions, the estimated code-intensity gradient $\theta_4$ is a statistically insignificant $0.015$ log points. Among junior positions, however, the coding-intensity slope is $0.086$ log points more negative than the experienced slope ($\beta_4 = -0.086$). Combining these terms, a one-standard-deviation increase in role code intensity is associated with a $0.071$ log-point larger decline in junior hiring. Column~(4) replaces the continuous score $x_{r}$ with an indicator for the four code-intensive roles above the median: the junior-experienced hiring differential is $0.227$ log points more negative in code-intensive roles than in less code-intensive roles, and the implied high-minus-low contrast in junior hiring is $-0.180$ log points. Because the two role groups sit 1.6 standard deviations apart on the score, the Column~(3) slope implies a high-low contrast of about $0.14$ log points, which is very close to the continuous estimate.

\afterpage{
\begin{table}[H]
\centering
\caption{Hiring Effects by Code Intensity and Seniority}
\label{tab:junior_code_interaction}
\begin{tablebox}
{
\def\sym#1{\ifmmode^{#1}\else\(^{#1}\)\fi}
\begin{tabular}{lcccc}
\toprule
Dependent variable & \multicolumn{4}{c}{Log quarterly hires} \\
\cmidrule(lr){2-5}
 & (1) & (2) & (3) & (4) \\
\midrule
Quarter 4 after contract review (\(\rho_4\))&      -0.012         &      -0.018         &      -0.022         &      -0.023         \\
            &     (0.036)         &     (0.042)         &     (0.042)         &     (0.041)         \\
\addlinespace
Code intensity (\(\theta_4\))&      -0.023         &                     &       0.015         &       0.048         \\
            &     (0.022)         &                     &     (0.023)         &     (0.041)         \\
\addlinespace
Junior (\(\gamma_4\))&                     &      -0.013         &       0.008         &       0.007         \\
            &                     &     (0.039)         &     (0.040)         &     (0.039)         \\
\addlinespace
Junior \(\times\) code intensity (\(\beta_4\))&                     &                     &      -0.086\sym{***}&      -0.227\sym{***}\\
            &                     &                     &     (0.030)         &     (0.061)         \\
\midrule
Unit of observation & Role & Role \(\times\) seniority & Role \(\times\) seniority & Role \(\times\) seniority \\
Firm \(\times\) role fixed effects & \(\checkmark\) &  &  &  \\
Role \(\times\) quarter fixed effects & \(\checkmark\) &  &  &  \\
Firm \(\times\) role \(\times\) seniority fixed effects &  & \(\checkmark\) & \(\checkmark\) & \(\checkmark\) \\
Role \(\times\) seniority \(\times\) quarter fixed effects &  & \(\checkmark\) & \(\checkmark\) & \(\checkmark\) \\
\midrule
Implied junior gradient (\(\theta_4+\beta_4\)) &  &  & -0.071\sym{***} & -0.180\sym{***} \\
 &  &  & (0.024) & (0.053) \\
Firms & 5,048 & 5,048 & 5,048 & 5,048 \\
Observations & 231,880 & 357,346 & 357,346 & 357,346 \\
\bottomrule
\end{tabular}
}

\end{tablebox}
\begin{minipage}{\textwidth}
\vspace{3pt}
\tablenotes{The table reports reduced form effects on log quarterly hires four quarters after scheduled contract review. Column (1) estimates Equation~\ref{eq:junior_code_interaction} at the role-by-firm sample. Columns (2)--(4) estimate Equation~\ref{eq:junior_code_interaction} on a common firm-by-role-by-seniority sample. Column (4) replaces it with an indicator for the four roles above the median, centered at one half, so its $\theta_4$ and $\beta_4$ are high-minus-low contrasts and its junior main effect is evaluated at the midpoint of the two groups. Junior positions are Revelio levels 1--2 and more experienced positions are levels 3--7. A firm enters a role's sample if its average pre-launch headcount in that role is at least one in both the junior and the experienced group. Standard errors, clustered by firm, are in parentheses. The implied junior gradient is the linear combination $\theta_4+\beta_4$ of the estimates above it, with its standard error from the estimated covariance. \sym{*}~$p<0.10$, \sym{**}~$p<0.05$, \sym{***}~$p<0.01$.}
\end{minipage}
\end{table}
}

\begin{figure}[H]
\centering
\caption{Hiring Effects by Role and Seniority}
\label{fig:rf_hiring_by_role_seniority_q4}
\includegraphics[width=.8\textwidth]{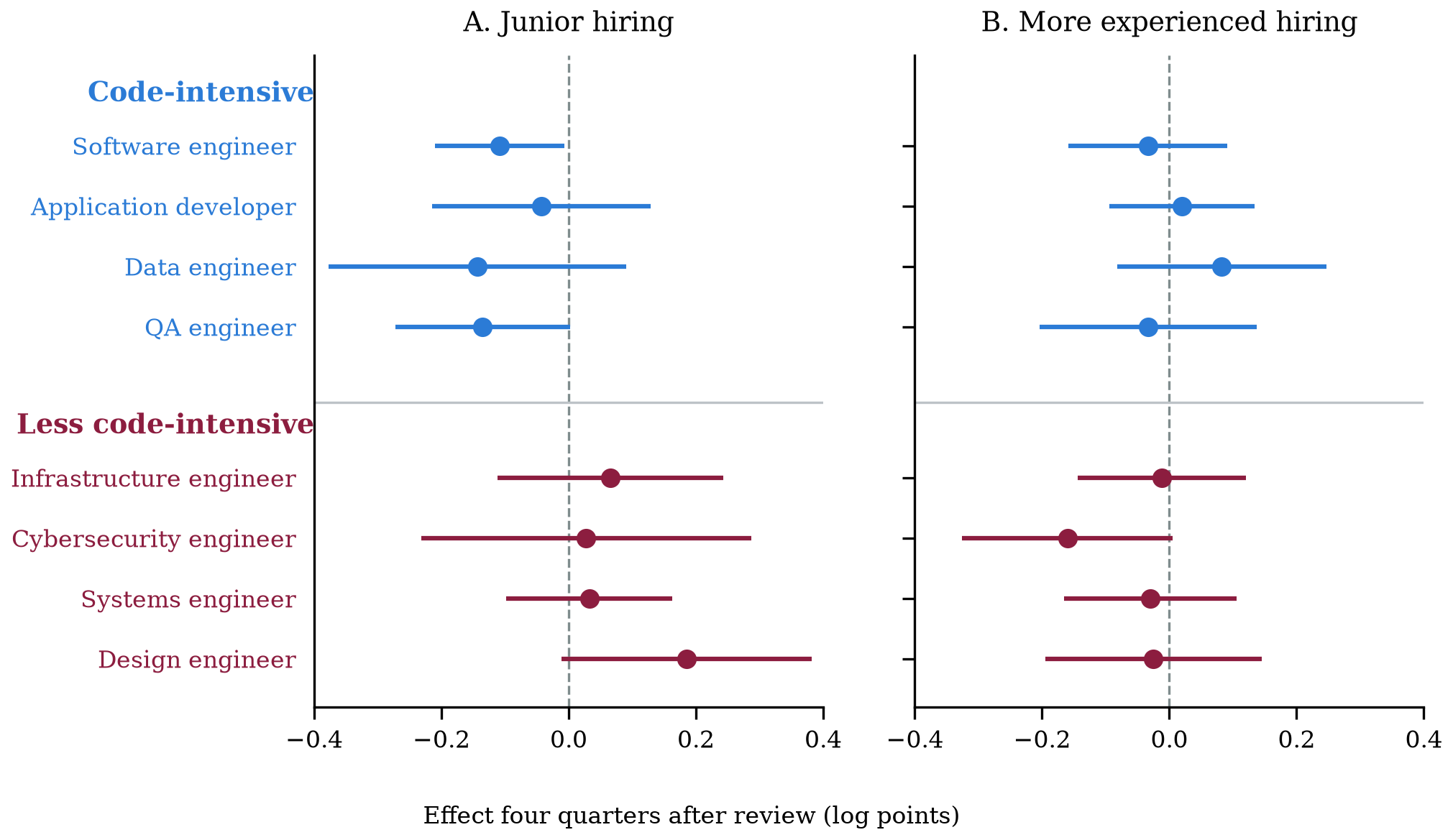}
\par\vspace{1ex}
\figurenotes{The figure reports role-specific reduced form effects four quarters after scheduled contract review. Panels A and B show log hire spells for junior (levels 1--2) and more experienced (levels 3--7) positions, estimated on positive-hiring firm--role--quarters. Roles are ordered by pre-GitHub Copilot code intensity. Horizontal whiskers are 95\% confidence intervals based on standard errors clustered by firm.}
\end{figure}

Figure~\ref{fig:rf_hiring_by_role_seniority_q4} plots the separate four-quarter reduced-form hiring estimates for each of the 16 role-by-seniority cells, ordered by role code intensity. While the individual cell estimates are noisier than the pooled regression, they show a consistent pattern across the role distribution rather than a single outlier occupation. In Panel~A, the point estimates for junior hiring are negative in all four code-intensive roles and positive in all four less code-intensive roles, whereas Panel~B shows no comparable ordering across roles for experienced hiring.

This divergence is robust across alternative measures and within-firm specifications. Although junior positions in code-intensive roles involve somewhat more coding than experienced positions in the same role (Appendix Figure~\ref{afig:coding_intensity_rolesen}), replacing the role-level score $x_{r}$ with cell-specific pre-launch coding-intensity scores $x_{rs}$ leaves the estimates nearly unchanged: the junior gradient is $-0.062$ log points, the experienced gradient is $0.020$ log points, and their difference is $-0.082$ log points (Columns~1 and~2 of Appendix Table~\ref{atab:gradient_robust}). 

To isolate how an individual firm adjusts the seniority mix within a role, we also re-estimate the model using the within-firm log junior-to-experienced hiring ratio, $\log(H_{irt}^{J}/H_{irt}^{E})$, among firms that hire both groups in role $r$ and quarter $t$, which differences out any firm-by-role-by-quarter shock that scales hiring proportionally across seniority levels. Four quarters after contract review, a one-standard-deviation increase in role code intensity reduces a firm's junior-to-experienced hiring ratio by 7.7\% ($-0.080$ log points, $p < 0.05$; Column~3 of Appendix Table~\ref{atab:gradient_robust}), and an indicator for the four code-intensive roles reduces the ratio by 15.5\% ($-0.169$ log points, $p < 0.05$; Column~4 and Appendix Figure~\ref{afig:rf_hiring_ratio_by_role_q4}).

Across role groups, contract review does not affect the probability of positive quarterly hiring or of hiring both junior and experienced workers in the same quarter, confirming that our intensive-margin and ratio estimates are not driven by extensive-margin selection (Appendix Table~\ref{atab:extensive_margin_main_results}). Finally, as in Section~\ref{sec:robustness}, these role-by-seniority estimates show no pre-launch placebo effects and are unchanged when adding firm-size-by-quarter and industry-by-quarter fixed effects (Appendix Tables~\ref{atab:placebo_2022_cycle_q0_q2} and~\ref{atab:additional_controls}).

\subsubsection{Implied Magnitudes and Compositional Shifts}
\label{sec:magnitudes}

\begin{figure}[H]
\centering
\caption{Implied Hiring Levels and Composition by Code Intensity and Seniority}
\label{fig:magnitudes}

\begin{subfigure}[t]{0.48\textwidth}
    \centering
    \includegraphics[width=\textwidth]{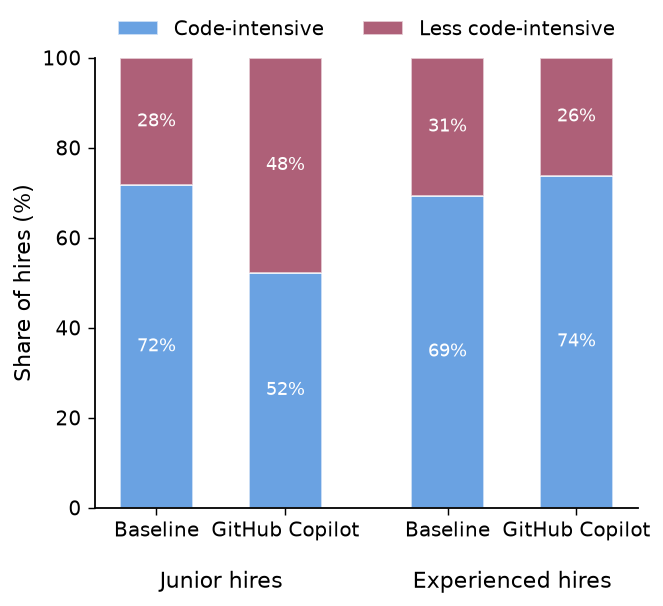}
    \subcaption{Composition by code intensity, within seniority}
\end{subfigure}
\hfill
\begin{subfigure}[t]{0.48\textwidth}
    \centering
    \includegraphics[width=\textwidth]{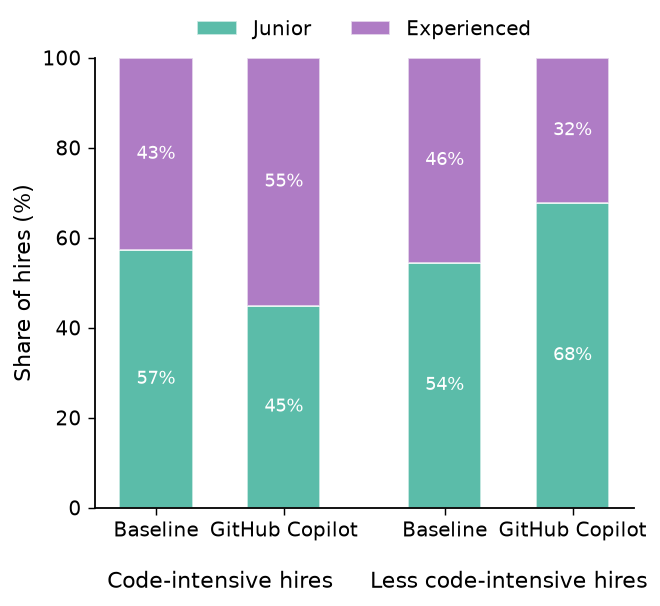}
    \subcaption{Composition by seniority, within code-intensity}
\end{subfigure}

\par\vspace{2ex}
\begin{subfigure}[t]{0.9\textwidth}
    \centering
    \includegraphics[width=\textwidth]{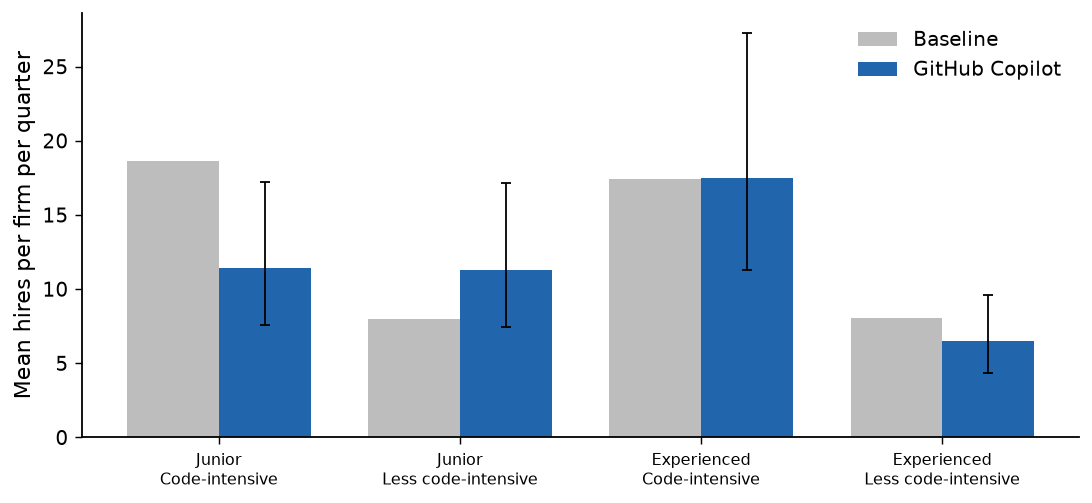}
    \subcaption{Quarterly hires per firm, by code intensity and seniority}
\end{subfigure}

\par\vspace{1ex}
\figurenotes{Baseline values are mean quarterly hire spells per sample firm in screened firm--role pairs in Q4 of 2022. The with ``GitHub Copilot'' values scale each baseline mean by $\exp(\hat\rho_g/\hat\pi)$, where $\hat\rho_g$ is group $g$'s reduced-form estimate four quarters after review from Table~\ref{tab:junior_code_interaction} column~(4), and $\hat\pi$ is the full-sample first stage at the same horizon. Panel A reports the code-intensive share of junior hires and of experienced hires; Panel B reports the junior share of code-intensive hires and of less code-intensive hires. Whiskers in Panel C are 95\% confidence intervals on $\hat\rho_g$, scaled the same way; they do not reflect first-stage uncertainty. Panels A and B show no intervals.}
\end{figure}

Because the hiring contraction is concentrated among junior positions in code-intensive roles, GitHub Copilot adoption reshapes both the entry paths of junior workers and the seniority structure of technical teams (Figure~\ref{fig:magnitudes}). At baseline, 72\% of all junior technical hires enter through code-intensive roles; with GitHub Copilot adoption, that share falls to 52\%, leaving entry-level hiring nearly evenly split between more and less code-intensive roles while the role mix of experienced hiring is largely unchanged (69\% to 74\% code-intensive; Panel~A). Viewed within role groups, the junior share of code-intensive hiring drops from 57\% to 45\%, whereas the junior share in less code-intensive roles rises from 54\% to 68\% (Panel~B). In level terms (Panel~C), junior hiring in code-intensive roles drops by about 7 spells per firm-quarter, while experienced hiring in these roles is fairly flat. Comparing the two ends of the coding-intensity distribution directly illustrates this divergence (Appendix Table~\ref{atab:latet_rolesen}; see Appendix Figure \ref{afig:swe_sys} for the monthly reduced form event studies). GitHub Copilot adoption reduces quarterly junior software engineering hiring by approximately 40\% ($-0.513$ log points), with no detectable decline among experienced software engineers ($-0.160$ log points) or systems and infrastructure engineers ($0.212$ log points).

Through the lens of our conceptual framework, a common change in the scale of software production cannot explain why junior and experienced hiring diverge within the same role. Instead, this divergence reflects differences across seniority levels in task-level productivity gains, job bundling, or task allocation. First, GitHub Copilot may generate larger direct labor-saving productivity gains on the coding tasks typically assigned to junior workers. Second, junior and experienced positions differ in how coding is bundled with other responsibilities, which explains why the seniority gap persists even when coding intensity is measured at the role-by-seniority cell level \citep{garicano_messy_2026}. Whereas junior positions in code-intensive roles center on discrete implementation tasks that can be specified, delegated, and evaluated independently at relatively low cost, experienced positions bundle code production with system architecture, domain judgment, and cross-team accountability. Because coding is more separable from the rest of the job in junior positions, GitHub Copilot may substitute more readily for junior hiring even if it improves productivity in both groups.

At the same time, the flat code-intensity gradient for experienced hiring does not imply that GitHub Copilot leaves experienced work unchanged or simply complements experienced labor. Even as GitHub Copilot saves experienced engineers time on their existing coding tasks, firms may expand the scope of work assigned to them, or change their time allocation. Experienced engineers may use GitHub Copilot to complete implementation work previously delegated to junior teammates, or spend more time reviewing, debugging, and integrating AI-generated code alongside their core architectural and coordination duties. These offsetting forces, labor savings on existing coding tasks alongside task reallocation from junior positions and new verification demands, could leave net experienced hiring approximately flat across the code-intensity distribution even as GitHub Copilot reshapes their day-to-day work.

\subsection{Experience, Credentials, and Internal Hiring among Software Engineers}

Although firms maintain their volume of experienced software engineering hires after adopting GitHub Copilot, stable hiring volume could mask changes in the types of engineers firms recruit. Within software engineering, the largest and most code-intensive occupation in our sample, we next examine whether firms adjust the educational credentials, prior experience, and internal sourcing of the workers they hire. At baseline, experienced software engineers average 8.7 years since earning their bachelor's degree, compared with 4.8 years among junior engineers, and are significantly more likely to hold an academic master's or doctoral degree (Appendix Table~\ref{atab:stock_educ_background}).

In principle, AI coding tools could either substitute for or complement accumulated general and firm-specific human capital. If GitHub Copilot makes programming knowledge and codebase conventions readily accessible on the job, firms may relax degree, experience, or internal-tenure requirements when filling positions. Conversely, if deeper technical training, industry experience, or familiarity with the firm's architecture makes engineers more effective at directing AI tools, evaluating generated code, and coordinating system design, firms could raise the bar on these qualifications among new hires.

\begin{table}[H]
\centering
\caption{Contract Review, Qualifications, and Sources of Software Engineering Hires}
\label{tab:hire_experience_sourcing}
\begin{tablebox}
\def\sym#1{\ifmmode^{#1}\else\(^{#1}\)\fi}
\begin{tabular}{lcc}
\toprule
Dependent variable & Junior SWE & Experienced SWE \\
 & (1) & (2) \\
\midrule
\multicolumn{3}{l}{\textit{A. Credentials and experience}} \\
\quad Graduate degree share & -1.83 & 4.69\sym{**} \\
 & (2.23) & (2.21) \\
\quad Share 5+ years since BA & -3.32 & 4.29\sym{*} \\
 & (3.05) & (2.39) \\
\quad Years since BA & -0.01 & 1.28\sym{**} \\
 & (0.40) & (0.51) \\
\addlinespace
\multicolumn{3}{l}{\textit{B. Hiring by source}} \\
\quad Internal fills & -0.103 & 0.111 \\
 & (0.085) & (0.085) \\
\quad External fills & -0.096\sym{*} & -0.140\sym{**} \\
 & (0.058) & (0.067) \\
\midrule
Firm FE & \(\checkmark\) & \(\checkmark\) \\
Quarter FE & \(\checkmark\) & \(\checkmark\) \\
\bottomrule
\end{tabular}

\end{tablebox}
\begin{minipage}{\textwidth}
\vspace{3pt}
\tablenotes{The table reports reduced form estimates of the effect of contract review, pooling event quarters 3 and 4. Junior software engineers are levels 1--2 in Column 1, and experienced software engineers are levels 3--7 in Column 2. Share outcomes are reported in percentage points, mean experience is reported in years, and hiring outcomes are reported in log points. Graduate degrees include academic master's degrees and doctoral degrees. Internal fills are new position spells for workers employed at the same firm in the year prior. Estimates use the \citet{borusyak_revisiting_2024} imputation estimator; Standard errors clustered by firm are in parentheses. \sym{*}~$p<0.10$, \sym{**}~$p<0.05$, \sym{***}~$p<0.01$.}
\end{minipage}
\end{table}

Panel~A of Table~\ref{tab:hire_experience_sourcing} tests these predictions by estimating the reduced form effect of contract review, pooling the third and fourth quarters after review, on the education and experience of software engineering hires. Among junior software engineering positions (Column~1), where total hiring falls sharply, we find no statistically detectable change in the educational credentials or prior experience of the workers who are hired. Among experienced positions (Column~2), by contrast, stable total hiring masks a shift toward candidates with greater general human capital. The share of experienced hires holding a graduate degree rises by 4.7 percentage points, and the share who are at least five years past their bachelor's degree rises by a similar magnitude. On average, newly hired experienced software engineers have 1.28 additional years of potential experience since completing their bachelor's degree. Rather than lowering qualification thresholds, firms raise the bar for experienced software engineering roles.

Panel~B examines whether firms also shift toward internal candidates who possess firm-specific human capital or whose capabilities are better known to the firm \citep{becker_investment_1962, li_hiring_2026}. We define a position fill as internal if the worker held a position at the same firm in the prior year, such as an internship or prior full time role, and as external otherwise. Among experienced software engineering positions, external hires fall by approximately 13\%, whereas internal fills do not show statistically significant change. Among junior positions, both internal and external fills decline by similar proportions, with no statistically distinguishable shift toward internal sourcing.

Together, Panels~A and~B reveal an asymmetry in how firms adjust their software engineering workforce across the seniority distribution. Firms reduce the quantity of junior software engineering hires without changing their observable qualifications, while maintaining the volume of experienced hires and upgrading their education, prior experience, and internal tenure. Although AI coding assistants may deliver large task-level productivity gains to novices in experimental settings, in practice they may complement accumulated general and firm-specific human capital.

\subsection{Geographic Composition of Technical Hiring}\label{sec:geography}
Beyond shifting the roles, seniority, and credentials of new hires, AI coding assistants could also reshape where firms locate technical work. Firms in our sample employ technical workers both domestically and offshore, with India the most common offshore location. During 2021–2022, approximately 58\% of firms employed at least one technical worker in India, while 83\% employed at least one worker there in any occupation. This concentration reflects India's emergence as a leading global hub for offshore software development. Beginning in the early 2000s, India developed a large export-oriented software services industry, supported by a substantial pool of English-speaking engineers, strong technical universities, and labor costs well below those in the United States and Europe \citep{arora_underdogs_2005, parthasarathy_changing_2013}.

In principle, AI coding assistants could disproportionately benefit offshore workers through several channels. First, LLMs can facilitate communication across language barriers, reducing coordination costs associated with geographically distributed engineering teams \citep{brynjolfsson_li_raymond_2025}. Second, coding assistants can help workers interpret technical specifications, navigate unfamiliar codebases, and understand internal systems. These capabilities may be particularly valuable for offshore engineers, who often have less access to the informal knowledge and institutional context available to colleagues at a firm's headquarters. Third, given substantial wage differentials between domestic and offshore engineers, AI tools that reduce these productivity and coordination disadvantages could make offshore labor an even more attractive substitute for domestic labor. 

On the other hand, tasks that can be performed offshore may be more easily codified and therefore more susceptible to automation. Moreover, firms may be more willing to reduce hiring outside their headquarters country, where reputational concerns, social pressures, and information frictions may impose fewer constraints on employment adjustments \citep{landier2009trade, bassanini2017not}.

To examine this margin, Appendix Table \ref{atab:india_dom_hiring_preindia} estimates the effects of contract review on technical hiring in firms' headquarters countries (``domestic'') and India, restricting the sample to the 4,509 firms with a presence in India during the pre-period. The reduced-form estimates are imprecise, partly reflecting the smaller sample and sparser hiring abroad, and do not reveal statistically significant differences in hiring responses across locations. Four quarters after contract review, the point estimate for India-based technical hiring is $-0.060$ log points ($\text{SE} = 0.189$), compared with $-0.029$ log points for domestic technical hiring. Although the estimates are too imprecise to detect differences across locations reliably, they provide no evidence of a shift toward offshore technical hiring following GitHub Copilot adoption. If anything, the point estimates suggest a larger slowdown in India.

\subsection{Contribution of GitHub Copilot to the Overall Decline in Technical Hiring}\label{sec:decomposition}

Finally, we benchmark our treatment effect estimates against the broader post-2022 contraction in technical recruiting. Between 2022 and 2024, our last full calendar year in the data, mean quarterly technical hiring across sample firms fell by 29\%, or 16.1 hires per firm-quarter (from 55.3 to 39.2). To evaluate how much of this aggregate slowdown can be attributed to GitHub Copilot adoption, we use our treatment effect estimates to impute counterfactual post-period hiring. Let $H_{i0}$ and $H_{i1}$ denote mean quarterly technical hires at firm $i$ in 2022 and 2024, and let $H_{i1}(0)$ denote the hires firm $i$ would have made in 2024 had it not adopted GitHub Copilot. We decompose the observed decline in mean quarterly hiring per firm, $\Delta$, into a component attributable to GitHub Copilot adoption and a residual component capturing other factors:
\begin{equation}
    \Delta = \frac{1}{N}\sum_i \big(H_{i0} - H_{i1}\big)
           = \underbrace{\frac{1}{N}\sum_i \big(H_{i1}(0) - H_{i1}\big)}_{\Delta_{\text{GitHub Copilot}}}
           + \underbrace{\frac{1}{N}\sum_i \big(H_{i0} - H_{i1}(0)\big)}_{\Delta_{\text{Other}}}.
    \label{eq:decomp_mag}
\end{equation}
For non-adopters, $H_{i1}(0) = H_{i1}$, so only adopters contribute to $\Delta_{\text{GitHub Copilot}}$. For a firm with an estimated log hiring effect $\beta_{i1}$, counterfactual hiring without adoption is $H_{i1}(0) = H_{i1}e^{-\beta_{i1}}$, and its forgone hiring is $H_{i1}(e^{-\beta_{i1}} - 1)$. We evaluate forgone hiring across the event quarters following each firm's 2023 contract review ($\ell = 0, \dots, 4$) and average across quarters and firms to obtain $\hat{\Delta}_{\text{GitHub Copilot}}$. See Appendix Section~\ref{asec:decomposition} for additional details and caveats.

\begin{figure}[H]
\centering
\caption{Counterfactual Hiring}
\label{fig:decomposition}
    \includegraphics[width=0.65\textwidth]{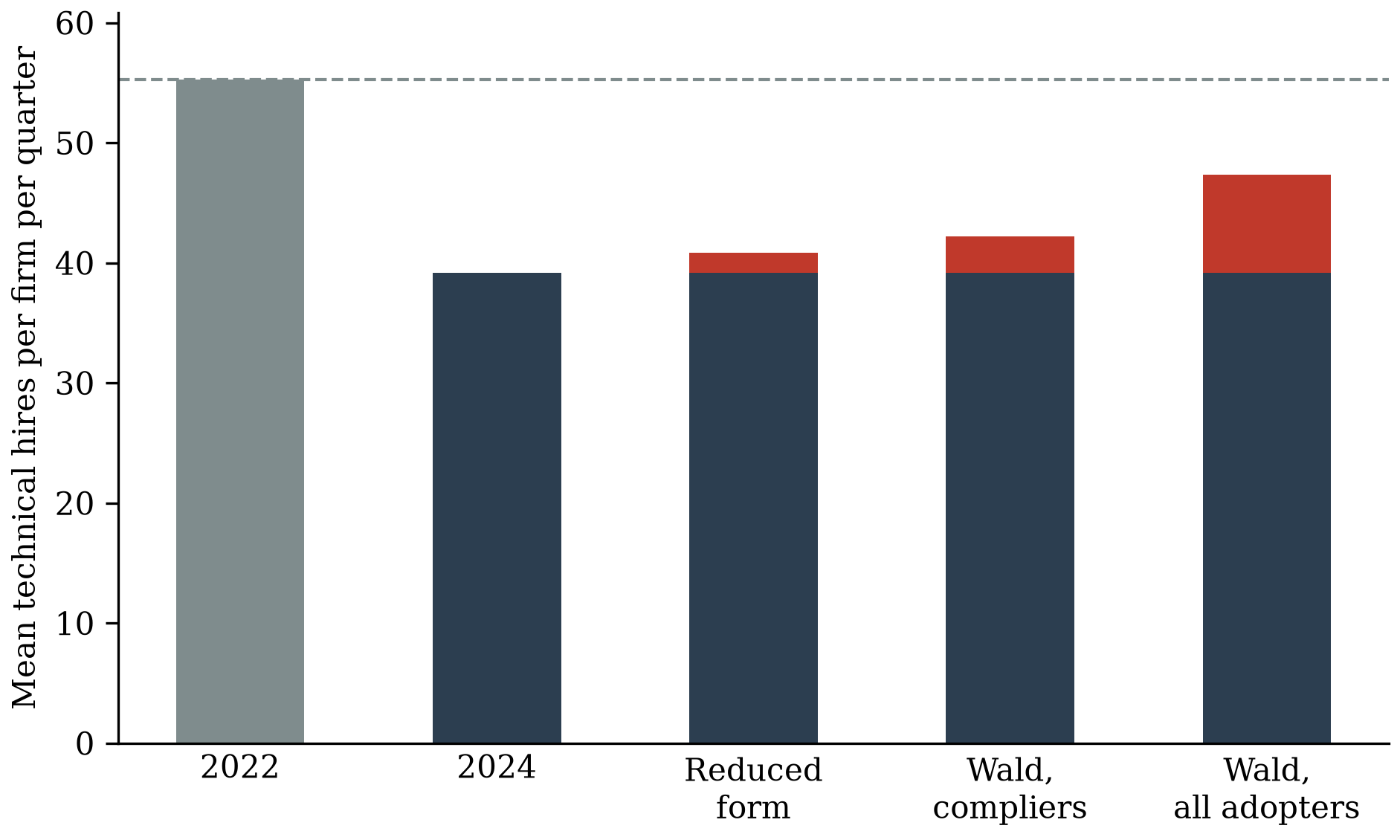}

\par\vspace{1ex}
\figurenotes{This figure plots results from a counterfactual exercise estimating how much of the overall decline in average quarterly technical hiring between 2022 and 2024 can be attributed to GitHub Copilot adoption. Calculation details are in Appendix Section~\ref{asec:decomposition}.}
\end{figure}

Figure~\ref{fig:decomposition} reports $\Delta_{\text{GitHub Copilot}}$ under three progressively stronger assumptions about which firms the estimated effects apply to:

\begin{enumerate}
    \item \textit{Reduced form counterfactual.} Applying the reduced-form coefficient $\hat{\rho}$ to all firms captures the intent-to-treat effect of reaching a scheduled contract review. This generates 1.7 forgone hires per firm-quarter or 10\% of the overall aggregate decline.
    \item \textit{Wald, compliers counterfactual.} Applying the Wald estimate $\hat{\theta}$ to adopting firms, scaled by the complier share of adopters, attributes 3 forgone hires per firm-quarter, or 19\% of the overall decline, to GitHub Copilot. Because adopting firms are larger and hire more workers at baseline than non-adopters, these estimates are larger than the reduced form benchmark.
    \item \textit{Wald, all adopters counterfactual.} This makes the stronger assumption that all adopters experience the same proportional hiring decline and path as compliers. This produces an estimated reduction to 8.1 hires per firm-quarter, or 51\% of the overall decline.
\end{enumerate}

A simple back-of-the-envelope calculation grounds this third benchmark: because our fourth-quarter treatment effect is similar in magnitude to the overall 2022 to 2024 hiring decline (29\%) and roughly half of the sample had adopted GitHub Copilot by 2024, applying the complier effect to all adopters accounts for about half of the aggregate slowdown, with half driven by other macroeconomic factors.

We emphasize that these calculations are illustrative benchmarks rather than exact counterfactuals for four reasons. First, treatment effects are estimated in event time over the year following each firm's 2023 contract review (spanning 2023Q1 through 2024Q4 across cohorts), whereas the aggregate decline $\Delta$ is measured between calendar years 2022 and 2024. The decomposition therefore treats the average effect over the first year post-review as the average effect in the 2024 post-period. Second, $\widehat{\rho}_{\ell}$ and $\widehat{\theta}_{\ell}$ are estimated in logs on unweighted firm-quarters with positive hiring, so applying them to firm hiring levels assumes treatment effects are homogeneous across firms. Third, the all-adopters benchmark (Appendix Equation~\eqref{eq:decomp_all}) does not have a clean upper bound. Because always-takers adopt earlier on average than compliers, they may have initially deployed earlier, less capable versions of the tool, or benefit more from the scale effects of AI adoption, leading to smaller hiring reductions. On the other hand, they could have larger hiring slowdowns. Fourth, the calculation is partial equilibrium: it treats the hiring of non-adopting firms as invariant to adoption elsewhere, whereas in equilibrium some candidates not hired by adopting firms may be absorbed by other employers in the sample. Finally, our design cannot separate which specific factors drove the remaining half to four-fifths of the hiring slowdown, such as monetary tightening, the correction of pandemic-era overhiring, changes in the tax treatment of software research and development, or AI adoption operating through other channels.

\section{Conclusion}
As AI systems become more capable, a central question is how firms adjust their workforces. Which roles will experience increases, decreases, or no change in demand? Will firms adjust on the hiring margin, by reallocating or separating from incumbent workers, or all three? We study these margins among large, technology-intensive multinational firms adopting GitHub Copilot, an early enterprise AI coding assistant, exploiting variation in the timing of pre-existing enterprise software contract reviews to identify causal effects. We find that the initial labor market adjustment occurs on the hiring margin rather than through separations. Rather than reducing recruitment uniformly, adopting firms cut entry-level hiring in code-intensive roles while maintaining the volume of experienced hires and raising the bar on their education, prior experience, and firm-specific tenure. The early impact of AI coding assistance thus appears less as broad job displacement and more as a narrowing of entry-level pipelines alongside greater demand for accumulated human capital. Thus, even when generative AI provides the largest task-level productivity boost to novices, those individual gains need not translate into greater labor demand.

Our findings also point to several open questions about how the labor market effects of AI will evolve across firms and over time. First, our estimates characterize large, established multinationals operating complex software systems. Smaller or younger firms facing fewer coordination constraints may reorganize production more rapidly, or respond to lower software development costs by expanding output and technical hiring more elastically. Second, our sample period captures the first wave of AI coding assistants, which operated primarily by autocompleting code within a developer's workflow. Newer agentic systems capable of executing multi-step engineering tasks autonomously could deepen substitution or extend it further up the seniority distribution. Conversely, if early corporate adjustments partly reflected expectations that outpaced realized productivity gains, or if shrinking the junior cohort eventually creates bottlenecks in the supply of senior talent, firms' hiring strategies may rebalance over a longer horizon.

Finally, our firm-level estimates do not capture wage adjustments or aggregate employment. Slower hiring and shifts in skill demand can alter compensation and the division of surplus between firms and workers even when total headcount is stable. Moreover, a contraction in a specific entry channel at large adopting firms need not translate one-for-one into lower aggregate technical employment \citep{raymondjmp, acemoglu_competing_2020}. Workers who would otherwise have joined these firms as junior software engineers may instead enter adjacent roles such as systems or infrastructure engineering, or take software engineering positions at smaller or later-adopting firms. On the other hand, firms who adopt AI may steal market share from other firms. Determining how these firm-level shifts in hiring and career ladders reshape market-wide employment, wages, and long-run labor demand remains a central question for future research.

\vspace{10pt}

\newpage

\clearpage
\begin{spacing}{0.8}
\bibliographystyle{aea}
\bibliography{Bibliography.bib}
\end{spacing}

\begin{appendix}

\renewcommand{\thefigure}{A.\arabic{figure}}
\setcounter{figure}{0}
\renewcommand{\thetable}{A.\arabic{table}}
\setcounter{table}{0}

\newpage
\clearpage
\noindent \textbf{\Large{Appendix Materials -- For Online Publication}}


\section{Background on GitHub and AI Coding Tools}

\subsection{GitHub}
\begin{figure}[H]
\centering
\caption{GitHub Repository and Pull Request Workflow}
\label{afig:github_demo}
\begin{subfigure}[t]{0.5\textwidth}
    \centering
    \includegraphics[width=\textwidth]{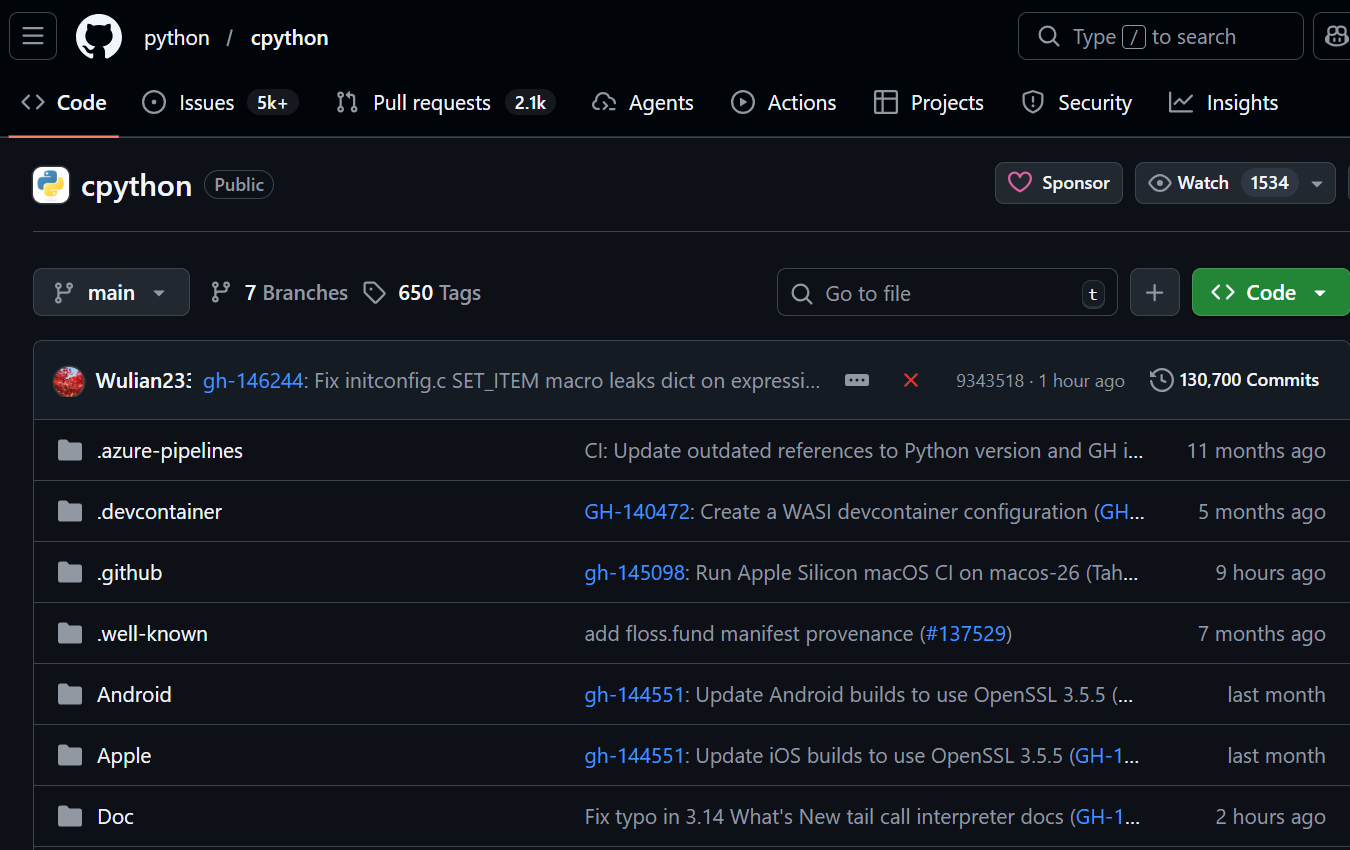}
    \subcaption{Public repository for the Python language}
\end{subfigure}
\begin{subfigure}[t]{0.5\textwidth}
    \centering
    \includegraphics[width=\textwidth]{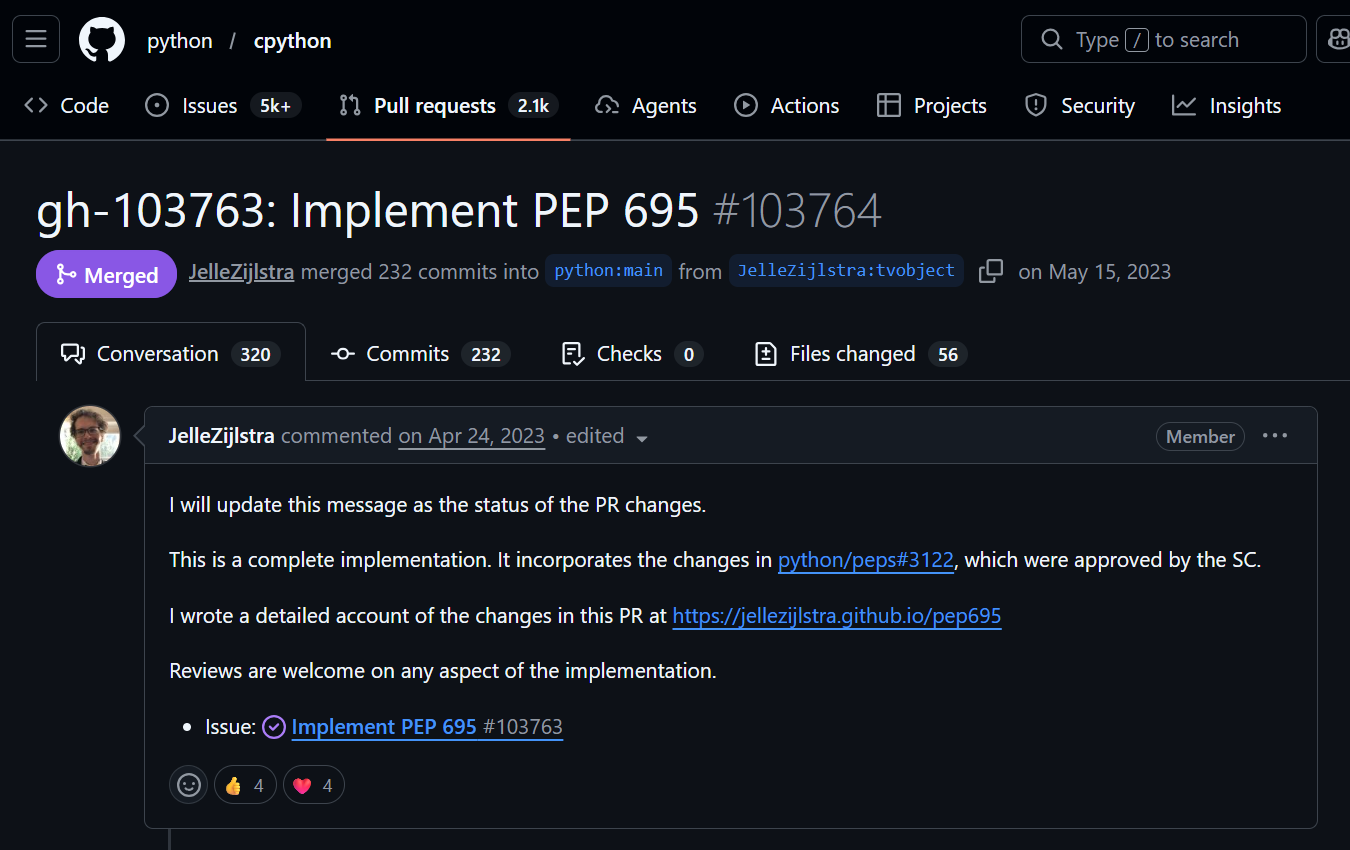}
    \subcaption{``Pull request'' seeking to update the Python repository}
\end{subfigure}

\par\vspace{1ex}
\figurenotes{Screenshots of the public Python-language repository and a pull request proposing a change to it.}
\end{figure}

Pull-based software development originated in the open-source community. In 2005, Linus Torvalds released Git, a distributed version control system that allowed developers to maintain independent copies of a codebase and propose changes for integration. Three years later, Tom Preston-Werner, Chris Wanstrath, and P.J. Hyett launched GitHub as a web-based collaboration layer built on top of Git, introducing a structured pull-request workflow through which developers submit, review, and merge proposed code changes (Appendix Figure~\ref{afig:github_demo}). The platform grew from 100{,}000 users in 2009 to more than 31 million by 2018.

In its early years, GitHub served primarily as an open-source collaboration platform offering free hosting for public repositories. Commercial adoption grew out of bottom-up developer demand: as software engineers increasingly used personal GitHub accounts for workplace projects in place of legacy internal tools, corporate customers requested an enterprise version with centralized administration and security controls, prompting the launch of GitHub Enterprise in 2011. Over the following decade, the pull-request workflow became the standard model for commercial software development at large organizations. Microsoft acquired GitHub in 2018, but GitHub continued to operate as an independent business unit with its own enterprise sales force and contracting systems \citep{nagle_frank_copilots_2024}.

\subsection{AI Coding Tools}\label{asec:codingtools}
Automated code assistance predates large language models. Rule-based tools such as Microsoft's IntelliSense, introduced in Visual Studio in 1997, provided syntax-aware identifier completion and error flagging. By 2006, code-completion tools were widely used by Java developers, although early empirical studies found that they primarily reduced documentation lookup rather than substantially accelerating code production \citep{murphy2006java, jiang2024analysis}. Later tools such as Tabnine introduced statistical and early deep-learning models to offer short line-by-line completions.

Large language models marked a step change in code generation capabilities. Because LLMs predict subsequent tokens conditional on surrounding context, programming languages are particularly well suited to their architecture: code follows strict syntactic and semantic rules, vast corpora of paired source code and natural-language documentation are publicly available for training, and compilers and test suites provide objective verification signals. These features supported rapid capability gains. On SWE-bench, an evaluation benchmark introduced in 2023 that tasks models with resolving real GitHub issues across 12 Python repositories, frontier model resolution rates rose from under 5\% at launch to nearly 80\% by early 2026 \citep{jimenez2024swebench, maslej2025artificial}.

\begin{figure}[H]
\centering
\caption{GitHub Copilot Capabilities}
\label{afig:copilot_autocomplete}
\begin{subfigure}[t]{0.65\textwidth}
    \centering
    \includegraphics[width=\textwidth]{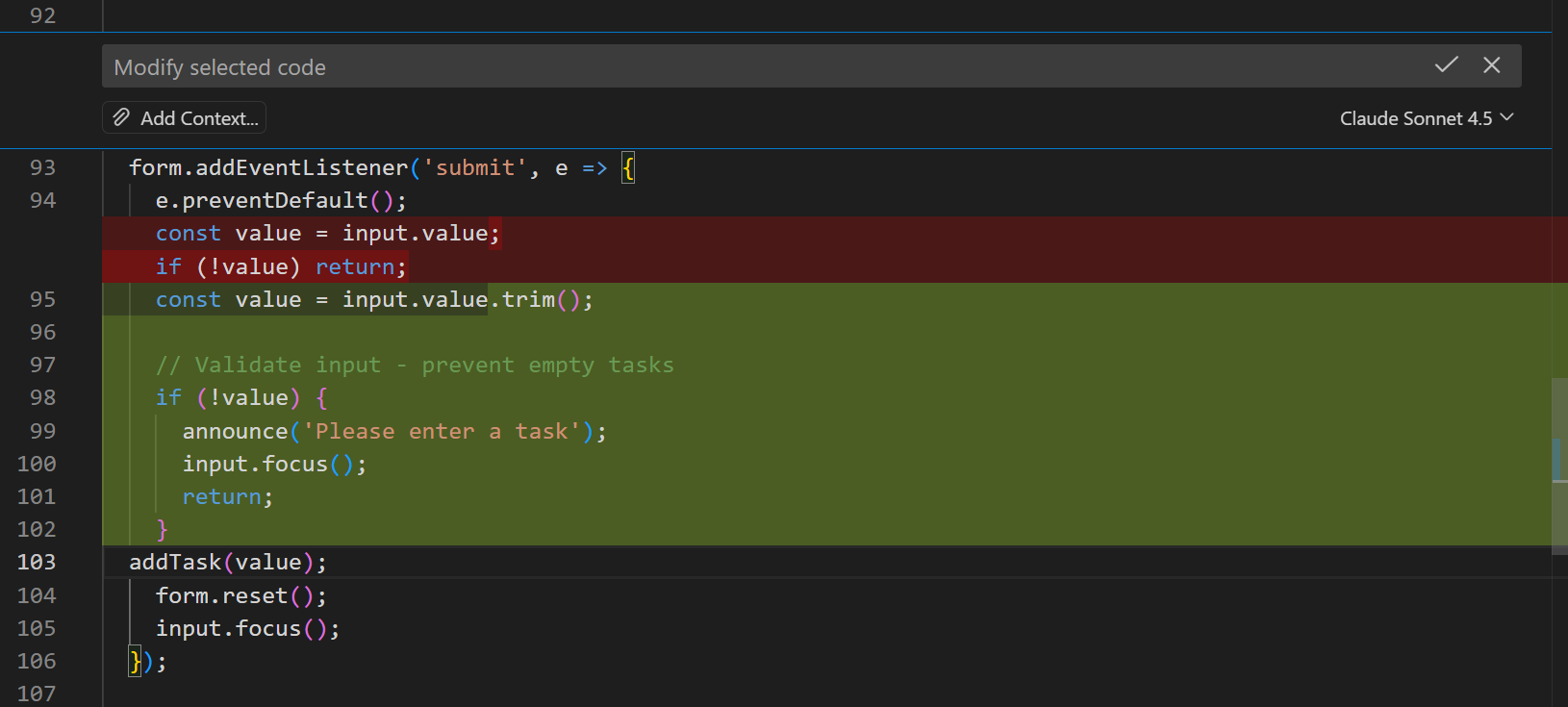}
    \subcaption{Code generation}
\end{subfigure}
\begin{subfigure}[t]{0.65\textwidth}
    \centering
    \includegraphics[width=\textwidth]{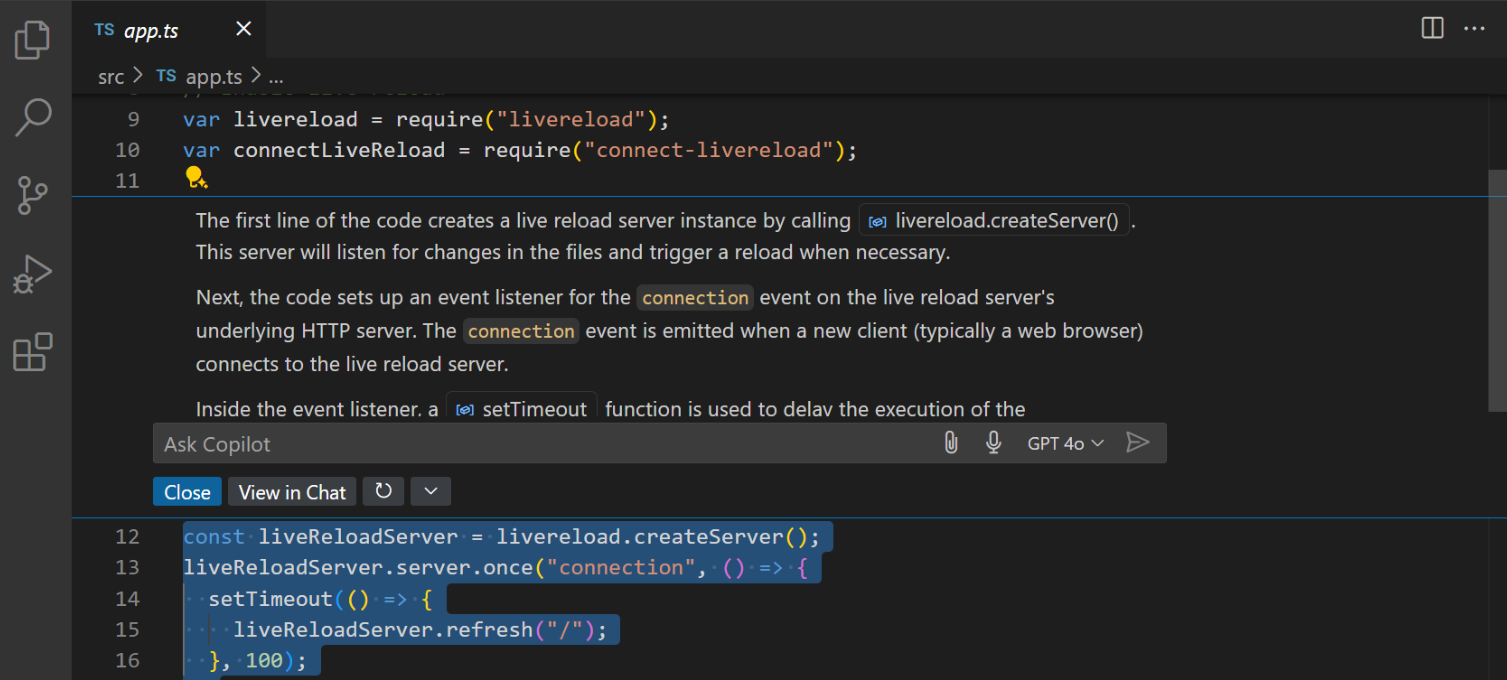}
    \subcaption{Chat}
\end{subfigure}

\par\vspace{1ex}
\figurenotes{Screenshots of GitHub Copilot's in-line code generation and chat interfaces.}
\end{figure}

GitHub Copilot first launched as a technical preview in June 2021 \citep{friedman_introducing_2021}. GitHub Copilot for Individuals became generally available in June 2022, followed by GitHub Copilot for Business for enterprise customers in early 2023. Developer uptake was rapid: in a 2023 industry survey, 56\% of professional developers reported using GitHub Copilot, more than four times the share of the next most common AI coding tool \citep{stackoverflow2023developersurvey}. The initial version of GitHub Copilot was powered by OpenAI Codex, a model trained on billions of lines of public source code. Subsequent releases expanded the product beyond in-line autocomplete to include an interactive chat interface (Appendix Figure~\ref{afig:copilot_autocomplete}), allowed users to select among multiple frontier models \citep{mckenna_github_2026}, and enabled enterprise customers to index private repositories to ground responses in their internal codebase.

A growing body of research examines the effects of GitHub Copilot at the individual developer level; \citet{ani_recent_2024} provide an early review. Controlled experiments and field studies evaluate its effects on task-level developer speed and output \citep{cui2026effects}, the security and quality of generated code \citep{fusecurity2025}, and computer science education \citep{learnngcopilot2025}.

\section{Data Construction, Measurement, and Validation}\label{asec:data_construction}

\subsection{Revelio Occupational and Seniority Measures}\label{asec:revelio_measures}

Workers on LinkedIn often describe similar jobs using different titles, while identical titles can correspond to different occupations across firms and industries. Revelio standardizes free-text job titles using a hierarchical clustering algorithm that incorporates job titles, text from workers' and employers' job descriptions, workers' skills, coworkers, and prior experience \citep{revelio_labs_revelio_2026}. For example, titles such as ``Software Developer,'' ``Backend Engineer,'' and ``Full Stack Developer'' are grouped into a common ``Software Engineering'' category. We use Revelio's 150-cluster job taxonomy, the most granular classification recommended for firm-by-year analysis \citep{revelio_labs_methods_2026}.

We also use Revelio's measure of worker seniority. Because titles convey different levels of seniority across firms and occupations, Revelio constructs a seniority score using an ensemble model that combines an individual's title, company, and industry with the duration and seniority of previous employment and age \citep{revelio_labs_methods_2026}. The resulting measure ranges from entry-level to senior executive positions.\footnote{The seniority categories range from entry level (e.g., account intern, paralegal, software engineer trainee), junior (e.g., Junior Software QA Engineer, Legal Adviser), associate (e.g., attorney, senior tax accountant), manager, director, executive, and senior executive (e.g., CEO, CFO, COO) \citep{revelio_labs_methods_2026}.}

From the underlying individual position spells, we construct a monthly panel of employment stocks, new hires, and separations at the firm level, as well as disaggregated panels at the firm-by-role and firm-by-role-by-seniority levels.

\subsection{Revelio Coverage and Measurement}\label{asec:revelio_coverage}

Online professional profiles over-represent white-collar and technical occupations, younger cohorts, and larger, more established firms. Coverage is particularly comprehensive in information technology and among professional, scientific, and technical workers in North America and Europe \citep{baird2024early}. These features make Revelio well suited to tracking the technical workforces of large GitHub customers. Because coverage is thinner in other regions and has grown over time with platform adoption, we restrict our sample to firms headquartered in the Americas and Europe.

In addition to raw spell counts, Revelio adjusts their data to correct for two sources of measurement error \citep{revelio_labs_methods_2026}. First, a cross-sectional sampling weight adjusts for differences in profile representation across occupations, industries, and countries. Second, a time-series adjustment accounts for reporting lags between when an employment transition occurs and when a worker updates their online profile. Both adjustments are calibrated against firm-level headcount benchmarks from regulatory filings. Throughout our analysis, we use these adjusted employment and flow series.

\subsection{GitHub--Revelio Firm Matching Procedure}\label{asec:matching_procedure}

We link GitHub administrative records to Revelio Labs using firm name, country, industry, and size indicators. As a first step, we apply Revelio's internal name-matching algorithm, which strips common legal suffixes (such as ``LLC,'' ``Inc.,'' and ``GmbH'') and administrative tags (such as ``Parent'') before performing an exact string match against standardized company names in Revelio's database. The algorithm excludes company profile pages with zero recorded employees and, when a match corresponds to a corporate subsidiary, aggregates employment to the ultimate parent firm.

This initial string-matching step yields a high-confidence match for approximately 64\% of GitHub accounts. The remaining accounts are not matched automatically because their recorded names are generic, abbreviated, or formatted differently across the two databases. Discrepancies typically arise when a firm operates under a distinct brand or division name on professional networking platforms (e.g., ``ABC Inc.'' versus ``ABC Science''), when non-English characters or transliterations differ across systems, or when the GitHub license is held by a subsidiary whose legal name does not appear as a standalone entity in Revelio. To resolve these cases, we combine an AI-assisted matching pipeline with human verification, using firm country, industry, and employee size to identify and validate the corresponding company profile page, as detailed in the next subsection. 

\subsection{LinkedIn Search Algorithm}\label{asec:agent_matching}

\begin{figure}[H]
\centering
\caption{Agent-Assisted Firm Matching: Search and Validation Workflow}
\label{afig:matching_search}
\begin{subfigure}[t]{0.5\textwidth}
    \centering
    \includegraphics[width=\textwidth]{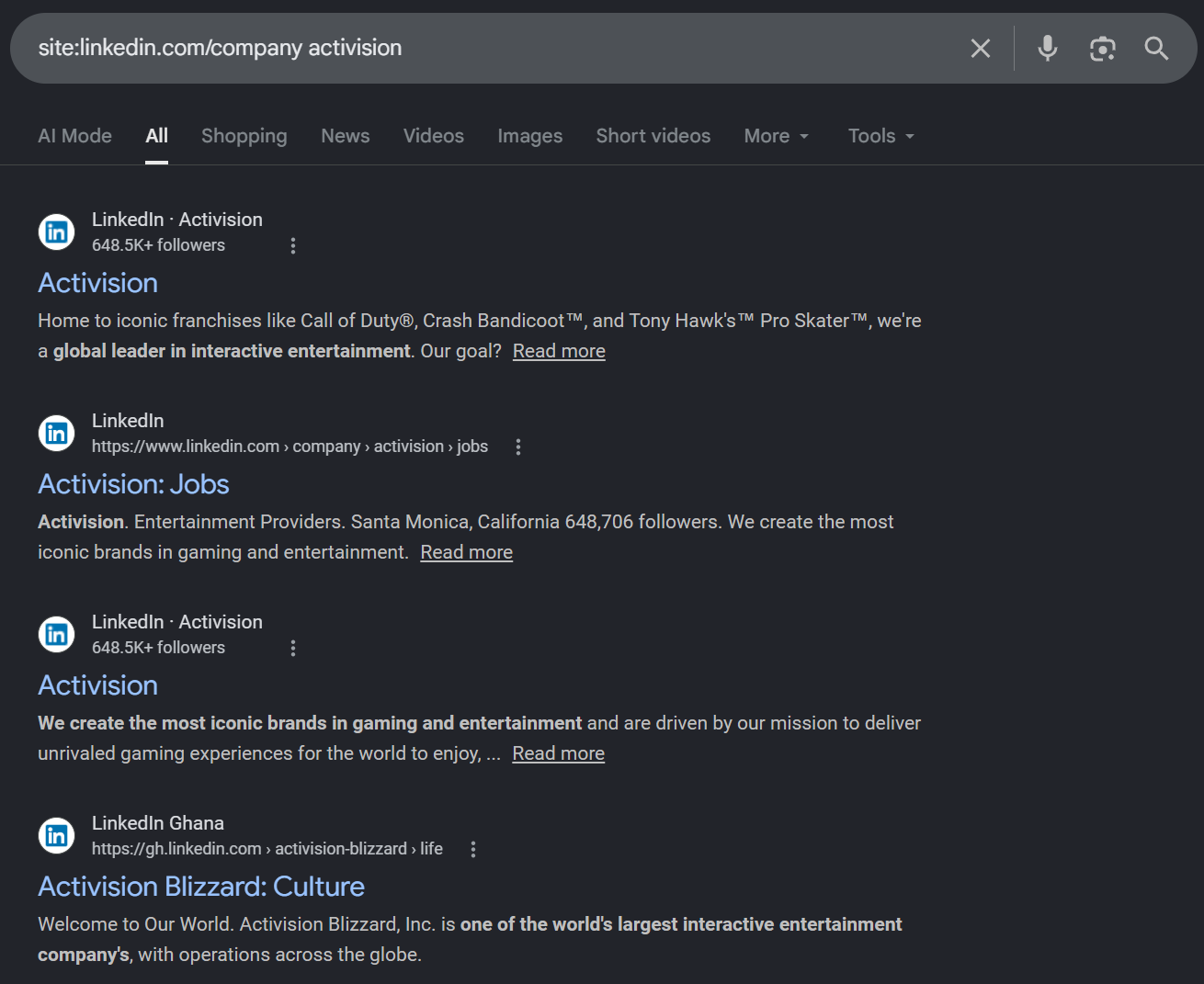}
    \subcaption{Using site-specific search for ``activision''}
\end{subfigure}
\begin{subfigure}[t]{0.5\textwidth}
    \centering
    \includegraphics[width=\textwidth]{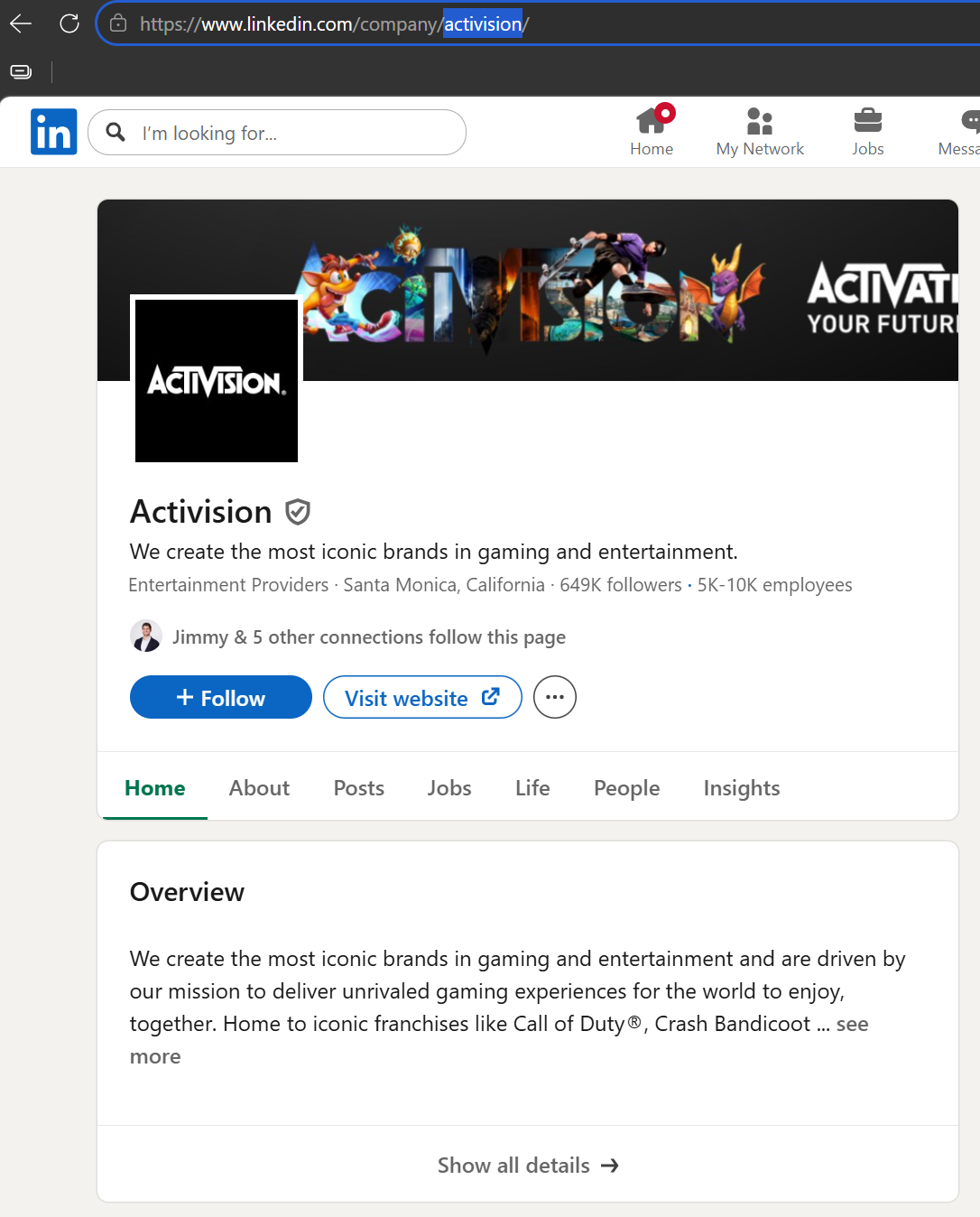}
    \subcaption{LinkedIn company profile page showing the slug \texttt{activision} in the URL}
\end{subfigure}

\par\vspace{1ex}
\figurenotes{Example of the firm-matching procedure in Section~\ref{asec:agent_matching}: a site-specific search for a firm's LinkedIn page and the resulting profile used to record its company identifier.}
\end{figure}

\begin{enumerate}
    \item \textbf{Search.} Use Google's site-specific search to search LinkedIn for the firm's Salesforce firm name and country: \texttt{site:linkedin.com/company ``COMPANY\_NAME'' COUNTRY}. If results are poor, search using less restrictive queries, including dropping exact search, removing company suffixes, reordering words, and using name stubs instead of full names.
    \item \textbf{Filter.} Among the results, identify the company page whose name closely matches the full Salesforce account name and whose LinkedIn-reported location matches the firm's account country. Additional suggestive criteria include whether the company page is ``claimed'' by an owner (and therefore not automatically generated) and larger follower and employee counts.
    \item \textbf{Record.} Record the page's LinkedIn slug from the URL (e.g., \texttt{linkedin.com/company/} \\ \texttt{activision}; Figure~\ref{afig:matching_search}), and note possible issues with the match, such as company renames, parent-subsidiary relationships, and mismatched countries.
\end{enumerate}

For firms not matched by Revelio's algorithm, we have an AI agent (Claude Sonnet 4.6) follow this algorithm to construct a candidate match. A second agent then scores the confidence of each proposed match as \textit{high}, \textit{medium}, \textit{low}, or \textit{no\_match} by evaluating the similarity between the firm's Salesforce name and the LinkedIn page name, whether the country matches, and whether the LinkedIn page has a plausible number of employees. We retain only high- and medium-confidence matches for the final sample.

\paragraph{RCID lookup.} Each matched LinkedIn slug is converted to a Revelio company identifier (RCID) by querying Revelio's \texttt{company\_mapping} table on WRDS, which maps LinkedIn URLs to RCIDs. Of the 2,803 high- and medium-confidence agent matches, 2,679 (96\%) yield an RCID, confirming that the matched LinkedIn pages correspond to firms in Revelio's employment database.

\paragraph{Duplicate resolution.} In some cases, multiple GitHub firms (typically a parent and subsidiary) match to the same LinkedIn page. We identify 82 such duplicate matches that are parent-subsidiary relationships and consolidate them.

\paragraph{URL repair.} A second human reviewer checked, in four batches, the firms whose LinkedIn slugs were dead, redirected, or otherwise inaccessible. The reviewer provided corrected slugs where possible, recovering matches for 46 of 59 firms with broken URLs. This URL repair step corrects link rot but does not independently re-evaluate the agent's firm identification.

\subsection{Match Quality Filters}\label{asec:match_quality}

We apply two pre-launch filters to screen out incorrect links between GitHub and Revelio and to ensure adequate profile coverage among matched firms. First, we exclude matched firms with zero recorded total employment or zero software engineers in Revelio prior to GitHub Copilot's launch. Second, we exclude matches in which the firm's average pre-launch total employment in Revelio is smaller than the number of GitHub licenses it purchased. Because each GitHub license corresponds to an individual user license, a firm's total workforce must be at least as large as its licensed count; violations of this condition indicate either an incorrect entity match or severe profile undercoverage.

These two criteria remove 1,215 matches, leaving 6,959 firms (80\% of matched accounts) with verified links and sufficient employment coverage. Excluded accounts are disproportionately public-sector or government organizations, or firms headquartered outside North America and Europe, where online profile coverage is sparser. Because both filters are defined strictly using pre-2023 data prior to GitHub Copilot's launch, they do not select firms on the basis of subsequent GitHub Copilot adoption or post-launch employment trajectories.

\subsection{Validation Sample and Match Accuracy}\label{asec:rev_matching_validation}

We evaluate the accuracy of our firm-matching pipeline through three validation exercises. The first two draw random samples of 100 firms from the algorithmic and AI-assisted pipelines, respectively, and compare the assigned match against the company profile page independently identified by a human reviewer using firm name, country, and industry. The third exercise conducts a larger-scale stress test of Revelio's name-matching algorithm using firms with known ground-truth identifiers.

\begin{enumerate}
    \item \textbf{Revelio algorithmic match (precision audit, N = 100).} The reviewer compared the Revelio-assigned company (RCID) to the firm they identified independently, recording whether the match was correct, incorrect (wrong firm), or whether no correct match should have existed. Because the audit draws only from firms Revelio matched, it measures precision but not recall.

    \item \textbf{AI agent-assisted match (precision and recall, N = 100).} The reviewer compared the agent's proposed LinkedIn slug to the slug they identified independently. Draws include firms the agent declined to match, so the sample supports both precision and recall.

    \item \textbf{Name-perturbation stress test (N = 995).} For 995 firms with known ground-truth RCIDs, we append legal suffixes (e.g., ``Ltd.,'' ``GmbH'') and parent qualifiers (e.g., ``\,--\,Parent'') to the firm name before sending it through Revelio's matching algorithm. This isolates the algorithm's sensitivity to name-cleanliness variation holding the target firm fixed.
\end{enumerate}

\noindent Table~\ref{atab:match_validation_summary} reports precision, recall, and F1 for each test. Revelio's algorithmic matches are correct for 83 of 100 sampled firms (precision 0.83). The AI agent achieves higher precision (0.93) and near-complete recall (0.99), for an F1 of 0.96: most incorrect cases are firm-level errors (typically a subsidiary matched to its parent), not missed matches. The name-perturbation test yields the same precision (0.93) as the agent pipeline but lower recall (0.76): the algorithm returns no match for 22.7\% of perturbed names. This reflects the algorithm's reliance on exact-match-after-cleaning and motivates the agent-assisted pipeline for firms with non-standard name representations.

\begin{table}[H]
\centering
\caption{Match Validation Summary: Precision, Recall, and F1}
\label{atab:match_validation_summary}
\begin{tabular}{lrrrr}
\toprule
Test & N & Precision & Recall & F1 \\
\midrule
Algorithmic audit & 100 & 0.83 & -- & -- \\
Agent-assisted audit & 100 & 0.93 & 0.99 & 0.96 \\
Algorithmic perturbation & 995 & 0.93 & 0.76 & 0.84 \\
\bottomrule
\end{tabular}

\begin{minipage}{\textwidth}
\vspace{3pt}
\tablenotes{The table reports precision, recall, and F1 for the three validation exercises in Section~\ref{asec:rev_matching_validation}: a precision audit of Revelio's algorithmic matches ($N=100$), a precision-and-recall audit of the AI-agent matches ($N=100$), and a name-perturbation stress test of the Revelio algorithm on firms with known identifiers ($N=995$). A human reviewer independently identified the correct LinkedIn company page in each exercise; precision is the share of proposed matches that are correct, and recall is the share of firms with a correct match that are matched.}
\end{minipage}
\end{table}

Appendix Table~\ref{atab:match_validation_error_types} decomposes matching errors into five mutually exclusive categories, each expressed as a percentage of the corresponding validation sample. Assignments to the wrong firm account for the vast majority of errors in the algorithmic audit (16.0\% of sampled firms) but fall sharply in the AI-assisted audit (7.0\%) and the perturbation test (5.3\%). Conversely, missed matches (false negatives) are rare in the AI-assisted audit (1.0\%) but account for 22.7\% of cases in the perturbation test, confirming that formatting noise primarily causes the rule-based algorithm to leave valid firms unmatched rather than to assign incorrect links.

\begin{table}[H]
\centering
\caption{Match Validation: Error Decomposition by Validation Test (\% of N)}
\label{atab:match_validation_error_types}
\begin{tabular}{lrrr}
\toprule
Error type & Algorithmic & Agent-assisted  & Algorithmic Perturbation \\
\midrule
Exact match & 83.0\% & 88.0\% & 72.0\% \\
Wrong firm & 16.0\% & 7.0\% & 5.3\% \\
Missed match & -- & 1.0\% & 22.7\% \\
Correctly unmatchable & -- & 4.0\% & -- \\
Should not have matched & 1.0\% & -- & -- \\
\bottomrule
\end{tabular}

\begin{minipage}{\textwidth}
\vspace{3pt}
\tablenotes{The table decomposes match errors into five categories for each validation exercise in Table~\ref{atab:match_validation_summary}, each expressed as a percentage of the exercise's sample size ($N=100$, $100$, and $995$).}
\end{minipage}
\end{table}

\subsection{Estimation Sample Construction and Summary Statistics}\label{asec:sample_construction}\label{asec:summary_stats}

Appendix Table~\ref{atab:sample_construction} summarizes the construction of our estimation sample from the universe of pre-existing GitHub accounts. To ensure that scheduled contract review timing is predetermined with respect to GitHub Copilot's launch, we begin with firms that held an active GitHub contract as of September 2022 with a scheduled post-launch contract review date fixed prior to the introduction of GitHub Copilot. Linking these accounts to Revelio Labs via our algorithmic and AI-assisted matching procedures and applying the pre-launch match-quality filters described in Appendix Section~\ref{asec:match_quality} yields 6,959 matched organizations with verified employment coverage.

From this matched sample, we apply two final sample restrictions. First, we exclude 599 public-sector, government, and nonprofit entities to focus on commercial firms, leaving 6,360 private-sector enterprises.\footnote{We identify public-sector and nonprofit entities using two-digit NAICS code 92 (``Public Administration''), keyword searches for government and non-commercial terms in organization names, and manual verification.} Second, to ensure consistent and representative profile coverage across countries, we restrict the sample to firms headquartered in the Americas or Europe, the regions where Revelio's coverage is most comprehensive.\footnote{Specifically, we retain firms headquartered in six geographic regions classified in Revelio: Northern America, Central America and the Caribbean, Southern America, Northern Europe, Western Europe, and Southern Europe. This restriction excludes 946 firms headquartered in Asia, Africa, the Middle East, or Oceania.} The resulting balanced panel comprises 5,414 firms observed monthly from January 2021 through June 2025. Appendix Table~\ref{atab:summary_stats} reports September 2022 baseline characteristics for the full sample and by whether firms adopt GitHub Copilot by June 2025.

\begin{table}[H]
\centering
\caption{Estimation Sample Construction}
\label{atab:sample_construction}
\begin{tablebox}
\begin{tabular}{lc}
\toprule
 & Firms \\
\midrule
Initial sample & 8{,}756 \\
Matched to Revelio (RCID) & 8{,}174 \\
\quad Algorithmic match & 5{,}558 \\
\quad Hand match & 2{,}644 \\
Match quality filter & 6{,}959 \\
Keep private firms & 6{,}360 \\
Restrict HQ to Americas + Europe & 5{,}414 \\
\bottomrule
\end{tabular}
\end{tablebox}

\begin{minipage}{\textwidth}
\vspace{3pt}
\tablenotes{The table reports the number of firms remaining after each sample restriction described in Appendix Section~\ref{asec:sample_construction}; the last row is the estimation sample. Indented rows split matched firms by match method, but 28 firms are matched by both methods. The private-firms row drops public-sector and nonprofit organizations identified by NAICS code or organization name, together with firms removed on manual review.}
\end{minipage}
\end{table}

\begin{table}[H]
\centering
\caption{Summary Statistics}
\label{atab:summary_stats}
\begin{tablebox}
    \begin{tabular}{lccc}
\toprule
 & Full sample & Eventual adopters & Non-adopters \\
 & (1) & (2) & (3) \\
\midrule
Headcount & 8,284 & 10,754 & 4,218 \\
 & (28,515) & (34,777) & (11,594) \\
Firm age & 47 & 45 & 50 \\
 & (45) & (45) & (45) \\
Technical share (\%) & 12.6 & 14.2 & 10.1 \\
 & (11.2) & (11.1) & (10.8) \\
SWE share (\%) & 5.7 & 6.9 & 3.7 \\
 & (7.1) & (7.2) & (6.5) \\
Junior share (\%) & 59.6 & 57.7 & 62.8 \\
 & (14.5) & (14.4) & (14.1) \\
Hiring rate (\%) & 2.55 & 2.80 & 2.15 \\
 & (1.45) & (1.52) & (1.23) \\
Separation rate (\%) & 1.92 & 2.03 & 1.73 \\
 & (0.83) & (0.82) & (0.82) \\
Headcount growth, 2021--2022 (\%) & 8.6 & 10.5 & 5.4 \\
 & (14.5) & (15.4) & (12.2) \\
SWE headcount growth, 2021--2022 (\%) & 13.4 & 15.6 & 9.6 \\
 & (31.0) & (29.0) & (33.8) \\
Contract-review cycle (months) & 12.1 & 11.9 & 12.3 \\
 & (2.0) & (1.4) & (2.4) \\
U.S.\ headquartered (\%) & 61.6 & 63.2 & 58.8 \\
Information Technology (\%) & 24.8 & 29.7 & 16.6 \\
Financial Services (\%) & 12.9 & 12.6 & 13.3 \\
\midrule
Firms & 5,414 & 3,368 & 2,046 \\
\bottomrule
\end{tabular}
\end{tablebox}

\begin{minipage}{\textwidth}
\vspace{3pt}
\tablenotes{The table reports firm-level summary statistics for the analysis sample described in Section~\ref{sec:data}. Headcount, firm age, the employment shares, contract-review cycle, location and industry are measured in September 2022; hiring and separation rates are monthly averages over the pre-launch period; headcount growth is the September 2021 to September 2022 change in rounded headcount, winsorized at the 1st and 99th percentiles. Standard deviations are in parentheses. Column (1) covers the full sample; columns (2) and (3) split firms by whether they adopt GitHub Copilot by the end of the panel.}
\end{minipage}
\end{table}

\subsection{Employment Data Validation}\label{asec:employment_validation}

Our empirical analysis relies on Revelio Labs data to measure firm-level employment stocks, hiring and separation flows, and workforce composition across technical roles. We validate these measures against two independent external benchmarks: audited total headcount from regulatory filings and administrative records of GitHub Enterprise licenses. 

First, for the 19\% of sample firms that file Form 10-K reports covered in Compustat, we compare average annual employment in Revelio against total headcount reported in Compustat. The two series differ in several institutional respects: 10-K filings report formal payroll employees at fiscal year-end, whereas Revelio measures annual average active spells based on online profiles, which can include contractors and freelancers. Despite these conceptual differences, the two measures are strongly correlated in log levels (Pearson's $r = 0.756$, Spearman's $\rho = 0.705$). More importantly for our longitudinal research design, which relies on within-firm variation over time, year-over-year log employment changes in Revelio and Compustat also track each other closely, with a Pearson correlation of $0.576$ and a Spearman rank correlation of $0.598$ (0.593 among eventual GitHub Copilot adopters and 0.458 among non-adopters). The first two rows of Appendix Figure~\ref{afig:validation_combined} plot these level and growth comparisons. 

Second, we validate Revelio's measurement of technical workforces against firms' pre-launch GitHub license counts. Administrative license counts and Revelio role headcounts do not map one-to-one: Enterprise contracts often provision licenses in blocks rather than matching exact active headcount, not all software engineers require a license on a given enterprise instance, and firms frequently assign licenses to technical workers in adjacent roles outside core software engineering. Even so, pre-launch GitHub Enterprise licenses and Revelio software engineering headcount exhibit a strong positive correlation in log levels (Pearson's $r = 0.684$). At the median firm, the ratio of licensed GitHub Enterprise licenses to Revelio software engineers is approximately 2.4. When we expand the denominator to include adjacent technical roles that routinely interact with code repositories (specifically, application development, data analytics, quality assurance, and systems and infrastructure engineering), the log-level correlation remains similarly strong ($r = 0.673$) and the median ratio of GitHub Enterprise licenses to technical workers is $0.9$, confirming that Revelio captures the relevant population of technical staff at sample firms. The third row of Appendix Figure~\ref{afig:validation_combined} plots these comparisons. 

\clearpage
\begin{figure}[H]
\centering
\caption{Validation of Revelio Employment Measures}
\label{afig:validation_combined}

\begin{subfigure}[t]{0.31\textwidth}
    \centering
    \includegraphics[width=\textwidth]{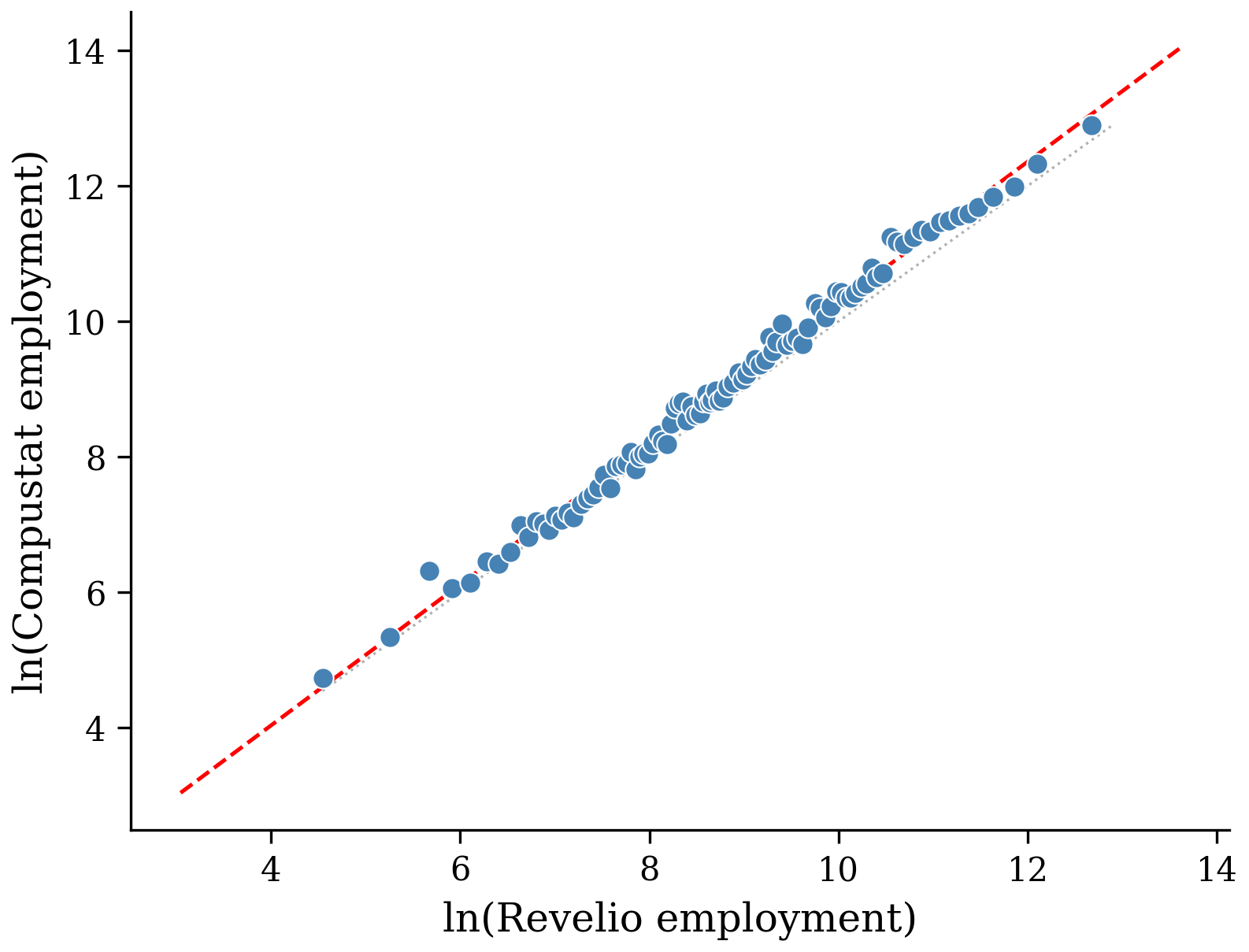}
    \subcaption{10-K/Revelio Levels: Full sample}
\end{subfigure}
\hfill
\begin{subfigure}[t]{0.31\textwidth}
    \centering
    \includegraphics[width=\textwidth]{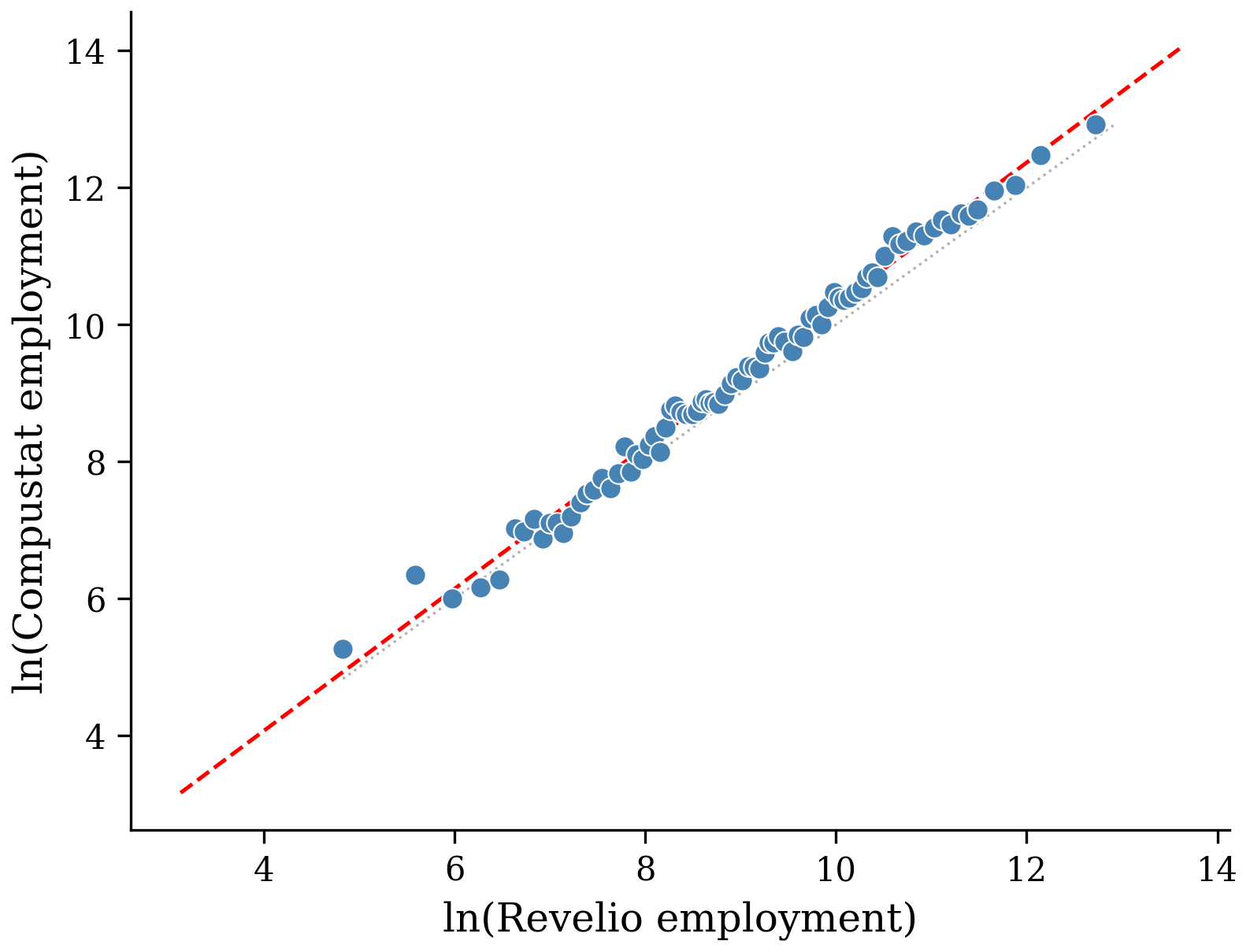}
    \subcaption{10-K/Revelio Levels: Adopters}
\end{subfigure}
\hfill
\begin{subfigure}[t]{0.31\textwidth}
    \centering
    \includegraphics[width=\textwidth]{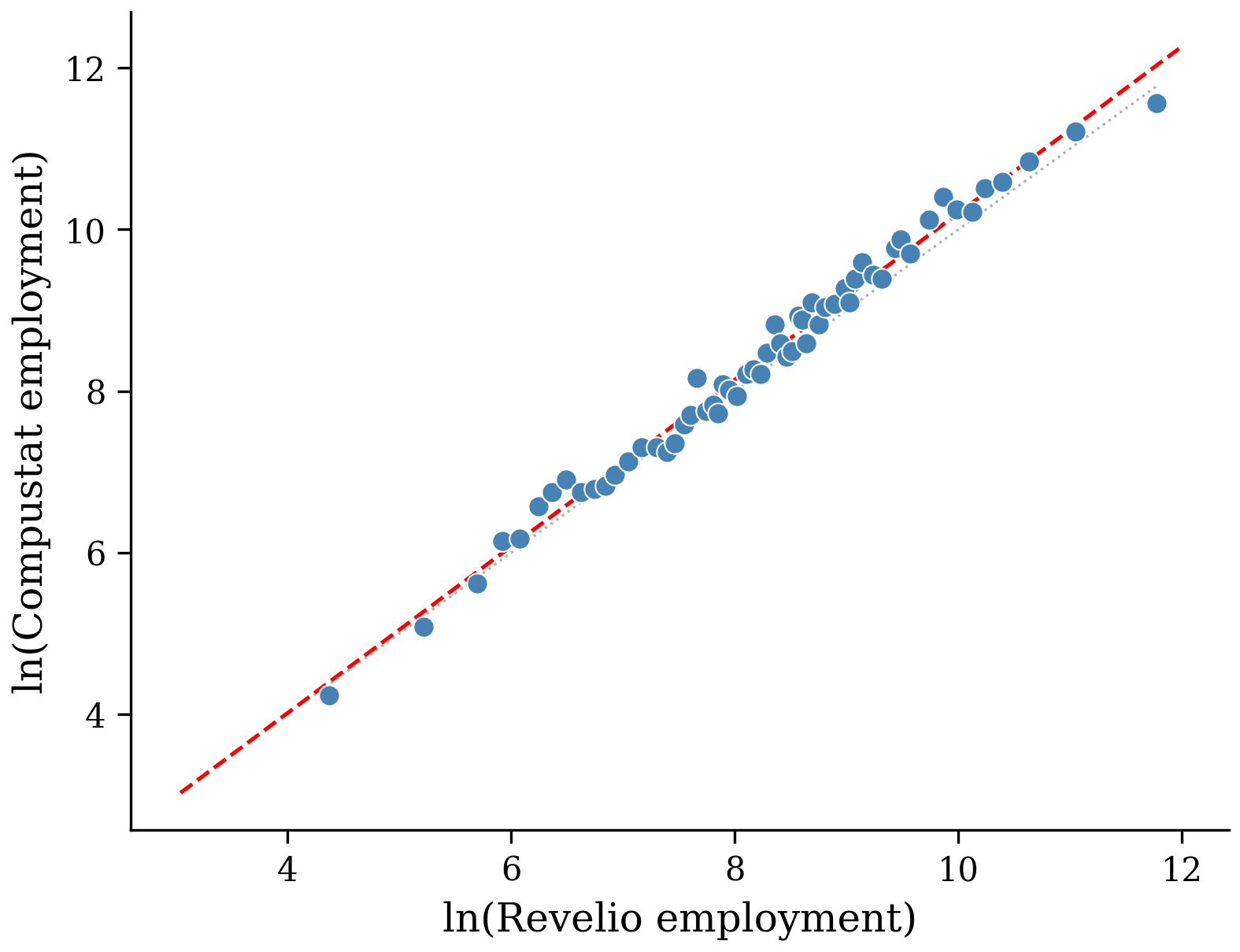}
    \subcaption{10-K/Revelio Levels: Non-adopters}
\end{subfigure}

\medskip

\begin{subfigure}[t]{0.31\textwidth}
    \centering
    \includegraphics[width=\textwidth]{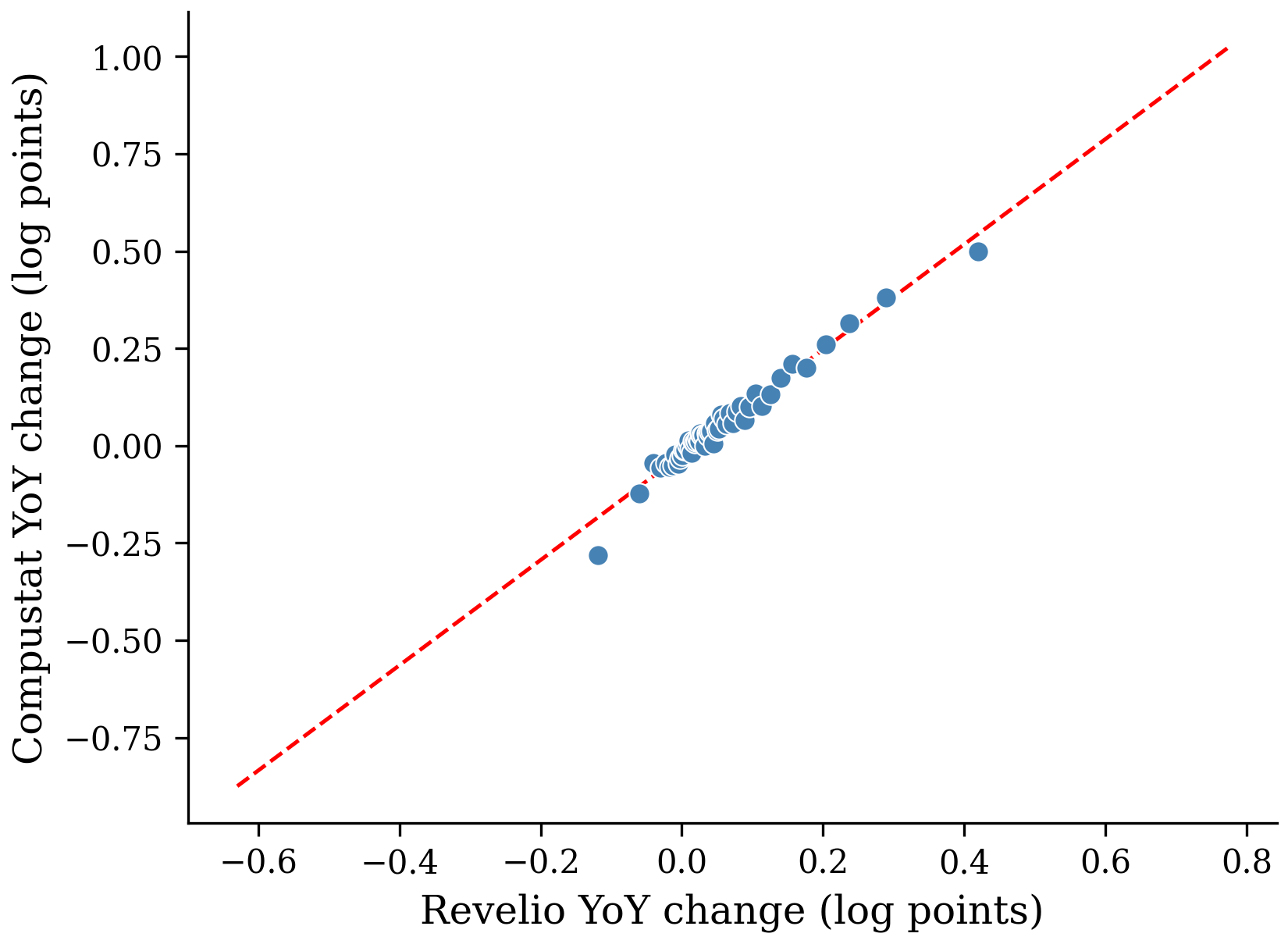}
    \subcaption{10-K/Revelio Flows: Full sample}
\end{subfigure}
\hfill
\begin{subfigure}[t]{0.31\textwidth}
    \centering
    \includegraphics[width=\textwidth]{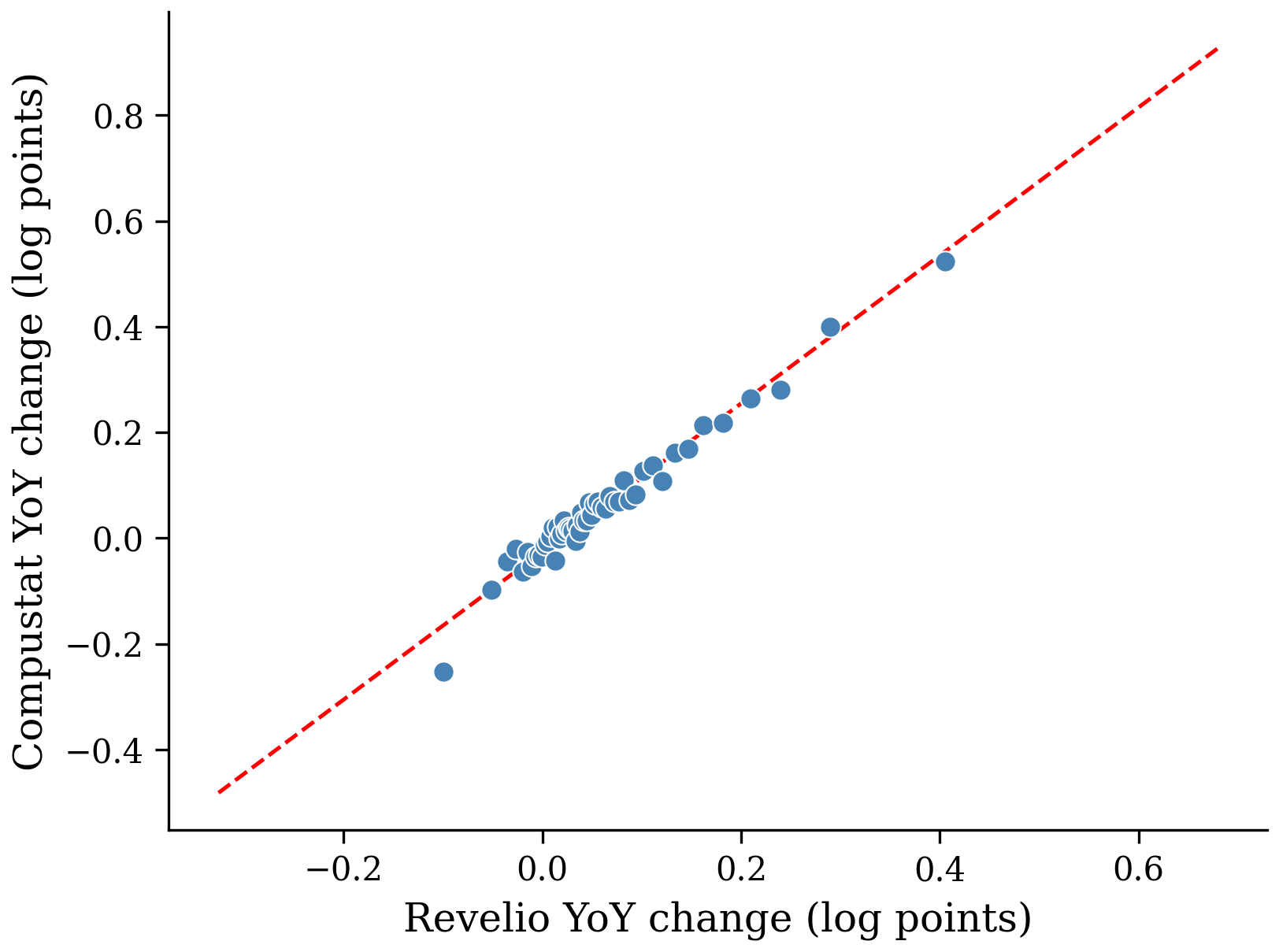}
    \subcaption{10-K/Revelio Flows: Adopters}
\end{subfigure}
\hfill
\begin{subfigure}[t]{0.31\textwidth}
    \centering
    \includegraphics[width=\textwidth]{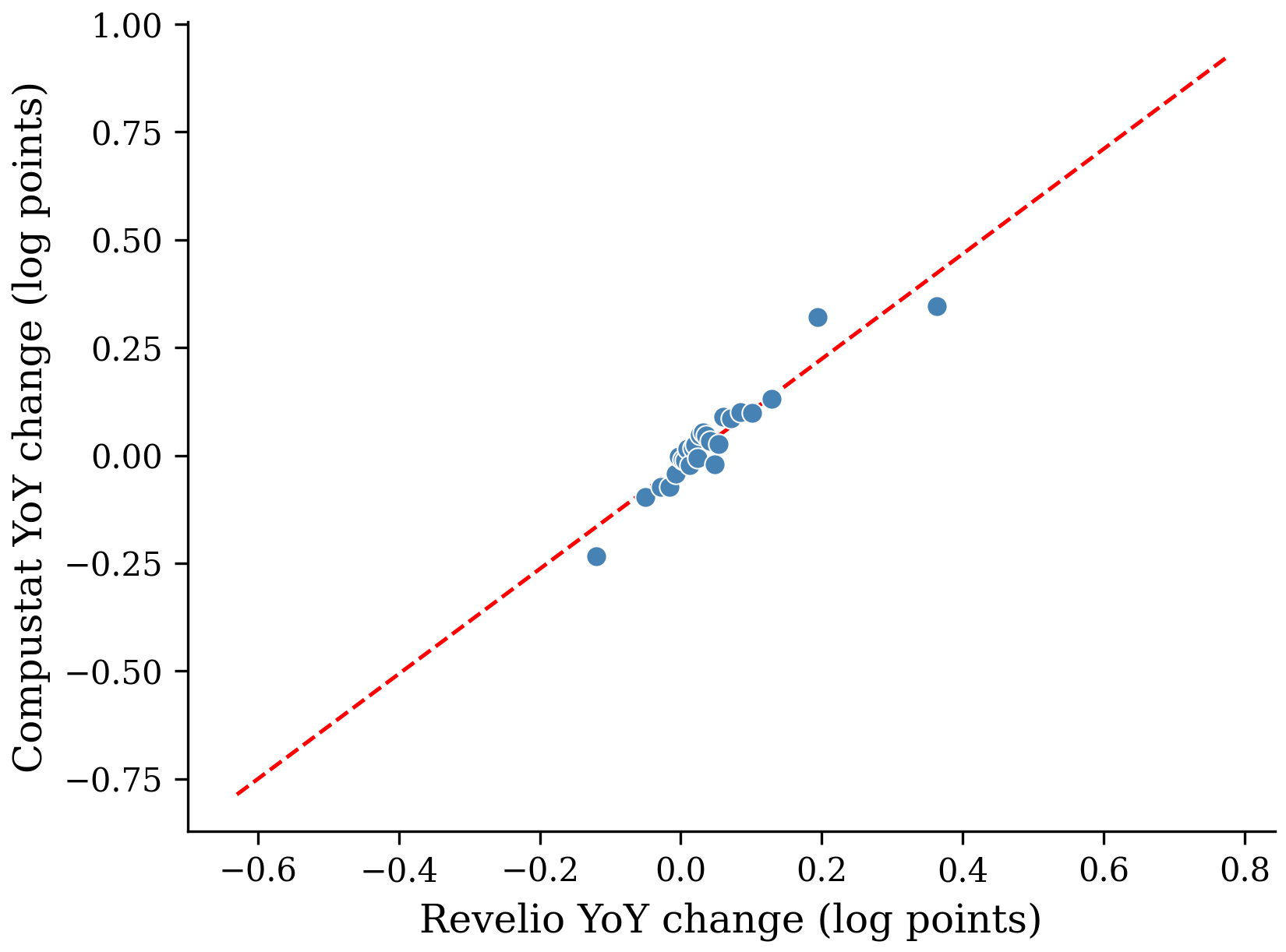}
    \subcaption{10-K/Revelio Flows: Non-adopters}
\end{subfigure}

\medskip

\begin{subfigure}[t]{0.31\textwidth}
    \centering
    \includegraphics[width=\textwidth]{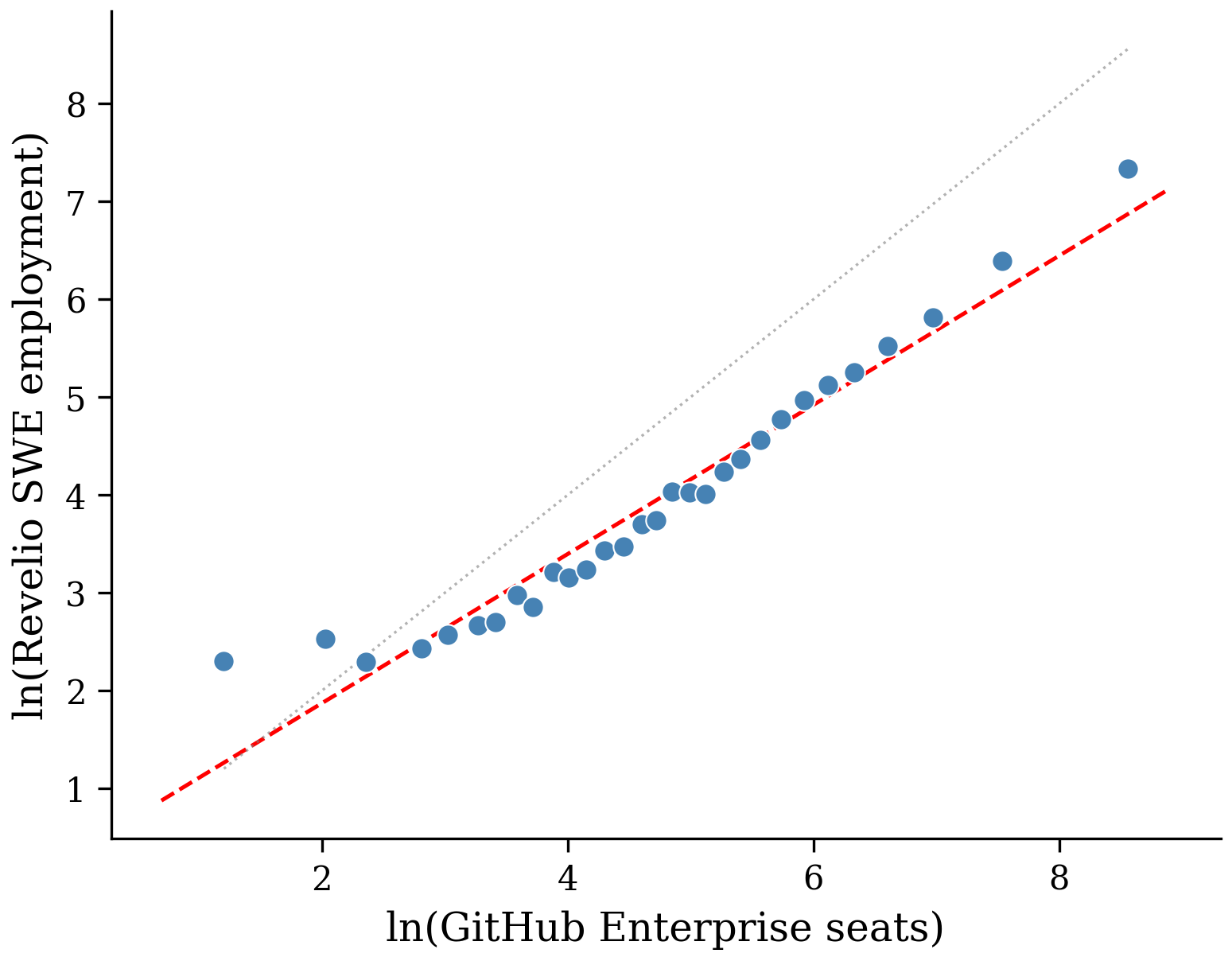}
    \subcaption{Licenses/SWE: Full sample}
\end{subfigure}
\hfill
\begin{subfigure}[t]{0.31\textwidth}
    \centering
    \includegraphics[width=\textwidth]{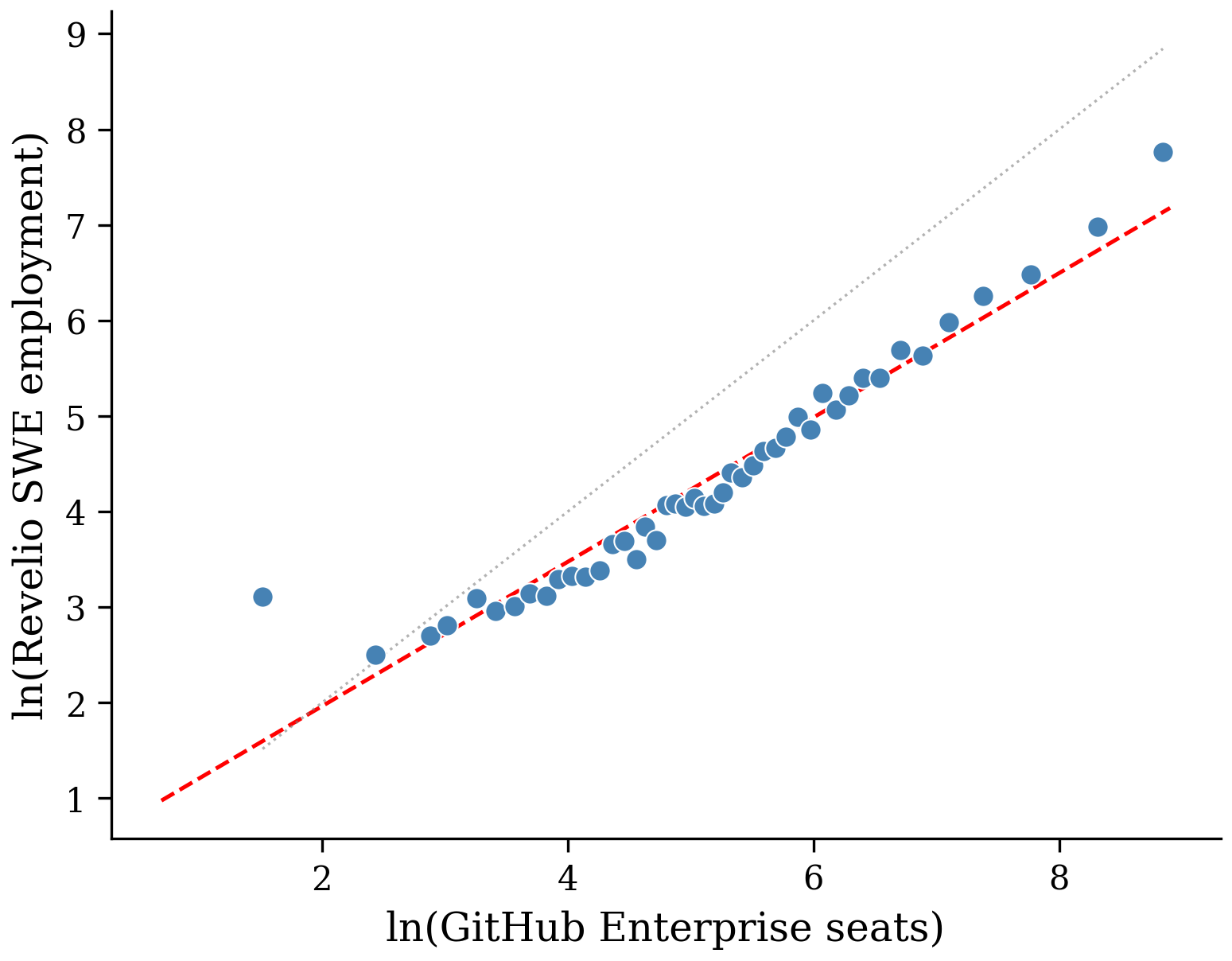}
    \subcaption{Licenses/SWE: Adopters}
\end{subfigure}
\hfill
\begin{subfigure}[t]{0.31\textwidth}
    \centering
    \includegraphics[width=\textwidth]{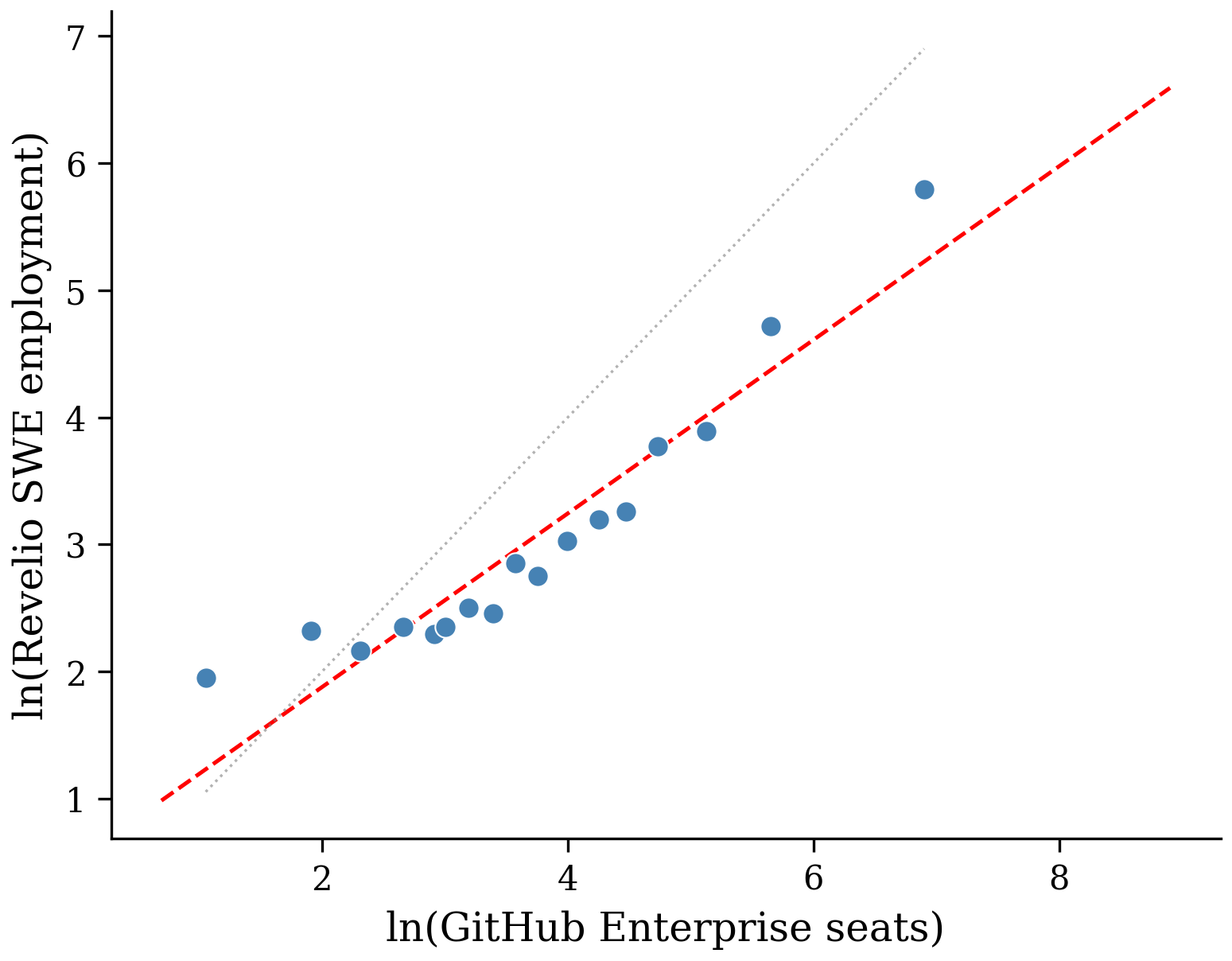}
    \subcaption{Licenses/SWE: Non-adopters}
\end{subfigure}

\par\vspace{1ex}
\figurenotes{Row 1 plots Revelio employment against Compustat 10-K employment in log levels; row 2 plots year-over-year log changes in the two measures; row 3 plots pre-period GitHub licenses against Revelio software-engineering employment in log levels. Each row splits the comparison by eventual GitHub Copilot adoption. The sample is described in Section~\ref{sec:data}.}
\end{figure}

\subsection{Coverage of the U.S. Technical Workforce}\label{asec:sample_coverage}

Although our estimation sample is not representative of the universe of U.S. employers, as GitHub customers are disproportionately large and technology-intensive, these firms account for a substantial share of aggregate U.S. technical employment. As of June 2025, sample firms employ 1,646,968 technical workers in the United States. As an external benchmark, total U.S. employment across the corresponding programming and computing occupations in the Bureau of Labor Statistics Occupational Employment and Wage Statistics (OEWS) is 4,347,980 \citep{bls_oews_2025}, implying that our sample covers approximately 38\% of the U.S. programming workforce. Because our estimation sample is restricted to private-sector firms whereas the OEWS benchmark includes both private- and public-sector employment, our sample's share of private-sector U.S. technical employment is even larger. Appendix Table~\ref{atab:bls_market} reports the national employment counts for each occupation included in the benchmark.

\begin{table}[H] 
\centering
\caption{U.S. Employment in Programming-Related Occupations, 2025}
\label{atab:bls_market}
\begin{tablebox}
\begin{tabular}{llr} 
\toprule
SOC code & Occupation & Employment (2025) \\
\midrule
11-3021 & Computer and Information Systems Managers & 670,570 \\
15-1211 & Computer Systems Analysts & 519,530 \\
15-1212 & Information Security Analysts & 190,650 \\
15-1243 & Database Architects & 67,140 \\
15-1244 & Network and Computer Systems Administrators & 314,340 \\
15-1251 & Computer Programmers & 92,230 \\
15-1252 & Software Developers & 1,687,890 \\
15-1253 & Software Quality Assurance Analysts and Testers & 186,740 \\
15-1254 & Web Developers & 70,190 \\
15-1255 & Web and Digital Interface Designers & 113,330 \\
15-1299 & Computer Occupations, All Other & 435,370 \\
\midrule
& \textbf{Total} & \textbf{4,347,980} \\
\bottomrule
\end{tabular}
\end{tablebox}

\begin{minipage}{\textwidth}
\vspace{3pt}
\tablenotes{U.S. national employment from BLS Occupational Employment and Wage Statistics, \href{https://data.bls.gov/oes/\#/area/0000000/2025}{OEWS 2025}. Excludes 15-2051 Data Scientists (262,440 workers) and 15-1241 Computer Network Architects (179,740 workers) which are also excluded from our definition of technical workers. Sample firms employ 1,646,968 U.S.-based technical workers in June of 2025, implying $\approx$38\% coverage.}
\end{minipage}
\end{table}

\subsection{Code Intensity Prompt}\label{asec:code_intensity_prompt}

We score each position description once with Claude Sonnet 4.6 at temperature zero. The model sees only the description text, inserted where the prompt reads \texttt{<position description text>}; it does not see the job title, occupation, seniority, employer, location, or date. The prompt below reproduces the prompt's instructions and the definition of the code-intensity dimension verbatim and the required output format. 

\begin{lstlisting}[basicstyle=\fontfamily{lmtt}\selectfont\scriptsize, breaklines=true, columns=fullflexible, frame=single]
You are scoring the work described in an anonymized job-position description on five dimensions. Score ONLY what the text describes; do not infer from writing style, language, or any place names what kind of person wrote it.

The dimension is a SHARE of the described work, scored as a continuous number in [0, 1]:

[...]
copilot_domain: the share of the described work inside GitHub Copilot's 2023 capability surface: writing, completing, testing, documenting, explaining, or debugging code (1.0 = all inside; 0.0 = none).
[...]

Position description:
---
<position description text>
---
\end{lstlisting}


\clearpage
\section{Descriptive Patterns and Identification Diagnostics}\label{asec:descriptive_identification}\label{asec:ols_adoption}\label{asec:first_stage_diagnostics}\label{asec:balance_appendix}

\begin{figure}[H]
\centering
\caption{GitHub Enterprise License-Count Increases and Decreases by Months to Scheduled Contract Review}
\label{afig:seat_change_magnitudes_mtnr}

\begin{subfigure}[t]{0.49\textwidth}
    \centering
    \includegraphics[width=\textwidth]{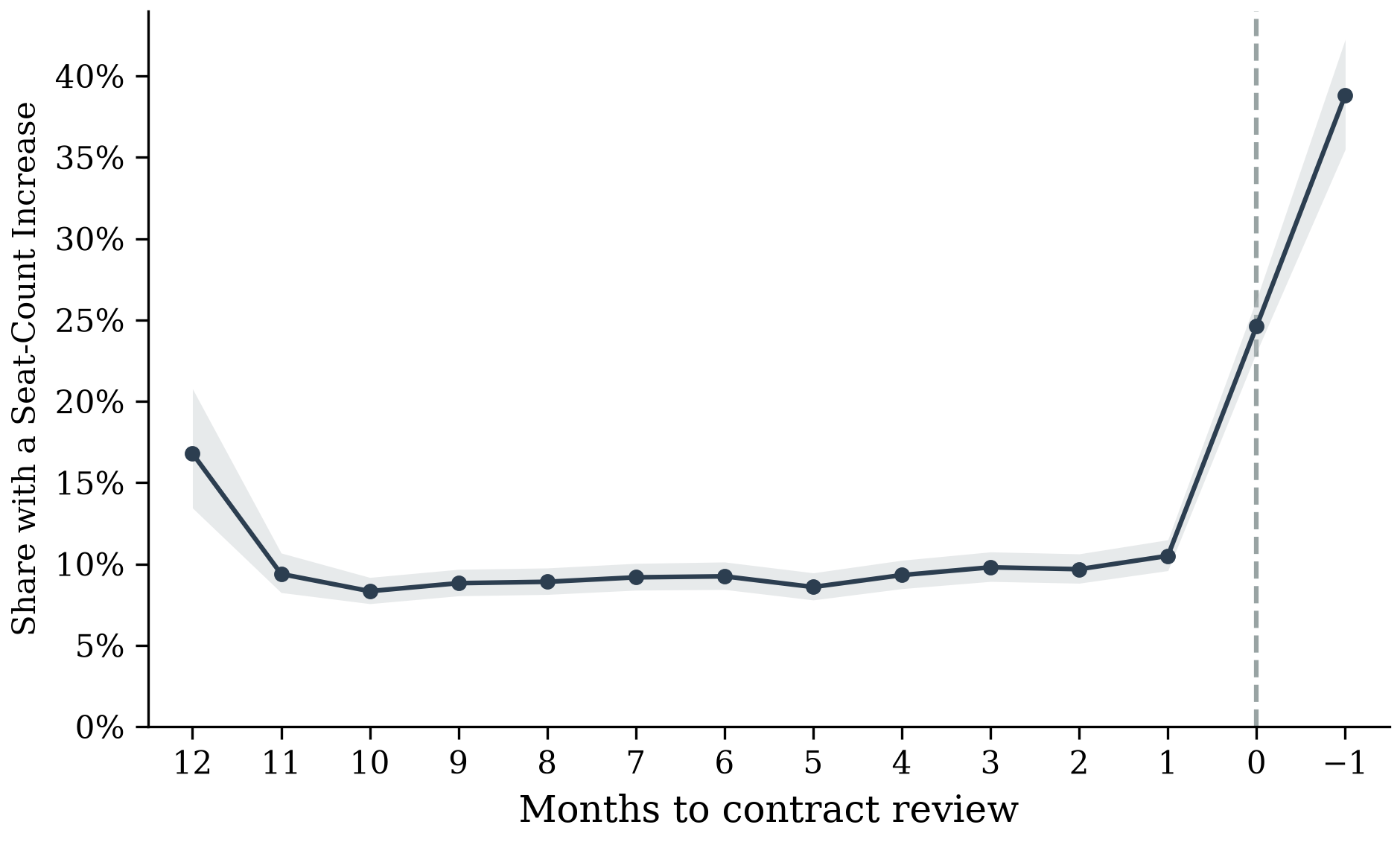}
    \subcaption{GitHub license increases}
\end{subfigure}
\hfill
\begin{subfigure}[t]{0.49\textwidth}
    \centering
    \includegraphics[width=\textwidth]{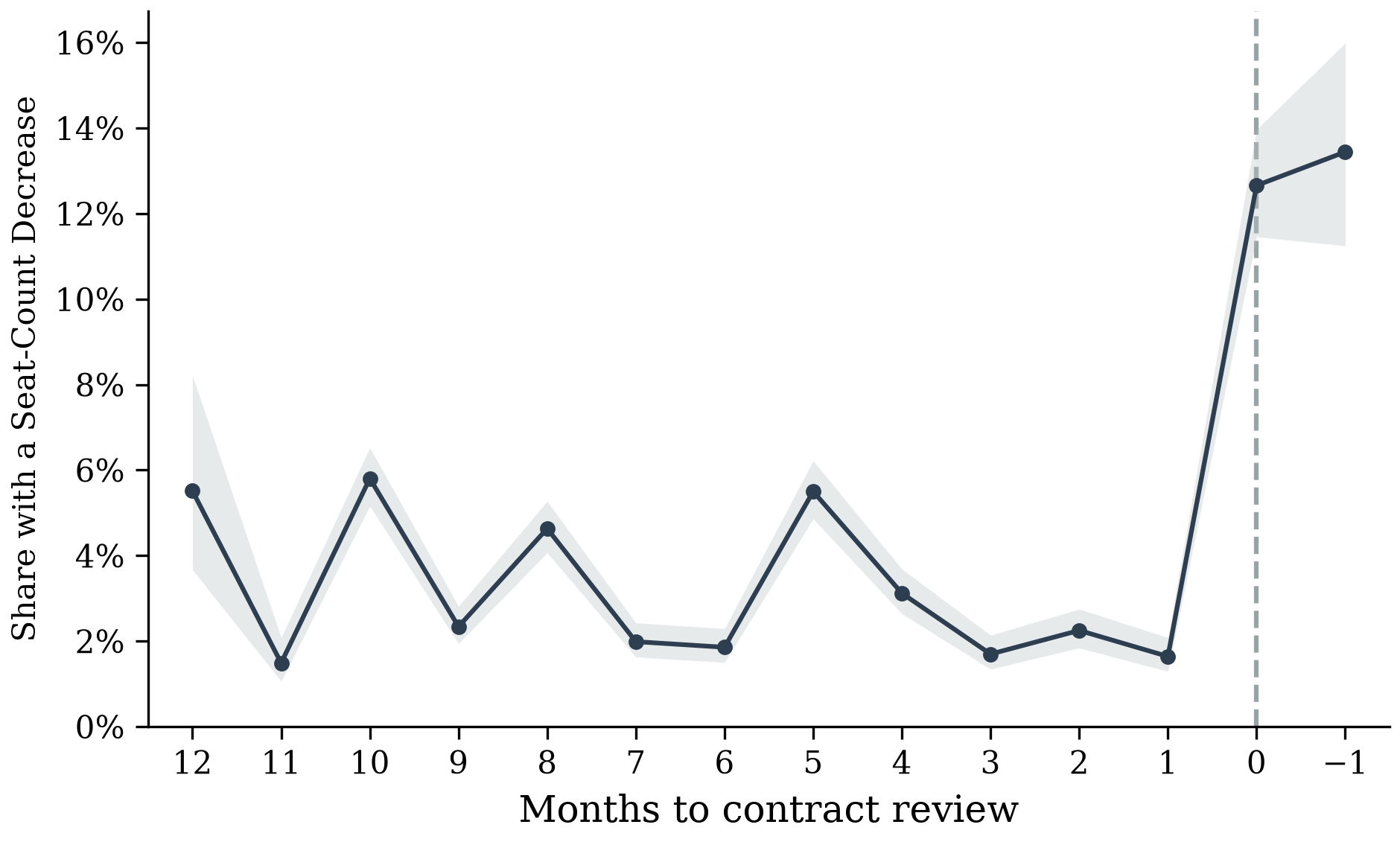}
    \subcaption{GitHub license decreases}
\end{subfigure}

\par\vspace{1ex}
\figurenotes{In the pre-GitHub Copilot era, Panel A plots the share of firm-months in which the firm's winsorized GitHub Enterprise license count is above the previous month's; Panel B plots the share in which it is below. Month $0$ is the review month and month $-1$ is the month after it.}
\end{figure}

\clearpage
\begin{figure}[H]
\centering
\caption{GitHub Copilot Adoption Hazard Ratio by Months to Scheduled Contract Review}
\label{afig:hazard_mtnr_idur}

    \centering
    \includegraphics[width=.8\textwidth]{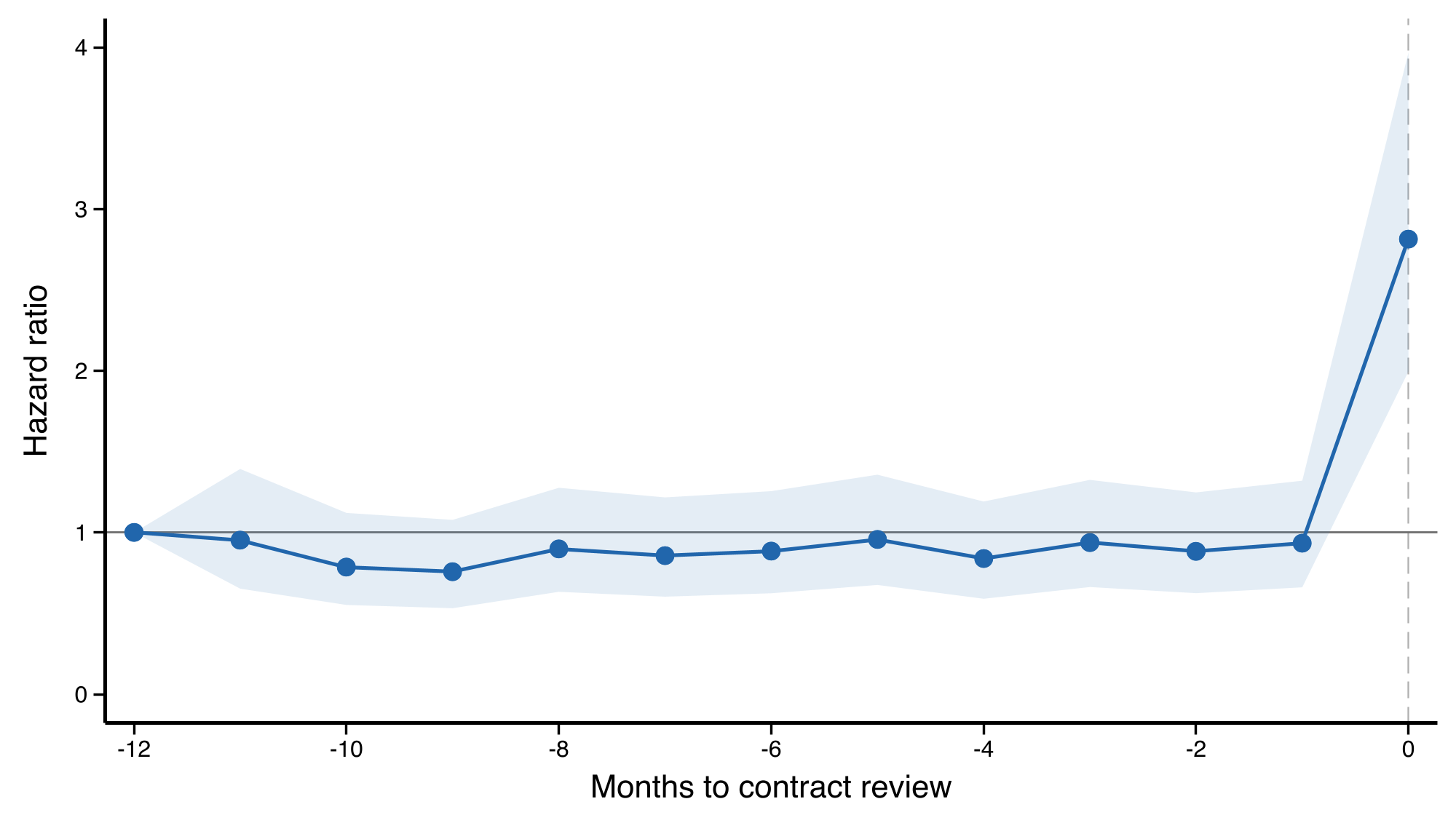}

\par\vspace{1ex}
\figurenotes{This figure reports hazard ratios from a complementary log-log model with indicators for months to review and months since GitHub Copilot's launch; $12$ months to contract review is the reference category. Standard errors are clustered by firm.}
\end{figure}

\clearpage
\begin{table}[H]
\centering
\caption{GitHub Copilot Adoption and Employment: OLS Estimates}
\label{atab:ols_adoption}
\begin{tablebox}
{
\def\sym#1{\ifmmode^{#1}\else\(^{#1}\)\fi}
\begin{tabular}{lccccc}
\toprule
Dependent variable & \multicolumn{3}{c}{Log headcount} & Log technical hires & Log technical separations \\
\cmidrule(lr){2-4}
 & All & Technical & SWE &  &  \\
 & (1) & (2) & (3) & (4) & (5) \\
\midrule
Firm FE + Month FE &  0.011\sym{***} &  0.018\sym{***} &  0.019\sym{***} & -0.051\sym{***} & -0.012 \\
 & (0.002) & (0.004) & (0.005) & (0.009) & (0.008) \\
\addlinespace
+ Industry \(\times\) month FE &  0.017\sym{***} &  0.028\sym{***} &  0.029\sym{***} & -0.037\sym{***} & -0.005 \\
 & (0.003) & (0.004) & (0.005) & (0.009) & (0.008) \\
\addlinespace
+ Size quartile \(\times\) month FE &  0.015\sym{***} &  0.023\sym{***} &  0.023\sym{***} & -0.029\sym{***} &  0.002 \\
 & (0.003) & (0.004) & (0.005) & (0.009) & (0.008) \\
\midrule
DV mean &  7.316 &  4.664 &  3.464 &  1.563 &  1.428 \\
Firms &     5,414 &     5,414 &     5,282 &     5,192 &     5,208 \\
Observations &      191,242 &      191,147 &      182,162 &      114,934 &      113,632 \\
\bottomrule
\end{tabular}
}

\end{tablebox}

\begin{minipage}{\textwidth}
\vspace{3pt}
\tablenotes{Each entry is the coefficient on an indicator equal to one from the month a firm first adopts GitHub Copilot onward, from a two-way fixed-effects regression that replaces the event-time indicators of Equation~\eqref{eq:ols_event_study} with a single post-adoption indicator. The first row includes firm and calendar-month fixed effects; the second adds two-digit NAICS industry $\times$ calendar-month fixed effects; the third further adds pre-launch headcount quartile $\times$ calendar-month fixed effects. Standard errors, clustered by firm, are in parentheses. Effects are in log points. The sample is adopter firm-months within twelve months of adoption and all non-adopter firm-months. \sym{*}~$p<0.10$, \sym{**}~$p<0.05$, \sym{***}~$p<0.01$.}
\end{minipage}
\end{table}

\begin{table}[H]
\centering
\caption{Eventual Adopters Around ChatGPT Launch: Difference-in-Differences Estimates}
\label{atab:post_launch_adopter_did}
\begin{tablebox}
{
\def\sym#1{\ifmmode^{#1}\else\(^{#1}\)\fi}
\begin{tabular}{lccc}
\toprule
Dependent variable & Log technical hires & Log technical separations & Log technical headcount \\
 & (1) & (2) & (3) \\
\midrule
Post-November 2022 \(\times\) eventual adopter&      -0.101\sym{***}&      -0.006         &       0.067\sym{***}\\
            &     (0.014)         &     (0.011)         &     (0.007)         \\
\midrule
Firm FE     &\(\checkmark\)         &\(\checkmark\)         &\(\checkmark\)         \\
Month FE    &\(\checkmark\)         &\(\checkmark\)         &\(\checkmark\)         \\
DV mean     &       1.691         &       1.522         &       4.868         \\
Firms       &       5,272         &       5,278         &       5,414         \\
Observations&     192,425         &     187,526         &     292,261         \\
\bottomrule
\end{tabular}
}

\end{tablebox}

\begin{minipage}{\textwidth}
\vspace{3pt}
\tablenotes{Each entry is the coefficient on Post-November 2022 $\times$ eventual adopter from a two-way fixed-effects regression that replaces the event-time indicators of Equation~\eqref{eq:chatgpt_event_study} with a single Post-November 2022 $\times$ eventual adopter indicator. Standard errors, clustered by firm, are in parentheses. Effects are in log points. The data are the balanced firm-by-month panel, January 2021 to June 2025. \sym{*}~$p<0.10$, \sym{**}~$p<0.05$, \sym{***}~$p<0.01$.}
\end{minipage}
\end{table}

\clearpage
\begin{table}[H]
\centering
\caption{Balance by Adoption and Contract-Review Timing}
\label{atab:balance_regs}
\begin{tablebox}
\begin{tabular}{lcc}
\toprule
 & Eventual Adopter vs. Non-Adopter & Early vs. Later Contract Review \\
\midrule
Hire Rate &     0.303 &    -0.067 \\
 & (    0.018) & (    0.019) \\
Separation Rate &     0.222 &    -0.019 \\
 & (    0.017) & (    0.017) \\
SWE Hire Rate &     0.163 &    -0.019 \\
 & (    0.011) & (    0.010) \\
SWE Separation Rate &     0.117 &     0.011 \\
 & (    0.010) & (    0.009) \\
Tech Growth 2021-2022 &     0.210 &    -0.064 \\
 & (    0.028) & (    0.028) \\
SWE Growth 2021-2022 &     0.147 &    -0.063 \\
 & (    0.030) & (    0.029) \\
Ln(Headcount) &     0.334 &     0.241 \\
 & (    0.027) & (    0.027) \\
SWE Share of Employment &     0.424 &     0.017 \\
 & (    0.027) & (    0.027) \\
Junior Share &    -0.347 &     0.021 \\
 & (    0.027) & (    0.027) \\
U.S. Headquartered &     0.082 &    -0.024 \\
 & (    0.028) & (    0.027) \\
U.K. Headquartered &     0.025 &    -0.017 \\
 & (    0.028) & (    0.028) \\
Information Technology &     0.304 &    -0.010 \\
 & (    0.026) & (    0.027) \\
Financial Services &    -0.023 &     0.027 \\
 & (    0.028) & (    0.027) \\
\bottomrule
\end{tabular}

\end{tablebox}

\begin{minipage}{\textwidth}
\vspace{3pt}
\tablenotes{The table reports the coefficients underlying Figure~\ref{fig:balance_coefplots}. Each entry is the coefficient, with standard error below, on a group indicator from a separate regression of a standardized pre-treatment characteristic. Column (1) contrasts eventual adopters with non-adopters; column (2) contrasts early- and late-scheduled contract review firms, split at June 2023. Time-varying characteristics use pre-launch firm-months with calendar-month fixed effects; time-invariant characteristics use a launch-month cross-section; the 2021--22 growth rows are the December 2021 to December 2022 change in rounded headcount. Standard errors, clustered by firm, are in parentheses.}
\end{minipage}
\end{table}

\clearpage
\section{Additional Employment and Hiring Results}\label{asec:composition_appendix}\label{asec:withinfirm}\label{asec:india_domestic_hiring}

\begin{table}[H]
\centering
\caption{Quarterly Wald Estimates for Technical Hires and Separations}
\label{atab:tech_flows_quarterly_latet}
\begin{tablebox}
{
\def\sym#1{\ifmmode^{#1}\else\(^{#1}\)\fi}
\begin{tabular}{lcccc}
\toprule
 & Reduced form & First stage & LATET & 95\% interval \\
Quarters to contract review (\(\ell\)) & (1) & (2) & (3) & (4) \\
\midrule
\multicolumn{5}{l}{\textit{Panel A. Technical hires}} \\
\quad \(+1\) & \(-0.021\) & \(0.057\)\sym{***} & \(-0.365\) & [-1.351, 0.354] \\
 & (0.020) & (0.015) & & \\
\quad \(+2\) & \(-0.050\) & \(0.107\)\sym{***} & \(-0.467\) & [-1.205, 0.083] \\
 & (0.031) & (0.024) & & \\
\quad \(+3\) & \(-0.063\) & \(0.161\)\sym{***} & \(-0.391\) & [-1.020, 0.046] \\
 & (0.039) & (0.032) & & \\
\quad \(+4\) & \(-0.073\) & \(0.215\)\sym{***} & \(-0.340\) & [-0.948, 0.088] \\
 & (0.051) & (0.039) & & \\
\midrule
\multicolumn{5}{l}{\textit{Panel B. Technical separations}} \\
\quad \(+1\) & \(0.031\)\sym{*} & \(0.057\)\sym{***} & \(0.556\) & [-0.108, 1.583] \\
 & (0.018) & (0.015) & & \\
\quad \(+2\) & \(0.018\) & \(0.107\)\sym{***} & \(0.163\) & [-0.348, 0.740] \\
 & (0.025) & (0.024) & & \\
\quad \(+3\) & \(0.026\) & \(0.161\)\sym{***} & \(0.161\) & [-0.285, 0.623] \\
 & (0.033) & (0.032) & & \\
\quad \(+4\) & \(0.034\) & \(0.215\)\sym{***} & \(0.160\) & [-0.323, 0.635] \\
 & (0.049) & (0.039) & & \\
\midrule
Bootstrap draws & \multicolumn{4}{c}{1,000} \\
\bottomrule
\end{tabular}
}

\end{tablebox}

\begin{minipage}{\textwidth}
\vspace{3pt}
\tablenotes{Columns (1), (3) and (4) are in log points; column (2) is in probability units. Column (1) is the reduced-form coefficient at each horizon; column (2) is the first-stage effect of contract review on the probability of having adopted GitHub Copilot at the same horizon; column (3) is their ratio (Equation~\ref{eq:wald_did}). Column (4) reports the 95 percent percentile interval from 1{,}000 firm-block bootstrap draws. Quarter 0 is omitted because the first stage in the review quarter is small, making the ratio unstable. Standard errors, clustered by firm, are in parentheses. \sym{*}~$p<0.10$, \sym{**}~$p<0.05$, \sym{***}~$p<0.01$.}
\end{minipage}
\end{table}

\begin{table}[H]
\centering
\caption{Effects of Contract Review on Technical Workforce Outcomes}
\label{atab:tech_flows_monthly}
\begin{tablebox}
{
\def\sym#1{\ifmmode^{#1}\else\(^{#1}\)\fi}
\begin{tabular}{lccccc}
\toprule
Dependent variable & \multicolumn{2}{c}{Technical hires} & \multicolumn{2}{c}{Technical separations} & Log technical headcount \\
\cmidrule(lr){2-3} \cmidrule(lr){4-5}
 & Log & Any & Log & Any & \\
Months to contract review (\(\ell\)) & (1) & (2) & (3) & (4) & (5) \\
\midrule
\(-12\)     &      -0.006         &      -0.005         &      -0.003         &      -0.001         &      -0.001         \\
            &     (0.011)         &     (0.005)         &     (0.010)         &     (0.005)         &     (0.003)         \\
\addlinespace
\(-11\)     &       0.015         &      -0.014\sym{***}&       0.007         &       0.016\sym{***}&      -0.001         \\
            &     (0.011)         &     (0.005)         &     (0.010)         &     (0.005)         &     (0.003)         \\
\addlinespace
\(-10\)     &       0.003         &       0.002         &       0.025\sym{**} &       0.003         &      -0.003         \\
            &     (0.012)         &     (0.006)         &     (0.011)         &     (0.006)         &     (0.004)         \\
\addlinespace
\(-9\)      &       0.003         &      -0.002         &       0.011         &       0.000         &      -0.003         \\
            &     (0.013)         &     (0.006)         &     (0.012)         &     (0.006)         &     (0.004)         \\
\addlinespace
\(-8\)      &       0.015         &       0.001         &       0.028\sym{**} &      -0.001         &      -0.005         \\
            &     (0.014)         &     (0.006)         &     (0.013)         &     (0.006)         &     (0.005)         \\
\addlinespace
\(-7\)      &       0.004         &      -0.006         &       0.030\sym{**} &       0.008         &      -0.006         \\
            &     (0.015)         &     (0.006)         &     (0.014)         &     (0.007)         &     (0.006)         \\
\addlinespace
\(-6\)      &      -0.003         &       0.002         &       0.010         &       0.010         &      -0.008         \\
            &     (0.016)         &     (0.007)         &     (0.015)         &     (0.007)         &     (0.006)         \\
\addlinespace
\(-5\)      &       0.013         &      -0.004         &       0.018         &       0.002         &      -0.010         \\
            &     (0.018)         &     (0.007)         &     (0.015)         &     (0.007)         &     (0.007)         \\
\addlinespace
\(-4\)      &       0.007         &      -0.006         &       0.008         &      -0.008         &      -0.011         \\
            &     (0.019)         &     (0.008)         &     (0.016)         &     (0.008)         &     (0.008)         \\
\addlinespace
\(-3\)      &       0.009         &      -0.001         &       0.020         &      -0.010         &      -0.012         \\
            &     (0.020)         &     (0.008)         &     (0.017)         &     (0.008)         &     (0.008)         \\
\addlinespace
\(-2\)      &       0.018         &      -0.009         &      -0.001         &      -0.006         &      -0.013         \\
            &     (0.022)         &     (0.008)         &     (0.018)         &     (0.008)         &     (0.009)         \\
\addlinespace
\(-1\)      &       0.012         &       0.001         &       0.004         &       0.002         &      -0.015         \\
            &     (0.023)         &     (0.009)         &     (0.019)         &     (0.009)         &     (0.010)         \\
\addlinespace
\(0\)       &      -0.008         &      -0.003         &       0.007         &      -0.002         &      -0.004         \\
            &     (0.012)         &     (0.006)         &     (0.012)         &     (0.006)         &     (0.003)         \\
\addlinespace
\(+1\)      &      -0.001         &      -0.015\sym{**} &       0.012         &      -0.001         &      -0.007\sym{*}  \\
            &     (0.014)         &     (0.006)         &     (0.013)         &     (0.006)         &     (0.004)         \\
\addlinespace
\(+2\)      &       0.007         &      -0.007         &       0.028\sym{**} &      -0.007         &      -0.008\sym{*}  \\
            &     (0.016)         &     (0.006)         &     (0.014)         &     (0.006)         &     (0.005)         \\
\addlinespace
\(+3\)      &      -0.011         &      -0.008         &       0.020         &      -0.003         &      -0.007         \\
            &     (0.017)         &     (0.007)         &     (0.015)         &     (0.007)         &     (0.005)         \\
\addlinespace
\(+4\)      &      -0.006         &      -0.005         &       0.020         &       0.002         &      -0.008         \\
            &     (0.020)         &     (0.008)         &     (0.016)         &     (0.007)         &     (0.006)         \\
\addlinespace
\(+5\)      &      -0.021         &      -0.014         &       0.030         &      -0.002         &      -0.010         \\
            &     (0.023)         &     (0.009)         &     (0.019)         &     (0.008)         &     (0.007)         \\
\addlinespace
\(+6\)      &      -0.030         &      -0.013         &       0.035         &      -0.002         &      -0.009         \\
            &     (0.025)         &     (0.010)         &     (0.022)         &     (0.009)         &     (0.009)         \\
\addlinespace
\(+7\)      &      -0.042         &      -0.019\sym{*}  &       0.014         &       0.004         &      -0.010         \\
            &     (0.028)         &     (0.011)         &     (0.023)         &     (0.010)         &     (0.010)         \\
\addlinespace
\(+8\)      &      -0.040         &      -0.014         &       0.039         &      -0.013         &      -0.009         \\
            &     (0.030)         &     (0.012)         &     (0.025)         &     (0.011)         &     (0.012)         \\
\addlinespace
\(+9\)      &      -0.052         &      -0.012         &       0.031         &      -0.005         &      -0.010         \\
            &     (0.033)         &     (0.014)         &     (0.028)         &     (0.011)         &     (0.013)         \\
\addlinespace
\(+10\)     &      -0.038         &      -0.018         &       0.025         &       0.007         &      -0.011         \\
            &     (0.034)         &     (0.015)         &     (0.035)         &     (0.013)         &     (0.015)         \\
\addlinespace
\(+11\)     &      -0.081\sym{**} &      -0.023         &       0.058         &       0.012         &      -0.013         \\
            &     (0.035)         &     (0.016)         &     (0.039)         &     (0.014)         &     (0.017)         \\
\addlinespace
\(+12\)     &      -0.079\sym{**} &      -0.010         &       0.029         &       0.011         &      -0.016         \\
            &     (0.038)         &     (0.017)         &     (0.041)         &     (0.015)         &     (0.018)         \\
\midrule
Firm FE     &\(\checkmark\)         &\(\checkmark\)         &\(\checkmark\)         &\(\checkmark\)         &\(\checkmark\)         \\
Month FE    &\(\checkmark\)         &\(\checkmark\)         &\(\checkmark\)         &\(\checkmark\)         &\(\checkmark\)         \\
Pre-trend joint \(p\)&       0.849         &       0.244         &       0.120         &       0.014         &       0.289         \\
DV mean     &       1.722         &       0.707         &       1.533         &       0.661         &       4.812         \\
Firms       &       5,339         &       5,414         &       5,357         &       5,414         &       5,414         \\
Observations&     149,565         &     221,346         &     143,324         &     221,346         &     221,298         \\
\bottomrule
\end{tabular}
}

\end{tablebox}

\begin{minipage}{\textwidth}
\vspace{3pt}
\tablenotes{The table reports reduced-form event-study estimates from the imputation estimator of \citet{borusyak_revisiting_2024} on the balanced firm-month panel with firm and calendar-month fixed effects. Columns (1) and (3) are log technical hires and separations; columns (2) and (4) are indicators for any such flow; column (5) is log technical headcount. Standard errors, clustered by firm, are in parentheses. Pre-trend joint p reports the $p$-value of a Wald test that the pre-review coefficients are jointly zero. \sym{*}~$p<0.10$, \sym{**}~$p<0.05$, \sym{***}~$p<0.01$.}
\end{minipage}
\end{table}

\clearpage
\begin{figure}[H]
\centering
\caption{Extensive Margin Effect of Contract Review on Technical Hiring and Separations}
\label{afig:technical_flows_extensive}

\begin{subfigure}[t]{0.49\textwidth}
    \centering
    \includegraphics[width=\textwidth]{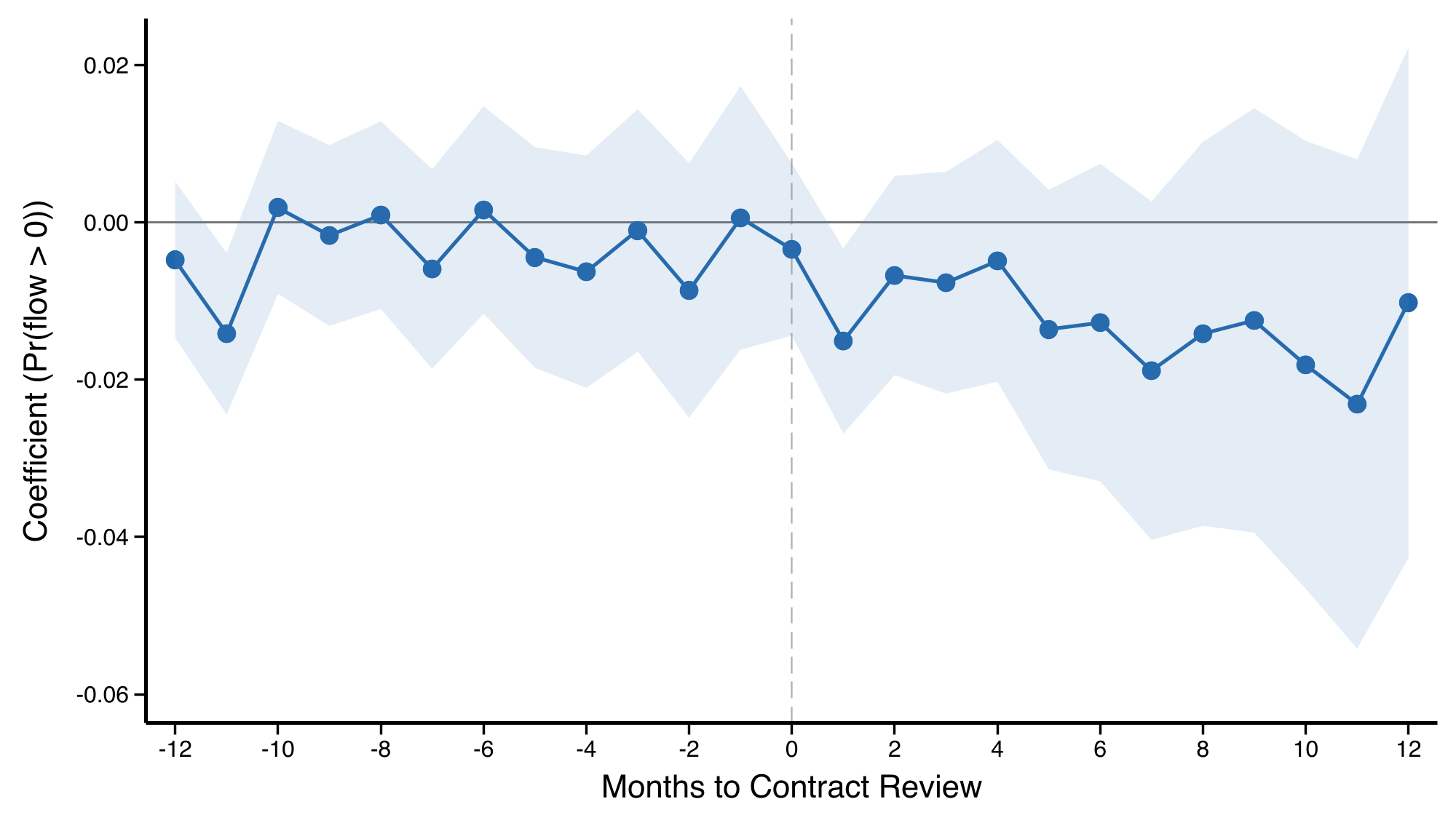}
    \subcaption{Hiring}
\end{subfigure}
\hfill
\begin{subfigure}[t]{0.49\textwidth}
    \centering
    \includegraphics[width=\textwidth]{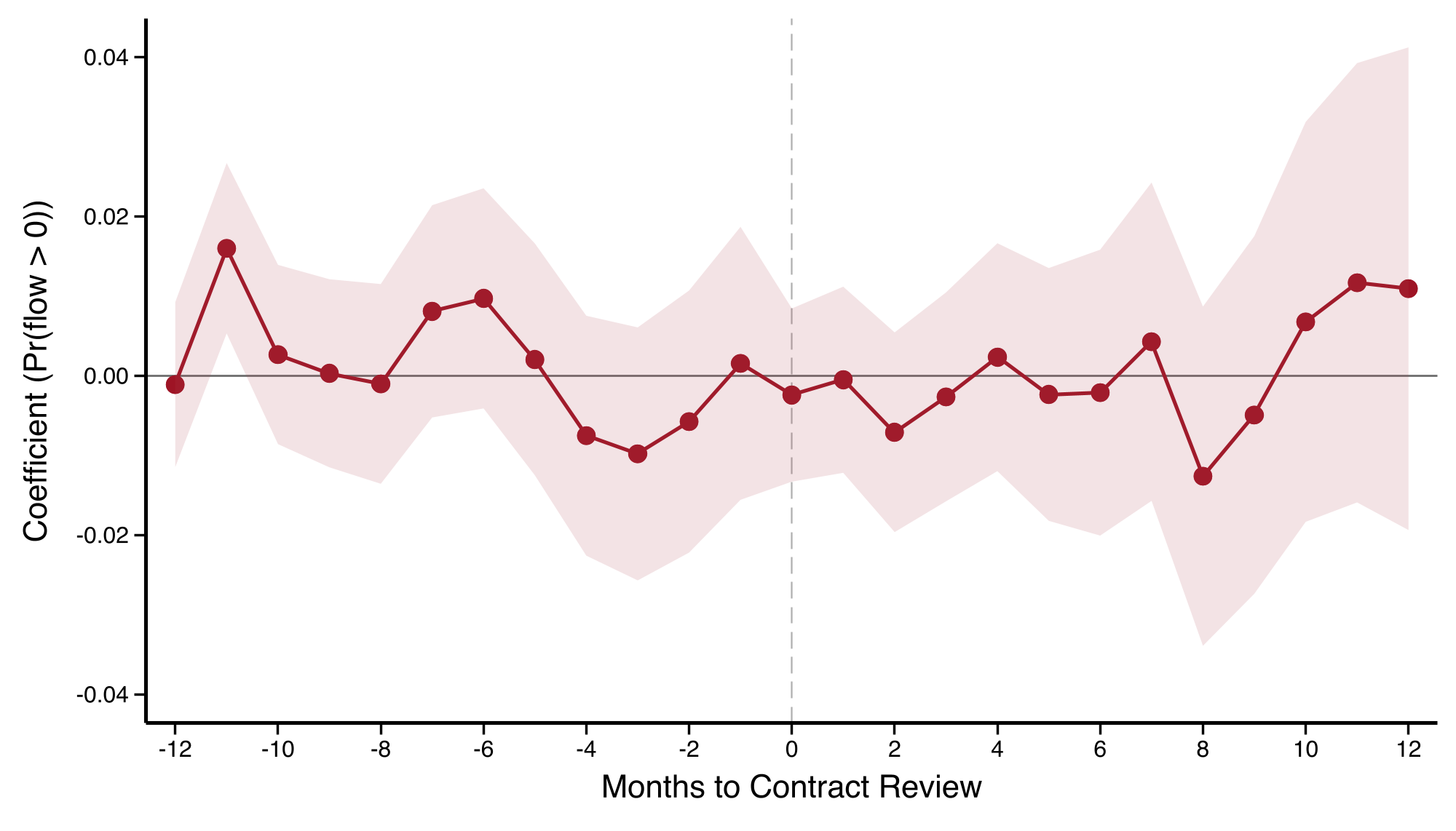}
    \subcaption{Separations}
\end{subfigure}

\par\vspace{1ex}
\figurenotes{The figure reports reduced-form coefficients and 95\% confidence intervals on indicators for any technical hiring (Panel A) and any technical separation (Panel B). Estimates use the imputation estimator of \citet{borusyak_revisiting_2024} on the balanced firm-month panel with firm and calendar-month fixed effects. Standard errors are clustered by firm.}
\end{figure}

\begin{table}[H]
\centering
\caption{Reduced-Form Effects on Technical Job Postings}
\label{atab:postings}
\begin{tablebox}
{
\def\sym#1{\ifmmode^{#1}\else\(^{#1}\)\fi}
\begin{tabular}{l*{3}{c}}
\toprule
Dependent variable & Log postings & Log posting duration & Any postings \\
Quarters to contract review (\(\ell\)) & (1) & (2) & (3) \\
\midrule
\(-4\) & \(-0.042\)\sym{*} & \(0.026\)\sym{*} & \(-0.008\) \\
 & (0.022) & (0.014) & (0.006) \\
\(-3\) & \(-0.050\) & \(0.045\)\sym{**} & \(-0.002\) \\
 & (0.034) & (0.020) & (0.010) \\
\(-2\) & \(-0.063\) & \(0.085\)\sym{***} & \(-0.004\) \\
 & (0.045) & (0.027) & (0.013) \\
\(-1\) & \(-0.072\) & \(0.098\)\sym{***} & \(0.006\) \\
 & (0.056) & (0.033) & (0.016) \\
\(0\) & \(0.017\) & \(-0.005\) & \(0.003\) \\
 & (0.018) & (0.013) & (0.005) \\
\(+1\) & \(0.012\) & \(0.012\) & \(0.003\) \\
 & (0.030) & (0.020) & (0.009) \\
\(+2\) & \(-0.011\) & \(0.021\) & \(0.000\) \\
 & (0.047) & (0.033) & (0.014) \\
\(+3\) & \(-0.080\) & \(0.042\) & \(-0.013\) \\
 & (0.067) & (0.041) & (0.018) \\
\(+4\) & \(-0.148\)\sym{*} & \(0.042\) & \(-0.021\) \\
 & (0.088) & (0.051) & (0.023) \\
\midrule
Firm FE & \(\checkmark\) & \(\checkmark\) & \(\checkmark\) \\
Quarter FE & \(\checkmark\) & \(\checkmark\) & \(\checkmark\) \\
Pre-trend joint \(p\) & 0.386 & 0.011 & 0.069 \\
DV mean & 3.677 & 3.266 & 0.857 \\
Firms & 4,822 & 4,661 & 5,414 \\
Observations & 43,297 & 40,782 & 53,363 \\
\bottomrule
\end{tabular}
}

\end{tablebox}
\begin{minipage}{\textwidth}
\vspace{3pt}
\tablenotes{The table reports quarterly reduced-form estimates from the imputation estimator of \citet{borusyak_revisiting_2024}, with firm and calendar-quarter fixed effects, on the 2022Q1--2025Q2 firm-quarter panel. Outcomes are: (1) log open technical postings, among firms with an open technical posting in at least two of the four 2022 quarters; (2) log mean days from posting to removal, among firms with a removed posting in at least two of the four 2022 quarters; and (3) an indicator for any open technical posting. Postings with an imputed removal date or removed within two days are dropped. Regressions are unweighted. Standard errors, clustered by firm, are in parentheses. \sym{*}~$p<0.10$, \sym{**}~$p<0.05$, \sym{***}~$p<0.01$.}
\end{minipage}
\end{table}

\clearpage
\begin{figure}[H]
\centering
\caption{Example LinkedIn Position Descriptions for Technical Roles}
\label{afig:description_examples}
\begin{subfigure}[t]{0.48\textwidth}
\centering
\includegraphics[width=\textwidth]{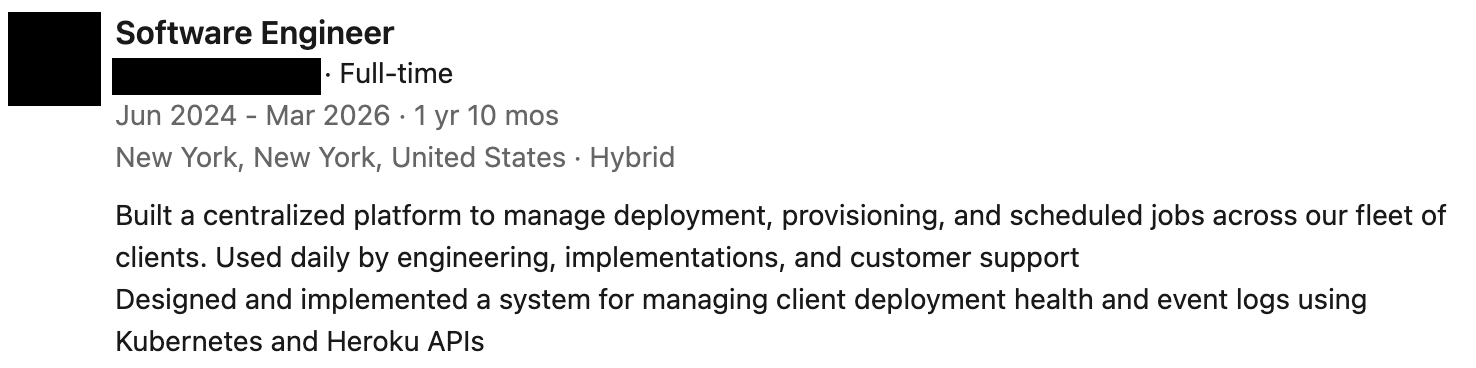}
\subcaption{Software Engineer}
\end{subfigure}
\hfill
\begin{subfigure}[t]{0.48\textwidth}
\centering
\includegraphics[width=\textwidth]{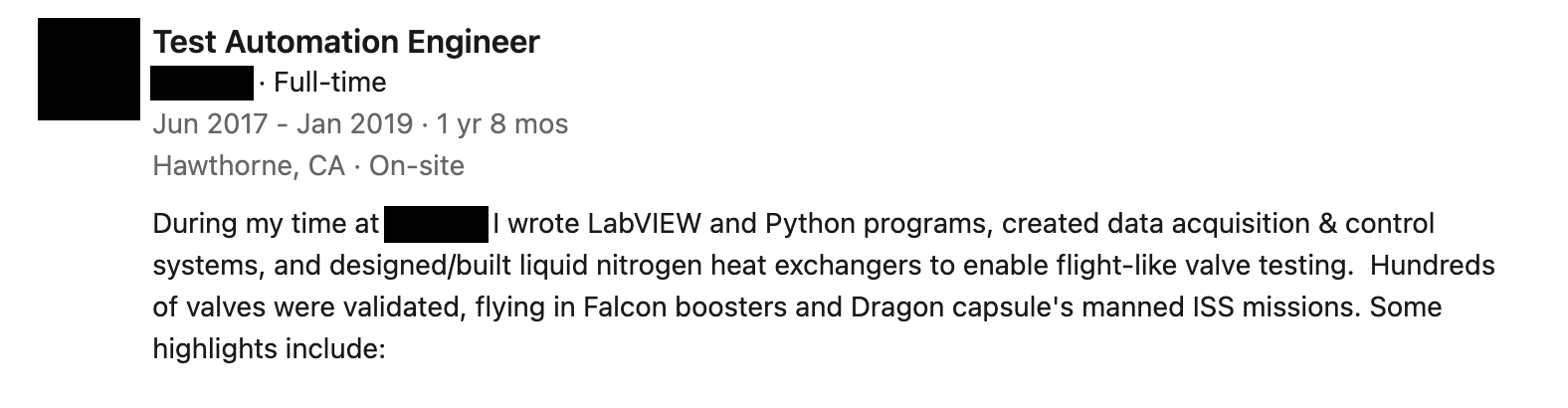}
\subcaption{Quality Assurance Engineer}
\end{subfigure}
\par\vspace{1ex}
\begin{subfigure}[t]{0.48\textwidth}
    \centering
    \includegraphics[width=\textwidth]{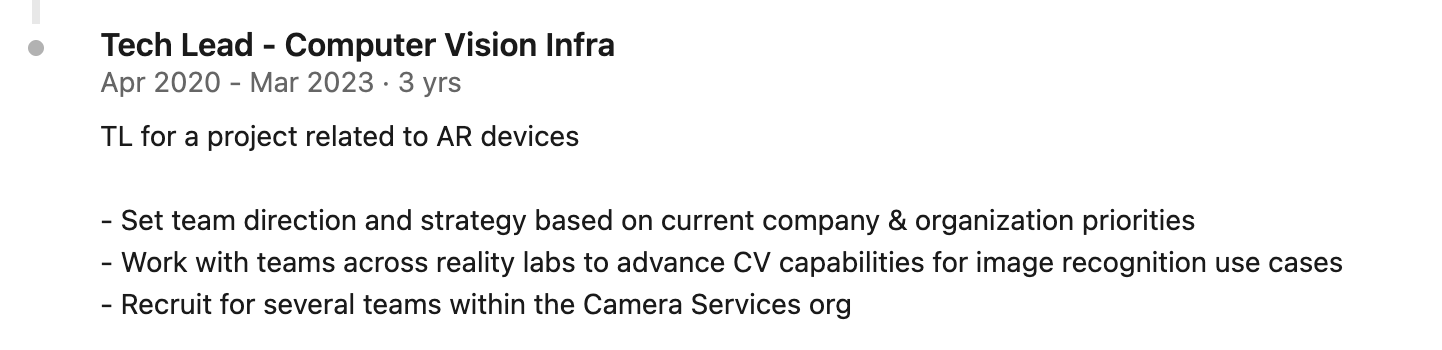}
    \subcaption{Infrastructure Engineer, Computer Vision}
\end{subfigure}
\hfill
\begin{subfigure}[t]{0.48\textwidth}
    \centering
    \includegraphics[width=\textwidth]{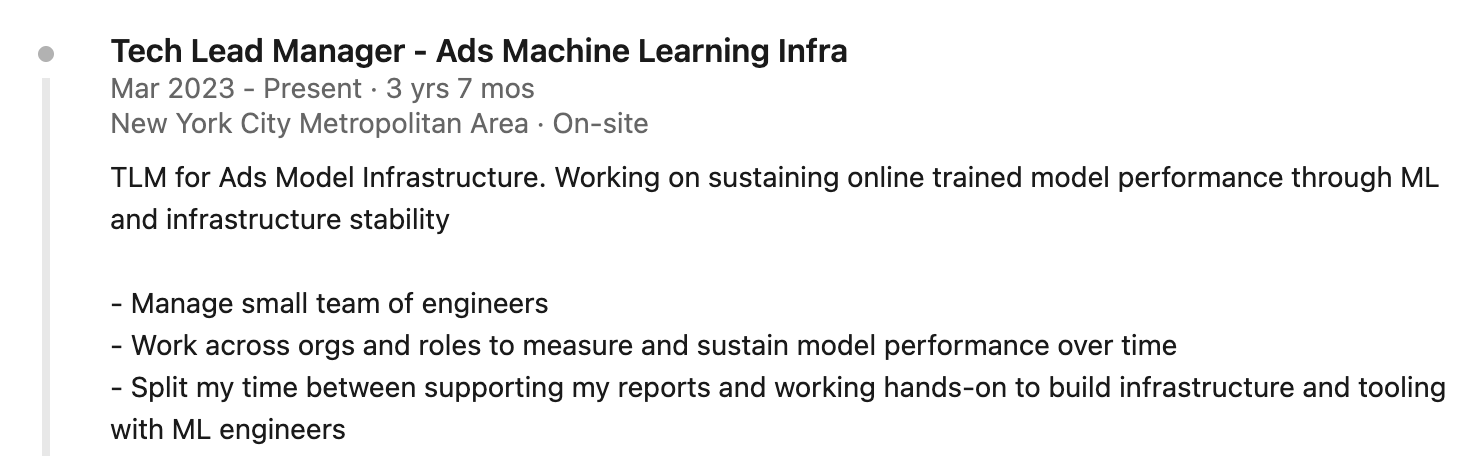}
    \subcaption{Infrastructure Engineer, Machine Learning}
\end{subfigure}

\par\vspace{1ex}
\figurenotes{Examples are drawn from firms outside the sample to preserve anonymity; employer names and logos are blacked out. The code intensity measure uses only the free-text description; titles, employers, dates, locations, and seniority are removed before scoring.}
\end{figure}

\begin{figure}[H]
\centering
\caption{Coding Intensity by Technical Role and Seniority}
\label{afig:coding_intensity_rolesen}
    \centering
    \includegraphics[width=.6\textwidth]{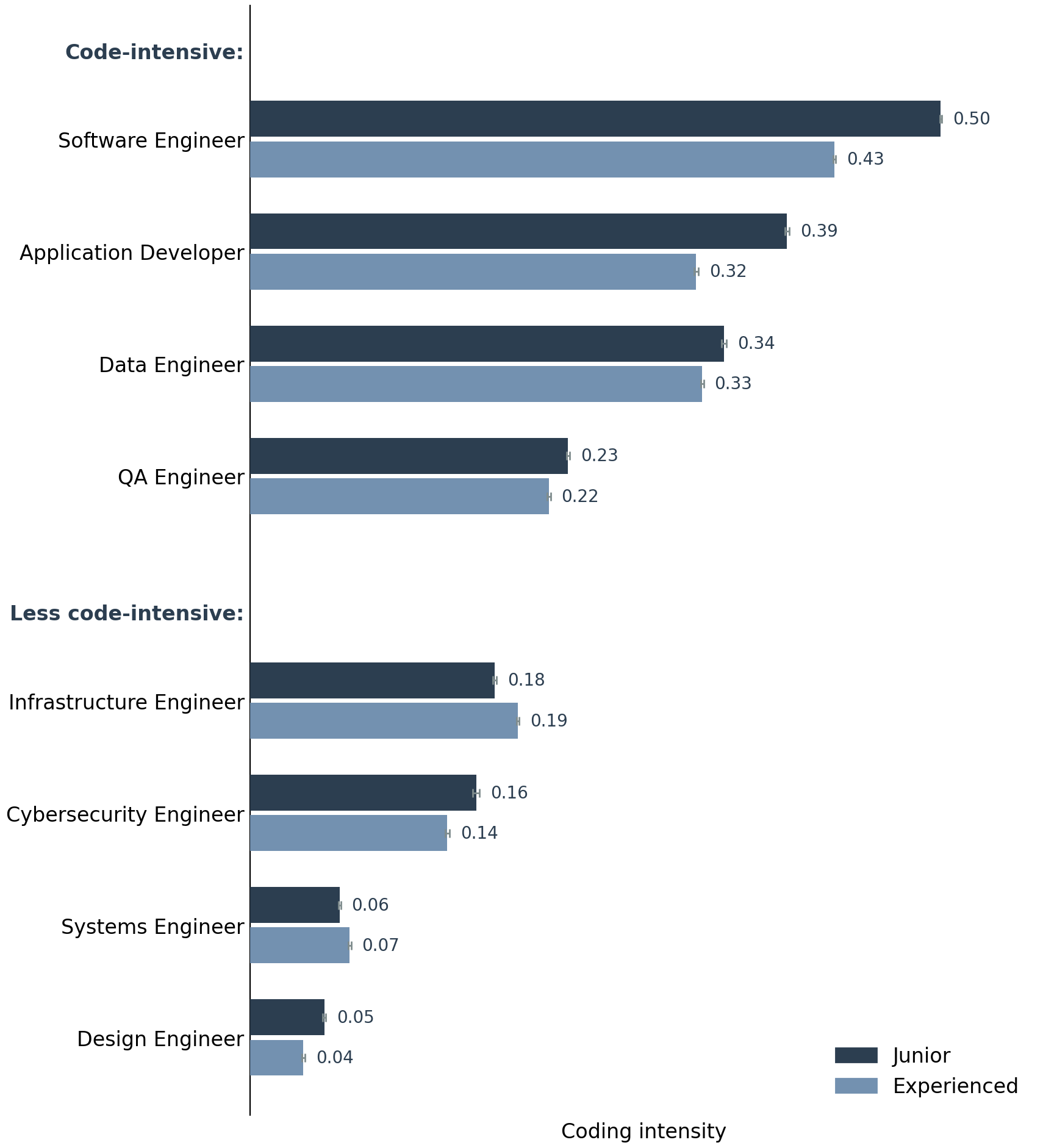}
    
\figurenotes{The figure reports unweighted mean code intensity scores across position descriptions beginning in 2018--2022, by role and seniority (junior positions are Revelio levels 1--2; experienced positions are levels 3--7). Intervals are normal 95 percent confidence intervals based on description-level variation.}
\end{figure}

\clearpage
\begin{table}[H]
\centering
\caption{Hiring Effects by Code Intensity and Seniority: Cell-Specific Scores and Within-Firm Ratios}
\label{atab:gradient_robust}
\begin{tablebox}
{
\def\sym#1{\ifmmode^{#1}\else\(^{#1}\)\fi}
\begin{tabular}{lcccc}
\toprule
Dependent variable & \multicolumn{2}{c}{Log quarterly hires} & \multicolumn{2}{c}{\shortstack{Log junior-to-experienced\\hiring ratio}} \\
\cmidrule(lr){2-3}\cmidrule(lr){4-5}
Code-intensity measure & \multicolumn{2}{c}{Role \(\times\) seniority cell} & Role score & Code-intensive role \\
 & (1) & (2) & (3) & (4) \\
\midrule
Quarter 4 after contract review (\(\rho_4\))&      -0.017         &      -0.021         &      -0.019         &      -0.026         \\
            &     (0.035)         &     (0.042)         &     (0.050)         &     (0.049)         \\
\addlinespace
Code intensity (\(\theta_4\))&      -0.029\sym{*}  &       0.020         &      -0.080\sym{**} &      -0.169\sym{**} \\
            &     (0.018)         &     (0.025)         &     (0.038)         &     (0.077)         \\
\addlinespace
Junior (\(\gamma_4\))&                     &       0.011         &                     &                     \\
            &                     &     (0.040)         &                     &                     \\
\addlinespace
Junior \(\times\) code intensity (\(\beta_4\))&                     &      -0.082\sym{***}&                     &                     \\
            &                     &     (0.030)         &                     &                     \\
\midrule
Firm \(\times\) role \(\times\) seniority FE & \(\checkmark\) & \(\checkmark\) & & \\
Role \(\times\) seniority \(\times\) quarter FE & \(\checkmark\) & \(\checkmark\) & & \\
Firm \(\times\) role FE & & & \(\checkmark\) & \(\checkmark\) \\
Role \(\times\) quarter FE & & & \(\checkmark\) & \(\checkmark\) \\
\midrule
Implied junior gradient (\(\theta_4+\beta_4\)) &  & -0.062\sym{***} & & \\
 &  & (0.021) & & \\
Firms & 5,048 & 5,048 & 4,490 & 4,490 \\
Observations & 357,346 & 357,346 & 125,856 & 125,856 \\
\bottomrule
\end{tabular}
}

\end{tablebox}
\begin{minipage}{\textwidth}
\vspace{3pt}
\tablenotes{The table reports reduced-form effects four quarters after scheduled contract review. Columns (1) and (2) use a separate pre-launch coding-intensity score for each of the 16 role-by-seniority cells, standardized across cells (Figure~\ref{afig:coding_intensity_rolesen}), on the firm--role--seniority--quarter sample of Table~\ref{tab:junior_code_interaction}. The implied junior gradient is $\theta_4+\beta_4$, with its standard error from the estimated covariance. Columns (3) and (4) report effects on the log junior-to-experienced hiring ratio within a role. In column (3), $\rho_4$ is the effect at mean code intensity and $\theta_4$ its change per standard deviation of the pre-launch role score. In column (4), code intensity is an indicator for the four code-intensive roles above the median. Standard errors, clustered by firm, are in parentheses. \sym{*}~$p<0.10$, \sym{**}~$p<0.05$, \sym{***}~$p<0.01$.}
\end{minipage}
\end{table}

\clearpage
\begin{table}[H]
\centering
\caption{Extensive Margin Estimates for Hiring by Role Group}
\label{atab:extensive_margin_main_results}
\begin{tablebox}
{
\def\sym#1{\ifmmode^{#1}\else\(^{#1}\)\fi}
\begin{tabular}{lcccccc}
\toprule
Dependent variable & \multicolumn{4}{c}{Any hiring} & \multicolumn{2}{c}{Both groups hire} \\
\cmidrule(lr){2-5} \cmidrule(lr){6-7}
 & Software eng. & Systems and infra. & More code-intensive & Less code-intensive & Junior and exp. SWE & SWE and systems/infra. \\
 & (1) & (2) & (3) & (4) & (5) & (6) \\
\midrule
Quarter 4 after contract review & 0.002 & -0.002 & -0.016 & 0.009 & -0.001 & -0.009 \\
 & (0.024) & (0.022) & (0.019) & (0.021) & (0.027) & (0.027) \\
\midrule
Firm FE & \(\checkmark\) & \(\checkmark\) & \(\checkmark\) & \(\checkmark\) & \(\checkmark\) & \(\checkmark\) \\
Quarter FE & \(\checkmark\) & \(\checkmark\) & \(\checkmark\) & \(\checkmark\) & \(\checkmark\) & \(\checkmark\) \\
Pre-trend joint \(p\) & 0.475 & 0.687 & 0.094 & 0.806 & 0.189 & 0.594 \\
DV mean & 0.712 & 0.647 & 0.792 & 0.697 & 0.540 & 0.551 \\
Firms & 5,182 & 5,130 & 5,360 & 5,202 & 4,561 & 4,932 \\
Observations & 71,741 & 70,980 & 74,230 & 71,995 & 63,029 & 68,189 \\
\bottomrule
\end{tabular}
}

\end{tablebox}
\begin{minipage}{\textwidth}
\vspace{3pt}
\tablenotes{The table reports reduced-form effects four quarters after scheduled contract review on indicators for any hiring in a role group (columns 1--4) and for hiring in both groups of a pair in the same quarter (columns 5--6), estimated with the imputation estimator of \citet{borusyak_revisiting_2024} on the quarterly firm panel with firm and calendar-quarter fixed effects. Regressions are unweighted. The pre-trend $p$-value jointly tests the four leads. Standard errors, clustered by firm, are in parentheses. \sym{*}~$p<0.10$, \sym{**}~$p<0.05$, \sym{***}~$p<0.01$.}
\end{minipage}
\end{table}

\clearpage
\begin{figure}[H]
\centering
\caption{Junior-to-Experienced Hiring Ratios by Role}
\label{afig:rf_hiring_ratio_by_role_q4}
\includegraphics[width=.85\textwidth]{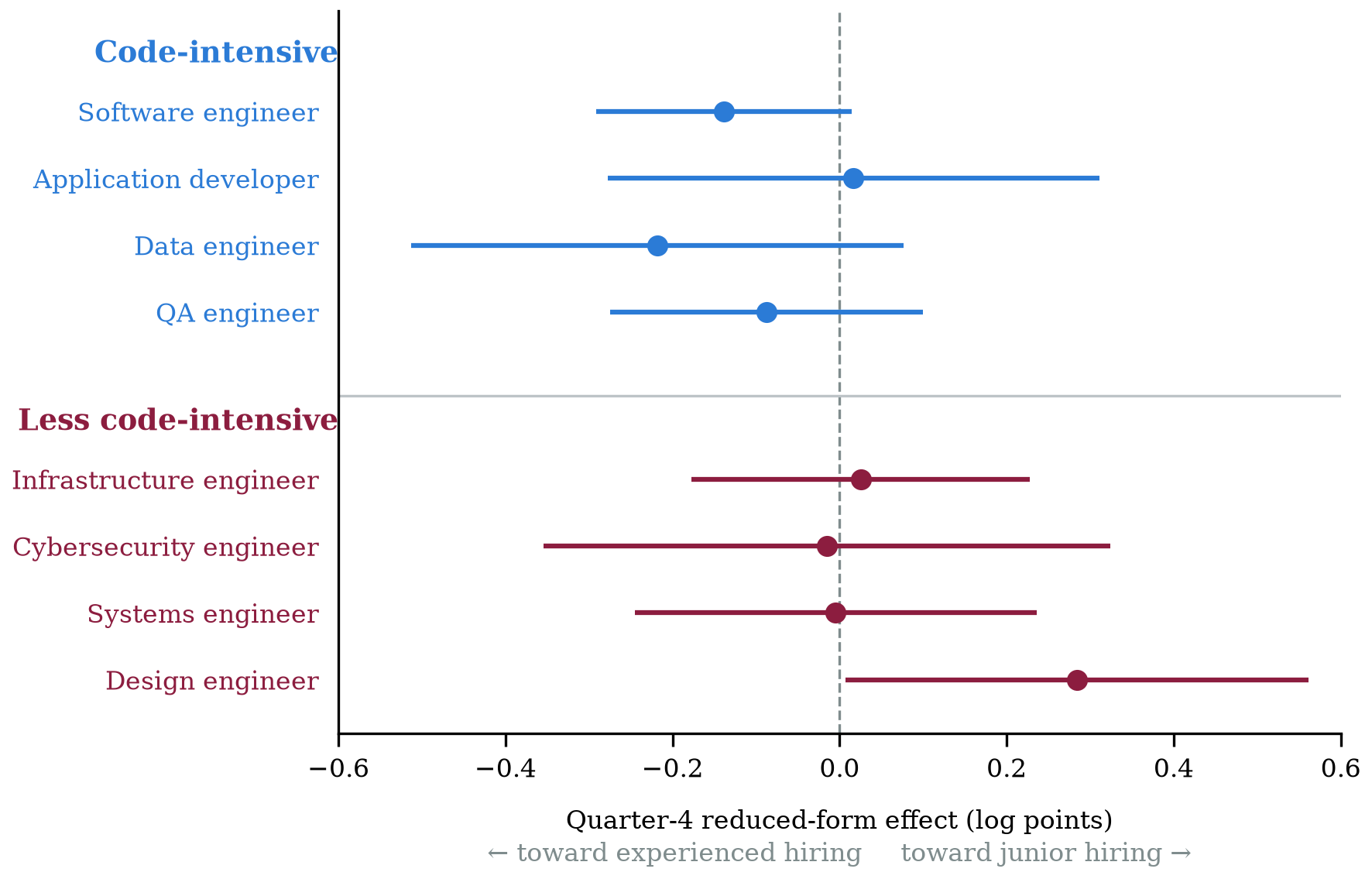}
\par\vspace{1ex}
\figurenotes{The figure reports role-specific reduced-form effects four quarters after scheduled contract review on the log junior-to-experienced hiring ratio. Negative estimates indicate a shift toward more-experienced hiring; positive estimates indicate a shift toward junior hiring. Roles are ordered by pre-GitHub Copilot code intensity. Points report role-specific estimates and horizontal whiskers are 95\% confidence intervals based on standard errors clustered by firm.}
\end{figure}

\clearpage
\begin{table}[H]
\centering
\caption{Wald Estimates for SWE and Systems/Infra Hiring}
\label{atab:latet_rolesen}
\begin{tablebox}
{
\def\sym#1{\ifmmode^{#1}\else\(^{#1}\)\fi}
\begin{tabular}{lcccc}
\toprule
 & \multicolumn{4}{c}{Log quarterly hires} \\
\cmidrule(lr){2-5}
 & \multicolumn{2}{c}{By role} & \multicolumn{2}{c}{Software engineering, by seniority} \\
\cmidrule(lr){2-3} \cmidrule(lr){4-5}
 & Software & Systems/ & Junior & Experienced \\
 & engineering & infrastructure & & \\
 & (1) & (2) & (3) & (4) \\
\midrule
\multicolumn{5}{l}{\textit{Panel A. Reduced form}} \\
Contract review (Q4) & \(-0.064\) & \(0.050\) & \(-0.108\)\sym{**} & \(-0.034\) \\
 & (0.048) & (0.055) & (0.052) & (0.064) \\
\midrule
\multicolumn{5}{l}{\textit{Panel B. First stage}} \\
First-stage effect on adoption & \(0.206\)\sym{***} & \(0.238\)\sym{***} & \(0.211\)\sym{***} & \(0.211\)\sym{***} \\
 & (0.042) & (0.039) & (0.047) & (0.047) \\
\midrule
\multicolumn{5}{l}{\textit{Panel C. Wald estimate}} \\
GitHub Copilot adoption & \(-0.310\) & \(0.212\) & \(-0.513\) & \(-0.160\) \\
 & \([-0.884,\ 0.186]\) & \([-0.247,\ 0.798]\) & \([-1.262,\ 0.008]\) & \([-0.830,\ 0.492]\) \\
\midrule
Firm FE & \(\checkmark\) & \(\checkmark\) & \(\checkmark\) & \(\checkmark\) \\
Quarter FE & \(\checkmark\) & \(\checkmark\) & \(\checkmark\) & \(\checkmark\) \\
Firms & 4,901 & 4,825 & 4,407 & 4,182 \\
Observations & 51,022 & 45,843 & 44,191 & 38,895 \\
\bottomrule
\end{tabular}
}

\end{tablebox}
\begin{minipage}{\textwidth}
\vspace{3pt}
\tablenotes{Each column reports estimates four quarters after scheduled contract review for log quarterly hires in the role or role-by-seniority group. Panel A reports the reduced-form effect of contract review in log points. Panel B reports the first-stage estimates in probability units; Panel C reports the Wald estimate in log points (Equation~\ref{eq:wald_did}); the bracketed interval is the 95\% percentile firm-block bootstrap interval. Both legs use the imputation estimator of \citet{borusyak_revisiting_2024} with firm and calendar-quarter fixed effects. Regressions are unweighted. Standard errors, clustered by firm, are in parentheses. \sym{*}~$p<0.10$, \sym{**}~$p<0.05$, \sym{***}~$p<0.01$.}
\end{minipage}
\end{table}

\clearpage
\begin{figure}[H]
\centering
\caption{Reduced-Form Hiring Estimates for Software Engineering and Systems/Infrastructure}
\label{afig:swe_sys}

\begin{subfigure}[t]{0.49\textwidth}
    \centering
    \includegraphics[width=\linewidth]{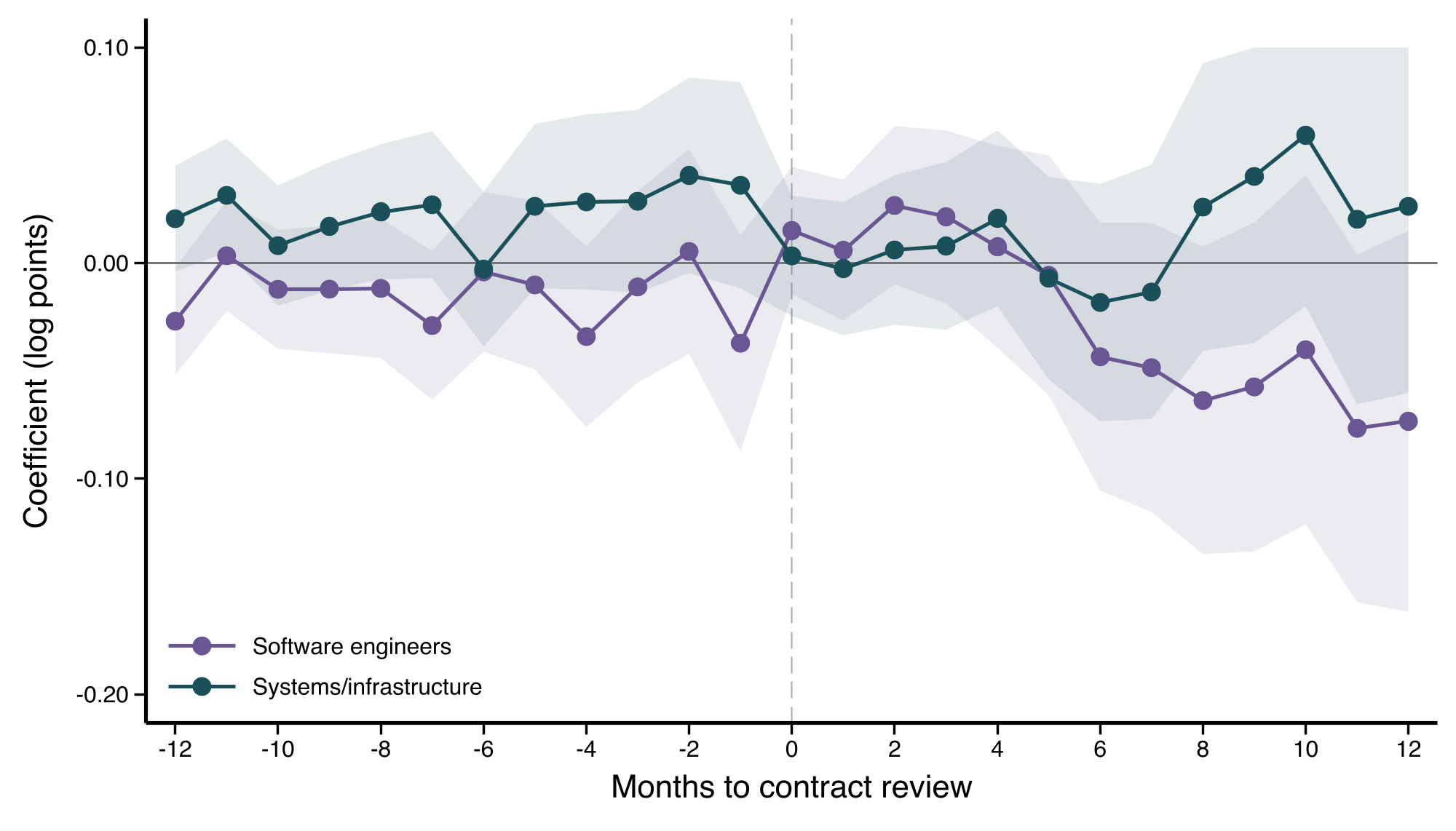}
    \subcaption{Software Engineering vs. Systems/Infrastructure}
\end{subfigure}
\hfill
\begin{subfigure}[t]{0.49\textwidth}
    \centering
    \includegraphics[width=\linewidth]{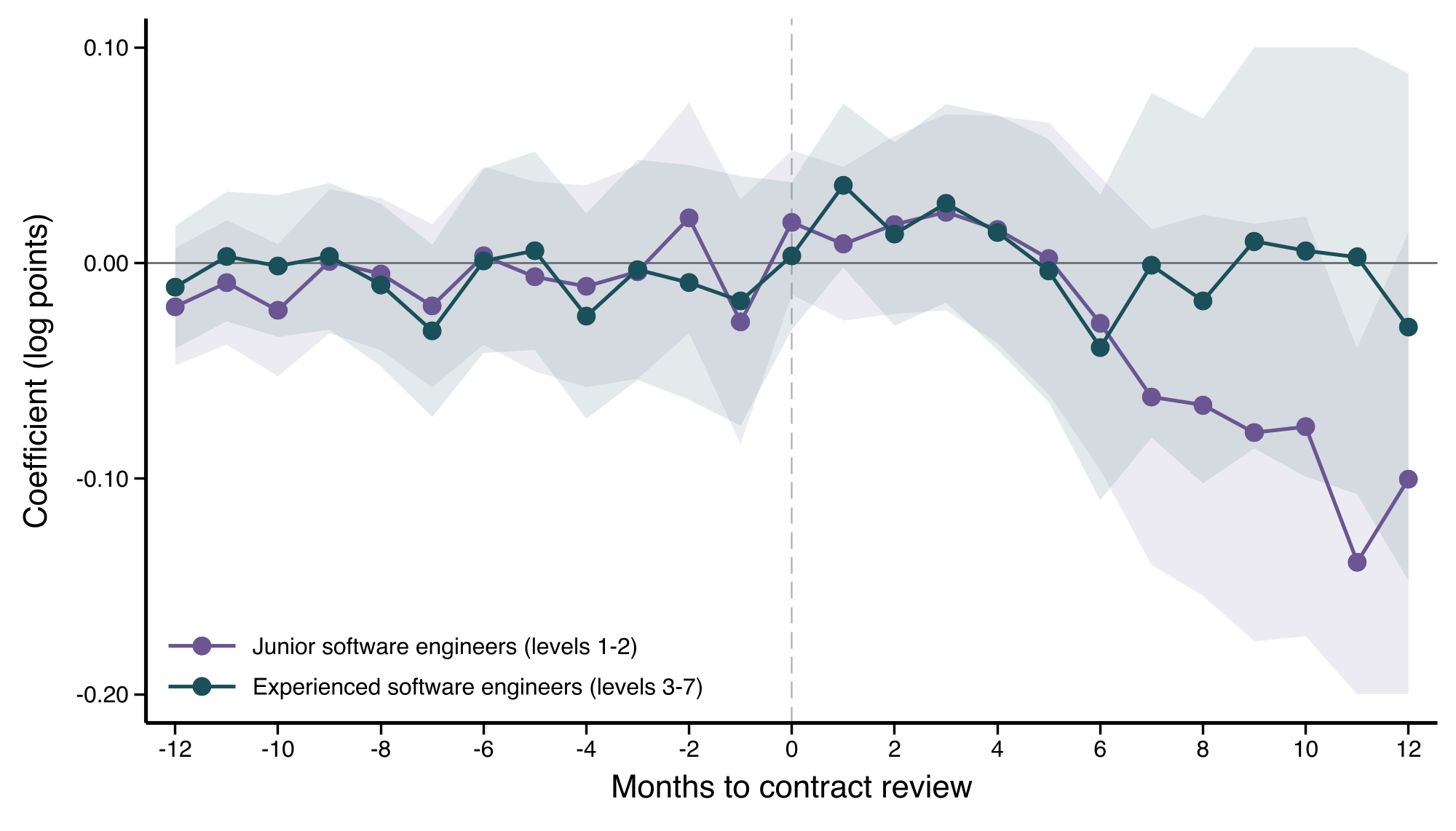}
    \subcaption{Within Software Engineering}
\end{subfigure}

\par\vspace{1ex}
\figurenotes{The figure reports monthly reduced-form event-study coefficients for log hire spells on firm-months with positive hiring. Panel A compares all software engineering positions with systems and infrastructure engineering positions; Panel B compares junior (levels 1--2) and experienced (levels 3--7) software engineering positions. Estimates use the imputation estimator of \citet{borusyak_revisiting_2024} with firm and calendar-month fixed effects. Bands are 95 percent confidence intervals; standard errors are clustered by firm.}
\end{figure}

\clearpage
\begin{table}[H]
\centering
\caption{Education and Experience of Technical Workers by Coding Intensity}
\label{atab:stock_educ_background}
\begin{tablebox}
\begin{tabular}{llccccr}
\toprule
 & & Years since BA & Academic master's (\%) & Doctorate (\%) & Engineering degree (\%) & Spells \\
\midrule
SWE & Junior & 4.8 & 26.9 & 1.6 & 93.6 & 1,518,836 \\
 & Experienced & 8.7 & 32.7 & 2.5 & 92.7 & 874,670 \\
Code-intensive & Junior & 5.0 & 25.9 & 1.6 & 92.6 & 2,361,788 \\
 & Experienced & 8.9 & 31.6 & 2.5 & 90.6 & 1,824,329 \\
Less code-intensive & Junior & 6.0 & 26.0 & 1.8 & 90.2 & 1,067,491 \\
 & Experienced & 10.4 & 32.4 & 3.0 & 86.2 & 916,679 \\
All technical & Junior & 5.2 & 26.0 & 1.6 & 91.9 & 3,429,279 \\
 & Experienced & 9.3 & 31.9 & 2.6 & 89.3 & 2,741,008 \\
\bottomrule
\end{tabular}

\end{tablebox}
\begin{minipage}{\textwidth}
\vspace{3pt}
\tablenotes{The sample is position spells held at sample firms in 2021 and 2022, unweighted, for software engineers, for all code-intensive and less code-intensive technical roles, and for all technical roles, split by seniority. Code-intensive roles are software engineers, application developers, and data and QA engineers; less code-intensive roles are infrastructure, systems, cybersecurity, and design engineers. Years since BA is the mean number of years between a worker's bachelor's degree and the start of the position. Degree shares are percentages of spells with the relevant education record: an academic master's (excluding MBAs and doctorates), a doctorate, or any degree in an engineering or information-technology field. The degree year is missing for 44 to 56 percent of spells, the degree level for 35 to 44 percent, and the field for 45 to 56 percent. Spells is the number of position spells.}
\end{minipage}
\end{table}

\clearpage
\begin{table}[H]
\centering
\caption{India-Based and Domestic Technical Hiring}
\label{atab:india_dom_hiring_preindia}
\begin{tablebox}
{
\def\sym#1{\ifmmode^{#1}\else\(^{#1}\)\fi}
\begin{tabular}{lcccc}
\toprule
Dependent variable & \multicolumn{2}{c}{Log technical hires} & \multicolumn{2}{c}{Any technical hires} \\
\cmidrule(lr){2-3} \cmidrule(lr){4-5}
 & India-based & Domestic & India-based & Domestic \\
Quarters to contract review (\(\ell\)) & (1) & (2) & (3) & (4) \\
\midrule
\(-4\)      &       0.023         &      -0.009         &       0.000         &       0.000         \\
            &     (0.026)         &     (0.012)         &     (0.005)         &     (0.005)         \\
\addlinespace
\(-3\)      &       0.020         &       0.007         &      -0.004         &      -0.006         \\
            &     (0.034)         &     (0.015)         &     (0.006)         &     (0.006)         \\
\addlinespace
\(-2\)      &       0.007         &       0.011         &      -0.008         &      -0.007         \\
            &     (0.045)         &     (0.020)         &     (0.008)         &     (0.008)         \\
\addlinespace
\(-1\)      &       0.004         &       0.005         &      -0.012         &      -0.006         \\
            &     (0.056)         &     (0.024)         &     (0.009)         &     (0.009)         \\
\addlinespace
\(0\)       &       0.005         &       0.009         &      -0.001         &       0.004         \\
            &     (0.034)         &     (0.015)         &     (0.005)         &     (0.006)         \\
\addlinespace
\(+1\)      &      -0.015         &      -0.001         &      -0.005         &       0.011         \\
            &     (0.055)         &     (0.022)         &     (0.008)         &     (0.009)         \\
\addlinespace
\(+2\)      &      -0.075         &      -0.022         &       0.012         &       0.016         \\
            &     (0.106)         &     (0.034)         &     (0.013)         &     (0.016)         \\
\addlinespace
\(+3\)      &      -0.079         &      -0.042         &       0.009         &       0.014         \\
            &     (0.128)         &     (0.045)         &     (0.017)         &     (0.020)         \\
\addlinespace
\(+4\)      &      -0.060         &      -0.029         &       0.029         &       0.016         \\
            &     (0.189)         &     (0.054)         &     (0.019)         &     (0.025)         \\
\midrule
Firm FE     &\(\checkmark\)         &\(\checkmark\)         &\(\checkmark\)         &\(\checkmark\)         \\
Quarter FE  &\(\checkmark\)         &\(\checkmark\)         &\(\checkmark\)         &\(\checkmark\)         \\
Pre-trend joint \(p\)&       0.869         &       0.568         &       0.643         &       0.744         \\
DV mean     &       2.367         &       2.331         &       0.275         &       0.872         \\
Firms       &       2,368         &       4,434         &       4,509         &       4,509         \\
Observations&      16,633         &      52,821         &      62,336         &      62,336         \\
\bottomrule
\end{tabular}
}

\end{tablebox}

\begin{minipage}{\textwidth}
\vspace{3pt}
\tablenotes{The sample is firms with at least one India-based employee in any role in some quarter from 2021 to 2022. India-based hires are hire spells into technical positions located in India; domestic hires are spells located in the firm's headquarters country. Columns (1) and (2) are in log points on firm-quarters with positive hiring in the cell; columns (3) and (4) are indicators for any hiring in the cell. All columns use the imputation estimator of \citet{borusyak_revisiting_2024} on the quarterly panel with firm and calendar-quarter fixed effects. The pre-trend joint $p$-value tests that the four lead coefficients are jointly zero. Regressions are unweighted. Standard errors, clustered by firm, are in parentheses. \sym{*}~$p<0.10$, \sym{**}~$p<0.05$, \sym{***}~$p<0.01$.}
\end{minipage}
\end{table}

\clearpage
\section{Placebo and Robustness Checks}

\begin{table}[H]
\caption{One-Year Contract-Review Placebo}
\label{atab:placebo_2022_cycle_q0_q2}
\centering
\begin{tablebox}
{
\def\sym#1{\ifmmode^{#1}\else\(^{#1}\)\fi}
\begin{tabular}{lcccc}
\toprule
Dependent variable & \shortstack{Log technical\\hires} & \shortstack{Log technical\\separations} & \shortstack{Log junior\\SWE hires} & \shortstack{Log experienced\\SWE hires} \\
Quarters to placebo review (\(\ell\)) & (1) & (2) & (3) & (4) \\
\midrule
\(-2\) & \(-0.042\)\sym{**} & \( 0.009\) & \(-0.018\) & \(-0.023\) \\
 & (0.018) & (0.018) & (0.022) & (0.024) \\
\addlinespace
\(-1\) & \(-0.049\)\sym{**} & \( 0.004\) & \(-0.021\) & \(-0.063\)\sym{**} \\
 & (0.023) & (0.023) & (0.028) & (0.031) \\
\addlinespace
\(0\) & \(-0.016\) & \( 0.006\) & \(-0.001\) & \( 0.001\) \\
 & (0.012) & (0.012) & (0.016) & (0.017) \\
\addlinespace
\(+1\) & \(-0.006\) & \( 0.026\) & \( 0.030\) & \(-0.017\) \\
 & (0.018) & (0.018) & (0.022) & (0.024) \\
\addlinespace
\(+2\) & \(-0.033\) & \( 0.045\) & \( 0.005\) & \(-0.058\) \\
 & (0.028) & (0.028) & (0.037) & (0.042) \\
\midrule
Firm FE & \(\checkmark\) & \(\checkmark\) & \(\checkmark\) & \(\checkmark\) \\
Quarter FE & \(\checkmark\) & \(\checkmark\) & \(\checkmark\) & \(\checkmark\) \\
Pre-trend joint \(p\) & 0.104 & 0.416 & 0.867 & 0.038 \\
Firms &        5,212 &        5,180 &        4,249 &        3,934 \\
Observations &       36,896 &       35,621 &       27,171 &       22,703 \\
\bottomrule
\end{tabular}
}

\end{tablebox}
\begin{minipage}{\textwidth}
\vspace{3pt}
\tablenotes{The sample comprises firm-quarters from 2021Q1 through 2022Q4. Placebo dates shift scheduled contract review back four quarters, retaining the baseline treated and control groups. Outcomes are log technical hires and separations, and junior and experienced software-engineer (SWE) hire spells. Estimates use the imputation estimator of \citet{borusyak_revisiting_2024} with firm and calendar-quarter fixed effects. Q0 denotes the quarter containing placebo review; later horizons cover fewer cohorts because outcomes end in 2022Q4 and so the sample is not fully balanced. The pre-trend joint $p$-value tests the three estimated leads. Standard errors, clustered by firm, are in parentheses. Effects are in log points. \sym{*}~$p<0.10$, \sym{**}~$p<0.05$, \sym{***}~$p<0.01$.}
\end{minipage}
\end{table}

\begin{figure}[H]
\centering
\caption{Accountant-Hiring Placebo: Event-Study Estimates}
\label{afig:accountant_placebo}
\begin{subfigure}[t]{0.49\textwidth}
    \centering
    \includegraphics[width=\linewidth]{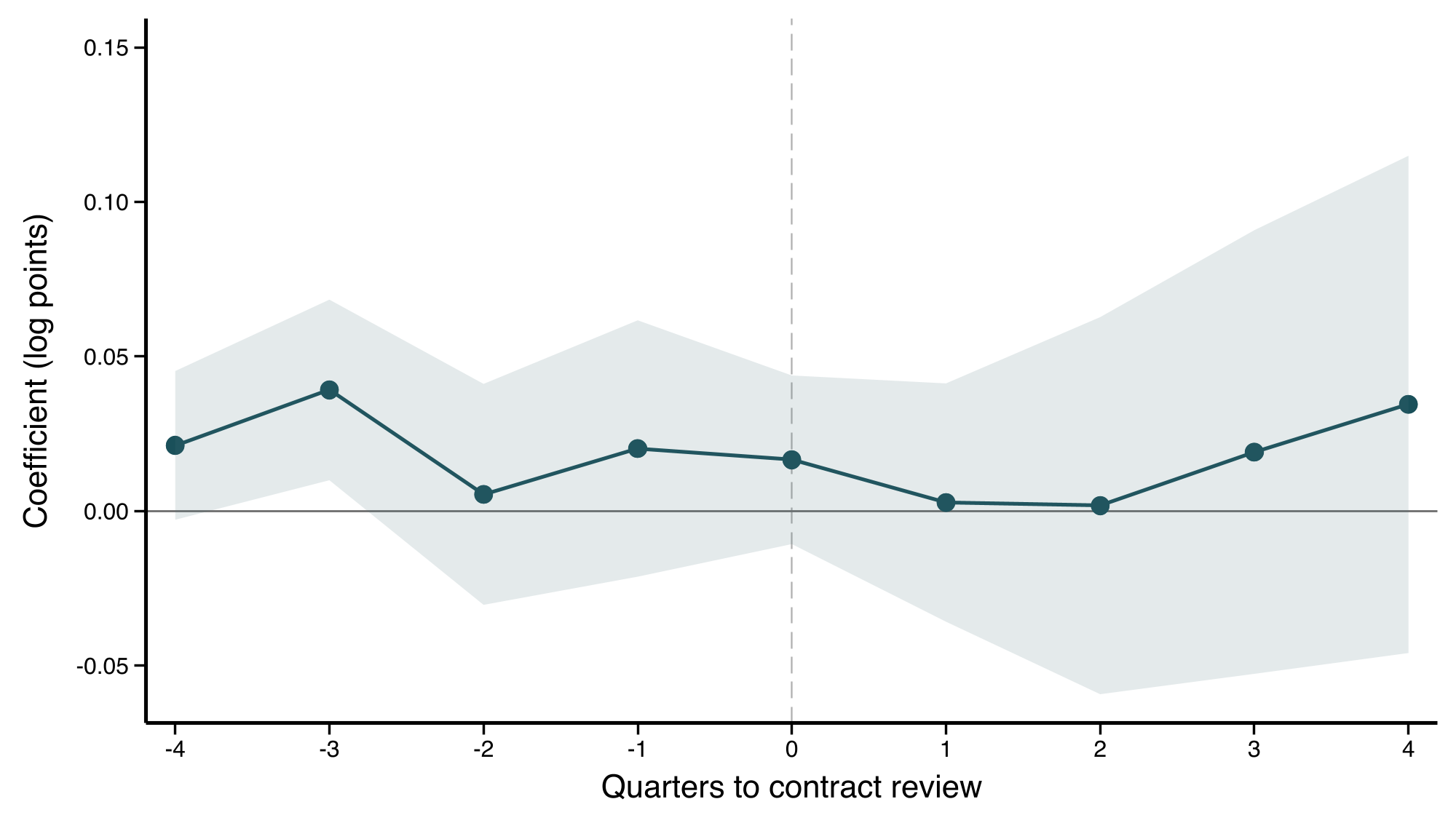}
    \subcaption{Log hires}
\end{subfigure}
\begin{subfigure}[t]{0.49\textwidth}
    \centering
    \includegraphics[width=\linewidth]{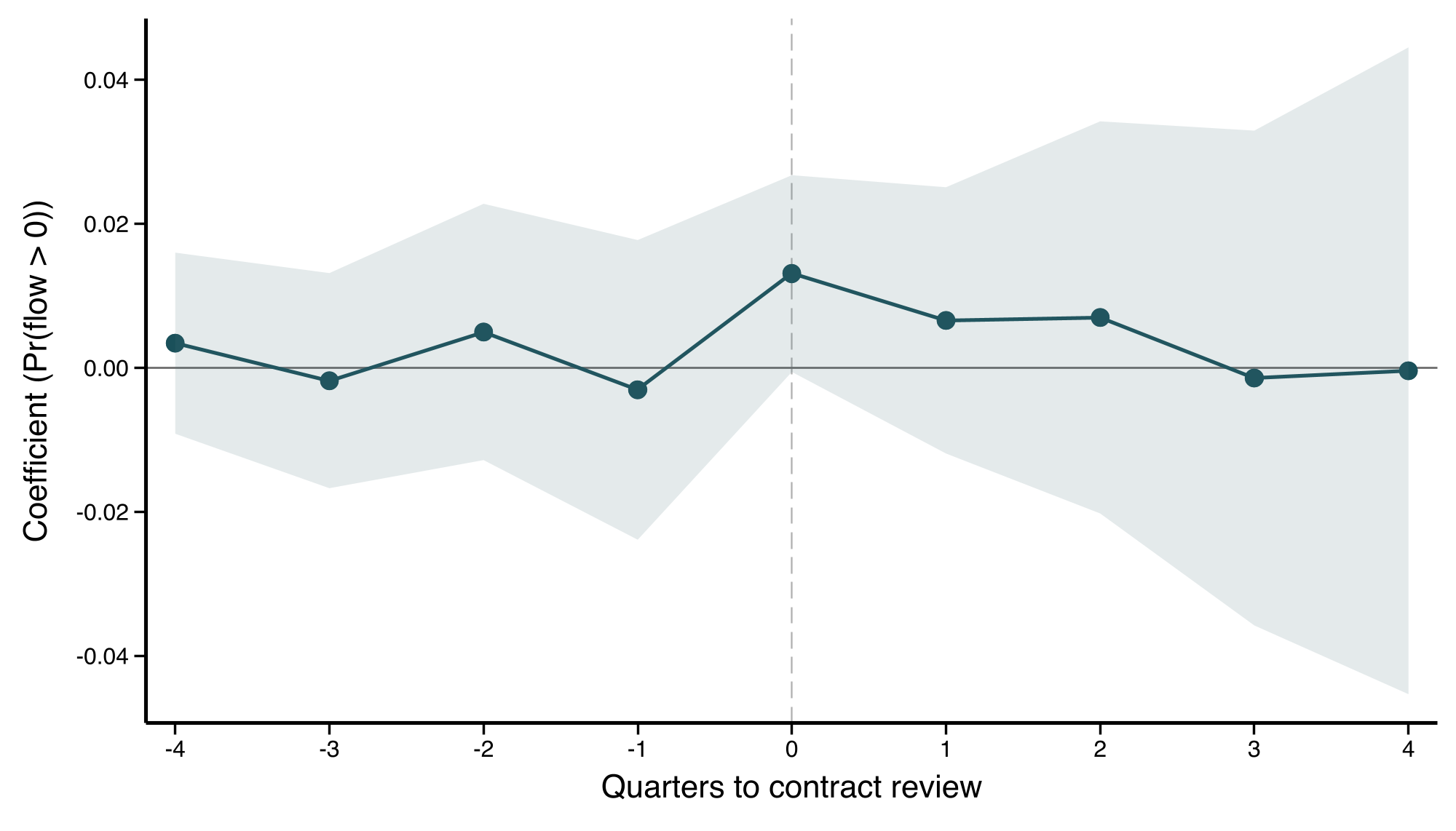}
    \subcaption{Any hiring}
\end{subfigure}

\par\vspace{1ex}
\figurenotes{The figure reports reduced-form effects on log accountant hire spells on firm-quarters with positive accountant hiring (Panel A) and an indicator for any accountant hiring (Panel B). Estimates use the imputation estimator of \citet{borusyak_revisiting_2024} with firm and calendar-quarter fixed effects. Bands are 95 percent confidence intervals; standard errors are clustered by firm.}
\end{figure}

\clearpage
\begin{table}[H]
\centering
\caption{Robustness to Size and Industry Trends}
\label{atab:additional_controls}
\begin{tablebox}
{
\def\sym#1{\ifmmode^{#1}\else\(^{#1}\)\fi}
\begin{tabular}{lccccccc}
\toprule
Dependent variable & \multicolumn{2}{c}{Technical} & \multicolumn{3}{c}{SWE} & \multicolumn{2}{c}{Hiring by code intensity} \\
\cmidrule(lr){2-3} \cmidrule(lr){4-6} \cmidrule(lr){7-8}
 & Log hires & Log separations & Log junior/experienced ratio & Log junior hires & Log experienced hires & More code-intensive & Less code-intensive \\
 & (1) & (2) & (3) & (4) & (5) & (6) & (7) \\
\midrule
\multicolumn{8}{l}{\textit{Baseline}} \\
Quarter 4 after contract review & \(-0.073\) & \(0.034\) & \(-0.139\)\sym{*} & \(-0.108\)\sym{**} & \(-0.034\) & \(-0.081\)\sym{*} & \(0.027\) \\
 & (0.051) & (0.049) & (0.078) & (0.052) & (0.064) & (0.047) & (0.057) \\
\midrule
\multicolumn{8}{l}{\textit{Firm size tercile \(\times\) quarter}} \\
Quarter 4 after contract review & \(-0.077\) & \(0.037\) & \(-0.139\)\sym{*} & \(-0.094\)\sym{*} & \(-0.048\) & \(-0.090\)\sym{*} & \(0.025\) \\
 & (0.051) & (0.046) & (0.075) & (0.054) & (0.062) & (0.046) & (0.056) \\
\midrule
\multicolumn{8}{l}{\textit{Industry \(\times\) quarter}} \\
Quarter 4 after contract review & \(-0.072\) & \(0.026\) & \(-0.116\) & \(-0.106\)\sym{**} & \(-0.053\) & \(-0.087\)\sym{*} & \(0.026\) \\
 & (0.050) & (0.050) & (0.080) & (0.052) & (0.065) & (0.046) & (0.057) \\
\midrule
\multicolumn{8}{l}{\textit{Firm size tercile \(\times\) quarter and industry \(\times\) quarter}} \\
Quarter 4 after contract review & \(-0.072\) & \(0.031\) & \(-0.148\)\sym{**} & \(-0.092\)\sym{*} & \(-0.058\) & \(-0.093\)\sym{**} & \(0.026\) \\
 & (0.049) & (0.048) & (0.074) & (0.054) & (0.063) & (0.044) & (0.056) \\
\bottomrule
\end{tabular}
}

\end{tablebox}

\begin{minipage}{\textwidth}
\vspace{3pt}
\tablenotes{The table reports reduced-form estimates from the imputation estimator of \citet{borusyak_revisiting_2024} on the quarterly panel. The baseline includes firm and calendar-quarter fixed effects; subsequent panels add quarter effects by pre-launch size tercile, by industry group, or by both. Columns (1)--(2) are log technical hires and separations. Columns (3)--(5) cover software-engineering hire spells and require mean pre-launch employment of at least one in both seniority groups; column (3) is the junior-to-experienced log hiring ratio. Columns (6)--(7) are log hire spells in the four more and four less code-intensive roles. Standard errors, clustered by firm, are in parentheses. \sym{*}~$p<0.10$, \sym{**}~$p<0.05$, \sym{***}~$p<0.01$.}
\end{minipage}
\end{table}

\clearpage
\section{Contribution of GitHub Copilot to the Decline in Technical Hiring}\label{asec:decomposition}

This section details the counterfactual calculations reported in Section~\ref{sec:decomposition} and Figure~\ref{fig:decomposition}. The sample consists of firms whose first scheduled post-launch contract review fell in 2023. We measure baseline pre-launch hiring $H_{i0}$ as firm $i$'s mean quarterly technical hires over 2022 and post-period hiring $H_{i1}$ as mean quarterly technical hires over 2024, yielding an observed decline of $\Delta = 16.1$ hires per firm-quarter.

To compute $\Delta_{\text{GitHub Copilot}}$, we evaluate forgone hiring firm-quarter by firm-quarter across event-time horizons $\ell = 0, \dots, 4$ relative to each firm's scheduled contract review quarter (the event clock of Equation~\ref{eq:reduced_form}) and take an average. In the review quarter itself ($\ell = 0$) we set $\widehat{\rho}_{0} = \widehat{\theta}_{0} = 0$, in the point estimates and in every bootstrap draw, because the first stage is small in that quarter and the Wald ratio is unstable there; the average runs over five quarters. Let $H_{i\ell}$ denote firm $i$'s observed technical hires $\ell$ quarters after its scheduled review and $D_{i\ell}$ its adoption indicator in that quarter. The three benchmarks in Figure~\ref{fig:decomposition} are computed as:
\begin{align}
\widehat{\Delta}^{RF}_{\text{GitHub Copilot}} &= \frac{1}{5}\sum_{\ell=0}^{4}\frac{1}{N}\sum_{i} H_{i\ell}\big(e^{-\widehat{\rho}_{\ell}}-1\big) \label{eq:decomp_rf}\\
\widehat{\Delta}^{C}_{\text{GitHub Copilot}} &= \frac{1}{5}\sum_{\ell=0}^{4}\frac{1}{N}\sum_{i} D_{i\ell}\,H_{i\ell}\big(e^{-\widehat{\theta}_{\ell}}-1\big)\,\frac{\widehat{\pi}_{\ell}}{\widehat{P}(D_{i\ell}=1)} \label{eq:decomp_compliers}\\
\widehat{\Delta}^{A}_{\text{GitHub Copilot}} &= \frac{1}{5}\sum_{\ell=0}^{4}\frac{1}{N}\sum_{i} D_{i\ell}\,H_{i\ell}\big(e^{-\widehat{\theta}_{k_{i\ell}}}-1\big) \label{eq:decomp_all}
\end{align}
In Equation~\eqref{eq:decomp_rf}, the horizon-specific reduced-form coefficient $\widehat{\rho}_{\ell}$ is applied to all firms. In Equation~\eqref{eq:decomp_compliers}, the horizon-specific Wald estimate $\hat{\theta}_{\ell} = \hat{\rho}_{\ell}/\hat{\pi}_{\ell}$ is applied to adopting firms and multiplied by the complier share of adopters at horizon $\ell$, $\widehat{\pi}_{\ell}/\widehat{P}(D_{i\ell}=1)$, where $\widehat{\pi}_{\ell}$ is the first-stage effect and $\widehat{P}(D_{i\ell}=1)$ is the sample adoption rate at horizon $\ell$. Because adopting firms are larger on average than non-adopters, reweighting the complier effect by adopter hiring levels yields a larger reduction than the unweighted reduced form (3.1 versus 1.7 hires per firm-quarter). In Equation~\eqref{eq:decomp_all}, the Wald estimate is applied to all adopting firms without scaling down by the complier share. This assumes that an always taker $k$ quarters after its own adoption experiences the same proportional hiring effect as a complier $k$ quarters after its review.

\end{appendix}

\end{document}